\documentclass[11pt]{article}
\usepackage{amssymb}
\def\one{1\hskip -.37em 1}     
\def\proj{\mathbb E}
\def\R{\mathbb R}
\def\Z{\mathbb Z}
\begin{document}
\begin{titlepage}
\begin{centering}
 
{\ }\vspace{2cm}
 
{\Large\bf The Quantum Geometer's Universe:}

\vspace{5pt}

{\Large\bf Particles, Interactions and Topology}

\vspace{2cm}

Jan Govaerts\\
\vspace{1.0cm}
{\em Institute of Nuclear Physics, Catholic University of Louvain}\\
{\em 2, Chemin du Cyclotron, B-1348 Louvain-la-Neuve, Belgium}\\
{\tt jan.govaerts@fynu.ucl.ac.be}

\vspace{2cm}

\begin{abstract}

\noindent
With the two most profound conceptual revolutions of XX$^{\rm th}$ century
physics, quantum mechanics and relativity, which have culminated
into relativistic spacetime geometry and quantum gauge field theory
as the principles for gravity and the three other known fundamental
interactions, the physicist of the XXI$^{\rm st}$ century has inherited 
an unfinished symphony: the unification of the quantum and the
continuum. As an invitation to tomorrow's quantum geometers who
must design the new rulers by which to size up the Universe at those
scales where the smallest meets the largest, these lectures review the 
basic principles of today's conceptual framework, and highlight by way 
of simple examples the interplay that presently exists between the
quantum world of particle interactions and the classical world of
geometry and topology.

\end{abstract}

\vspace{35pt}
 
To be published in the Proceedings of the\\
Second International Workshop on Contemporary Problems in Mathematical
Physics,\\
Institut de Math\'ematiques et de Sciences Physiques (IMSP), Universit\'e
d'Abomey-Calavi,\\
Cotonou, Republic of Benin\\
October 28$^{\rm st}$ - November 2$^{\rm th}$, 2001

\end{centering} 

\vspace{100pt}


\end{titlepage}

\section{Introduction}
\label{Sect1}

It is often said that the profound conceptual revolutions of
XX$^{\rm th}$ century physics may be ascribed to three
fundamental physical constants, namely Newton's constant $G_N$
characteristic of the gravitational interaction, light's velocity
in vacuum $c$ displaying the relativistic character of physical
reality, and Planck's constant $\hbar=h/2\pi$ as the hallmark
for the quantum character of the physical universe. All of these
constants have incessantly been used much like light beacons with which
to probe the as yet unexplored territories beyond the known
physical laws of our material world, grasping for this
ever unfulfilled dream of the ultimate unification of all of matter,
radiation and their interactions.

Each of these three constants on its own has led to its separate
conceptual revolution, even beyond the confines of the scientific
methods of physics, in ways that shall not be recalled here. However,
when considered in combination, these constants imply still further profound
conceptual revisions in our understanding of the physical world, which 
themselves stand out as the ge\-nuine unfinished revolutions of XX$^{\rm th}$ 
century physics. Indeed, even though the combinations of $G_N$ with $c$ on the 
one hand, and of $c$ and $\hbar$ on the other hand, have each led to a profound
new vision onto the material universe through the physicist's eye,
the formulation of a conceptual framework in which all three constants
play an equally important role is the wide open problem that confronts
physics in this XXI$^{\rm st}$ century.

As is well known, the marriage of $G_N$ and $c$ leads to a curved
spacetime whose geometry is dynamical and is governed by the energy-matter
distribution within it, a framework within which the gravitational 
interaction is the physical manifestation of any curvature in space and 
in spacetime. The most fascinating offsprings of this union are undoubtedly,
on the one hand, the cosmological theory of the history of our universe from 
its birth to its ultimate demise if ever, and on the other hand,
the prediction for regions of spacetime to be so much curled up by their
energy-matter content that even light can no longer escape from such
black holes. For instance, the value
\begin{equation}
r_0=2\frac{G_NM}{c^2}
\end{equation}
for the horizon of a neutral nonrotating black hole of mass $M$
displays the combined contribution of gravity and relativity.
These examples are but two specific outcomes of classical general relativity,
a relativistic invariant theory of gravity whose construction is based
on a simple geometrical thus physical principle: the description
of physical processes should be independent of the local spacetime
observer, namely, it should be independent of the choice
of local spacetime coordinate parametrization. The theory should be
invariant under arbitrary local coordinate transformations in
spacetime.\footnote{Einstein's theory of general relativity has furthermore 
inscribed into it the equivalence principle between inertial and 
gravitational mass.}
In other words, a gauge invariance principle is at work, leading
to a description of the gravitational interaction based on
a simple but powerful symmetry and thus geometry principle.

On the other hand, the marriage of $c$ and $\hbar$ leads naturally to 
the quantum field theory description of the elementary particles and
their interactions, at the most intimate presently accessible scales of 
space and energy, a fact made manifest by the value for their product,
\begin{equation}
\hbar\,c\simeq 197\ {\rm MeV}\cdot{\rm fm}\ .
\end{equation}
In fact, one offspring of this second union is the unification of
matter and radiation, namely of particles with their corpuscular propagating
properties and fields with their wavelike propagating properties. Particles,
cha\-rac\-te\-ri\-zed through their energy, momentum and spin values 
in correspondence
with the Poincar\'e symmetries of Minkowski spacetime in the absence of 
gravity, are nothing but the relativistic energy-momentum quanta of a field,
thereby implying a tremendous economy in the description of the
physical universe, accounting for instance at once in terms of a single
field filling all of spacetime for the indistinguishability of identical 
particles and their statistics. Furthermore, quantum relativistic 
interactions are then understood simply as couplings between the various 
quantum fields locally in spacetime, which translate
in terms of particles as diverse exchanges of the associated quanta.
Such a picture lends itself most ideally to a pertubative understanding
of the fundamental interactions, which has proved to be so powerful
beginning with quantum elec\-tro\-dy\-na\-mics, up to the modern
SU(3)$_C\times$SU(2)$_L\times$U(1)$_Y$ Standard Model of the
strong and electroweak interactions. Such a perturbative representation
of processes requires a renormalization procedure of the basic field
parameters---their normalizations, masses and couplings---,
and one has had to learn how to identify theories for which this
renormalization programme is feasable. In the course of time, a general
class of renormalizable field theories has been identified, all falling
again under the general spell of the gauge symmetry principle as did the
gravitational interaction! 

Even though the physical meaning ascribed to the renormalizability criterion 
has evolved in such a manner that these theories are nowadays viewed rather
as effective theories for some as yet unknown more fundamental description
becoming manifest and relevant at still higher energies,\cite{Wein1}
the fact remains that the gauge symmetry principle is again at
work at the most intimate level of the unification of the relativistic
quantum. But this time, this invariance under local transformations in
spacetime applies to some ``internal" space of degrees of freedom,
that fields and their quanta carry along and which are made physically manifest
through the different charges and quantum numbers that particles possess.
Hence, through countless experiments performed at ever increasing
energies and with ever increased technical sophistication, three
generations of quarks and leptons, the basic building blocks of matter, 
each such generation being comprised of two quarks, one charged lepton
and its associated neutrino, have been identified, and their reality
inscribed into the construction of the Standard Model. All interactions
among these six quarks and six leptons are governed by the gauge
symmetry principle, with SU(3)$_C\times$SU(2)$_L\times$U(1)$_Y$ symmetries
acting within internal space and independently, though in a continuous
fashion, at each point of spacetime. This local realization of the
symmetry requires the existence of gauge bosons, as the carriers of the
symmetry and thus of the interactions from one spacetime point to the next.
There are thus eight gluons for the strong interaction, the charged and neutral
massive electroweak gauge bosons $W^\pm$ and $Z_0$ for the charged and
neutral current weak interactions, and finally the photon for the
electromagnetic interaction. Only one member of the Standard Model family
has yet to be discovered experimentally, namely the so-called higgs particle
which should be responsible for a mechanism at the origin of the masses for
all quarks, leptons and massive gauge bosons. The higgs hunt is on at the most
powerful particle accelerators in the world, the last missing offspring
of the union of $c$ and $\hbar$.

Given the fundamental role played by symmetries, hence also geometry, in the
unifications of fundamental physics concepts achieved throughout the last
century, it is fair to characterize XX$^{\rm th}$ century
physics as the reign supreme of the symmetry principle, this principle being 
pushed into its most extreme realizations possible through the gauge symmetry
principle. This includes the possibility of supersymmetry, a symmetry that
relates bosonic and fermionic particles which, when rendered local in 
spacetime, leads to theories of supergravity that must necessarily include
a quantum gravitational sector. But it also appears that this symmetry 
principle has finally unveiled all its hidden physical secrets in the 
embodiement it has acquired within a field theory description of the 
universe, of its matter content and of its fundamental interactions. 
Even though the symmetry principle seems to have yielded all its potential, 
it proves not to be potent enough to bring order to a 
{\sl m\'enage \`a trois\/} in which all three fundamental constants 
$G_N$, $c$ and $\hbar$ would be living peacefully and happily together on 
equal terms, to bear many news fruits of their ultimate union.
As is well known, there does not yet exist a commonly accepted
theoretical formulation for a quantum theory of relativistic gravity
which would also include the other fundamental interactions and their
matter fields, all consistently expressed within a quantum framework.

Looking back at the brief and superficial highlights recalled above,
one realizes that the nonquantum relativistic description available for
the gra\-vi\-ta\-tional interaction is in fact the ideal realm of the
``relativistic continuum" reigning supreme, the utmost physical application 
as of today of the notion of differentiable structures in geometry. Likewise, 
the other component of the same story, namely the relativistic quantum field 
theory description of the elementary particles and their other fundamental 
interactions, is in fact the ideal realm of the ``relativistic quantum" 
reigning supreme, the utmost physical outcome of the ideas of quantization 
and its associated abstract algebraic structures. The fundamental problem 
that XXI$^{\rm st}$ century physics is to confront is that of the final 
marriage of the ``continuum" and the ``quantum", namely of identifying 
a mathematical formulation of what is referred to as ``quantum geometry", 
the new conceptualization of what the geometry of spacetime ought to be when
explored at the most extreme and smallest scales.

In terms of the three fundamental constants $G_N$, $c$ and $\hbar$, it is well
known how the quantum regime for relativistic gravity is characterized
by Planck's mass, length and time scales,
\begin{displaymath}
M_{\rm Pl}=\sqrt{\frac{\hbar c}{G_N}}\simeq 10^{19}\ {\rm GeV}/c^2\ \ ,\ \ 
L_{\rm Pl}=\frac{\hbar c}{M_{\rm Pl}c^2}\simeq 2\times 10^{-35}\ {\rm m}
\ \ ,\ \ 
\end{displaymath}
\begin{equation}
\tau_{\rm Pl}=\frac{L_{\rm Pl}}{c}\simeq 6\times 10^{-44}\ {\rm s}\ .
\end{equation}
Even though these values lie way beyond the reach of present day
accelerators, as well as of present day theories, processes at such scales 
must have taken place in the early universe, while from the conceptual point 
of view, the fundamental conflit between the classical relativistic realm of
the ``continuum" for gravity with the quantum relativistic realm
of the ``quantum" for particles and their other interactions, cries out
to the XXI$^{\rm st}$ physicist for a new conceptual revolution
that ought to resolve this basic mutual inconsistency of present day
physics principles. From that point of view, XXI$^{\rm st}$ century
physics will be the search for the Quantum Geometry Principle,
the inherited unfinished physics symphony of the XX$^{\rm th}$ century
composed so far according to the rules of the Symmetry Principle.

With the advent of M-theory,\cite{Pol} the nonperturbative embodiement of
superstring theories,\cite{GSW} and possibly also with the loop gravity 
programme,\cite{Loop}
we are most probably already getting the first glimpses of this
quantum geometry wai\-ting to be discovered by tomorrow's bright
young minds. Such a pursuit in search of the possible ultimate
unification, all at the same time, of matter and its interactions, and of 
geometry and the quantum, belongs to the best of scientific traditions finding 
its roots back in the earliest days of the human intellectual adventure. 
It should only be just that within all peoples of the world, as much
from developing as from developed countries, those whose
calling lies towards such an avenue should find an environment
within which to contribute on equal terms to this ultimate understanding
of our physical universe and its history. A workshop of this type
is an opportunity to highlight some of the issues surrounding this
unfulfilled quest, and hopefully entice bright new minds to dedicate
themselves to this adventure at the frontiers of physical concepts. 
The education to critical and scientific thinking that such a research 
activity requires can only benefit any society within which it is pursued, 
both in its human and intellectual aspirations as well as in its 
edu\-ca\-tio\-nal,
technological and economic development, bringing man always a little closer 
to the stars, the eternal yearning of his soul. Countless examples over human
history bear witness to this fact, and many of us today benefit in so many
ways from the fruits of this unswaying quest at the most abstract level as it 
has been pursued over centuries past.

These lecture notes do not, of course, have any pretence to outline what
quantum geometry ought to be, which, after all, is the XXI$^{\rm st}$ century 
quantum geometer's task! Rather, these lectures wish to present sort of a 
guided tour of the general principles of symmetry and quantum physics that 
have led to the relativistic quantum field theory description of the
elementary particles and their fundamental interactions, aiming at the
end towards illustrations of the fact that beyond the gauge symmetry
principle which seems to go\-vern all interactions, when it comes to
geometry---namely the ``continuum" and gravity---and the ``quantum",
topology is also called to play a vital role. In fact, one is very much
led to suggest that the problem of quantum gra\-vi\-ty should find a 
resolution only when considered together with all the other quantum
matter and interacting fields, while pure quantum gravity
is obli\-vious actually to any geometry, and would be governed only by the
rules of quantum topology. Indeed, this is the programme that was 
launched\cite{Wit1,Wit2} with the dis\-co\-ve\-ry of topological quantum 
field theories.\cite{Wit1,Wit2,TQFT}
Finally, these notes concentrate on the quantum field theory side of
the above story, assuming that the reader is most familiar already with
the views of classical continuum geometry as applied within the physical
context of the gravitational interaction and general relativity.
This is thus the spirit with which these notes are offered to the aspiring 
quantum geometers of the XXI$^{\rm st}$ century who are attending this 
Workshop.

Contents are organized as follows. Section~\ref{Sect2} discusses the
general rules of abstract canonical quantization, based on the
Hamiltonian formulation of a given dynamical system. These rules
are then applied to relativistic field theories in Section~\ref{Sect3},
to establish that such quantized theories provide a natural description
of quantum relativistic particles in Minkowski spacetime. Section~\ref{Sect4}
introduces then to interacting quantum field theories and, as a general
class of renormalizable theories in four dimensions, to general Yang-Mills
theories, possibly subjected to the Higgs mechanism of spontaneous
symmetry breaking. This discussion thus also serves as a motivation for
Section~\ref{Sect5} which addresses the general problem of the quantization
of systems subjected to constraints in phase space, which include any
gauge invariant system, following Dirac's general analysis of this 
issue.\cite{Dirac}
Rather than introducing then the ge\-ne\-ral methods of BRST quantization,
the recent and most efficient approach towards the quantization of constrained
systems based on the physical projector\cite{Klaud1} is also discussed. 
As an example
of its possible use, the quantization of 2+1-dimensional Chern-Simons
theory is briefly described in Section~\ref{Sect6}, which in fact is
one of the simplest examples of a topological quantum field theory.
Finally, Sections~\ref{Sect7} and \ref{Sect8} introduce to bosonic string
theory and its toroidal compac\-ti\-fication. These last three sections serve
as first witnesses to the necessity to develop a new mathematical
framework for quantum theories of gravity, whether they include
matter degrees of freedom or not, that should define the sought-for ``quantum
geometry" of the fundamental unification. Finally, further
comments are presented in the Conclusions.

Our conventions will be stated where appropriate. Notice also that
all the discussion will be confined to bosonic degrees of freedom
only, but that similar developments exist of course for systems
combining both bosonic and fermionic degrees of freedom. Suggestions for 
some exercises are also provided, some of which could in fact become 
PhD research topics on their own. Finally, no attempt has been made 
at providing an exhaustive bibliography, for which we apologize to anyone who
might feel her/his work is being overlooked. Rather, we hope that
references given would suffice to quickly identify further relevant
sources to any particular topic of interest.

\section{Abstract Canonical Quantization}
\label{Sect2}

This section briefly reviews\cite{JG1} the general rules of abstract canonical
quantization, starting from some action principle defining the actual
dynamics of a given physical system. This discussion is elaborated upon
much further in Prof. S.T. Ali's lectures in these Proceedings,
dedicated to the general pro\-blem of quantization. Prof. J.R. Klauder's
contribution is also directly related
to some of the issues addressed in this section. Furthermore, to
illustrate ex\-pli\-cit\-ly the general discussion through the simplest of
nontri\-vial examples, most points of relevance are also discussed in the
context of the one-dimensional harmonic oscillator.

\subsection{Dynamics}
\label{Subsect2.1}

Let us consider a given physical system of $N$ real degrees of freedom 
$q^n(t)$ $(n=1,2,\cdots,N)$, whose dynamics derives from the variational 
principle based on an action which is local in time and of the form
\begin{equation}
S[q^n]=\int dt\,L\left(q^n,\dot{q}^n\right)\ ,
\end{equation}
defined in terms of the Lagrange function $L(q^n,\dot{q}^n)$, where
$\dot{q}^n=dq^n/dt$. In the case of the one-dimensional harmonic oscillator, 
one simply has $N=1$ with
\begin{equation}
L(q,\dot{q})=\frac{1}{2}m\dot{q}^2-\frac{1}{2}m\omega^2q^2\ \ ,\ \ 
\omega,m>0\ .
\label{eq:LHO}
\end{equation}

It is well known how the variational principle leads to the following
Euler-Lagrange equations of motion,
\begin{equation}
\frac{d}{dt}\frac{\partial L}{\partial\dot{q}^n}-
\frac{\partial L}{\partial q^n}=0\ \ \ ,\ \ \ n=1,2,\cdots,N\ .
\end{equation}
In general, any given variation of the action also includes surface term
contributions---namely, total time derivative contributions for a
mechanical system---which one may also want to set to zero in order to
enforce the variational principle in a strong sense, namely such that the
action is an exact extremum for classical solutions. Depen\-ding on one's 
choice of boundary conditions, as motivated by specific physics 
considerations in order to solve the above equations of motion, such a 
requirement may be too restrictive. Whatever the case, a complete set of 
boundary conditions must be specified in order to determine a unique 
solution to the equations of motion of the system.

In the case of the simple harmonic oscillator, the Euler-Lagrange equation
of motion reads
\begin{equation}
m\ddot{q}=-m\omega^2\,q\ ,
\end{equation}
whose general solution may be expressed as follows for later
convenience,
\begin{equation}
q(t)=\frac{1}{\sqrt{2m\omega}}
\left[\alpha\,e^{-i\omega(t-t_0)}+\alpha^*\,e^{i\omega(t-t_0)}\right]\ .
\end{equation}
Here, $\alpha$ and $\alpha^*$ are complex integration constants (complex
conjugates of one another) which may be expressed in terms of whatever
choice of boundary conditions that might be contemplated, and $t_0$
is some specific time reference possibly associated to the choice of
boundary conditions.

\subsection{Hamiltonian formulation}
\label{Subsect2.2}

Since canonical quantization proceeds from the Hamiltonian formulation
of a dynamical system, let us now introduce the phase space description
of a general system as represented in Section~\ref{Subsect2.1}. Phase space
is thus spanned by the functions $q^n(t)$ and their conjugate momenta
$p_n(t)$ defined as
\begin{equation}
p_n=\frac{\partial L}{\partial\dot{q}^n}\ .
\end{equation}
As a matter of fact, phase space provides, within the Hamiltonian
formulation, the space of states which are accessible to the system
throughout its history. In addition to its parametrization in terms of
the local coordinates $(q^n,p_n)$, phase space also comes equipped with
a geometric symplectic structure which is related to the Poisson bracket
algebra of observables or functions defined over phase space. This Poisson
bracket structure is defined at a fixed reference time $t=t_0$, and Poisson
brackets are to be evaluated at equal time, namely for two observables
defined at the same instant in time, $t=t_0$. Given the algebraic 
properties of the Poisson bracket structure, its values are derived from 
those for the elementary phase space degrees of freedom $(q^n,p_n)$ of the 
system, given by
\begin{equation}
\{q^n(t_0),p_m(t_0)\}=\delta^n_m .
\label{eq:PoissonB}
\end{equation}
More generally, the Poisson bracket of any two observables $F(q^n,p_n)$ and
$G(q^n,p_n)$ is given by\footnote{Throughout these notes, implicit summation
is understood for repeated indices.}
\begin{equation}
\{F,G\}=\frac{\partial F}{\partial q^n}\frac{\partial G}{\partial p_n}\ -\
\frac{\partial F}{\partial p_n}\frac{\partial G}{\partial q^n}\ .
\end{equation}
As is well known, Poisson brackets obey a series of important algebraic 
pro\-per\-ties by which they are characterized, among which their antisymmetry
as well as the Jacobi identity. Furthermore, according to Darboux's theorem,
there always exists a local choice of phase space parametrization for
which the Poisson brackets take the canonical form in (\ref{eq:PoissonB}),
in which the associated coordinates are also referred to as being canonical.

Finally, besides this kinematics information encoding specific
properties of the system, its dynamics is generated through the Poisson
bracket from a specific observable, namely the canonical Hamiltonian defined
through the Legendre transform of the Lagrange function,
\begin{equation}
H_0(q^n,p_n)=\dot{q}^np_n-L(q^n,\dot{q}^n)\ .
\end{equation}
Usually at this point, it is said that this definition applies to
those systems for which the relations $p_n(q^n,\dot{q}^n)$ defining
the conjugate momenta may all be inverted in a unique fashion in
terms of the generalized velocities $\dot{q}^n(q^n,p_n)$, namely
under the condition of a regular Hessian for $L$,
\begin{equation}
{\rm det}\,\frac{\partial^2 L}{\partial\dot{q}^{n_1}
\partial\dot{q}^{n_2}}\ne 0\ .
\label{eq:regular}
\end{equation}
Systems for which this is feasible are called ``regular", or rather
described by a ``regular Lagrangian", and otherwise ``singular" in the
case of a singular Hessian.\footnote{Singular systems are the topic of 
Section~\ref{Sect4}.} However, whether this latter condition is met or 
not does not affect the fact that the quantity $H_0$ introduced above is 
always\cite{JG1} a function of $q^n$ and $p_n$, namely a function defined over 
phase space. Actually, the difference between a regular and a singular 
Lagrangian is that in the latter case, there exist specific constraints 
among the phase space degrees of freedom, which imply that time evolution 
of the system may be generated by a Hamiltonian $H(q^n,p_n)$ more general
than simply the canonical one $H_0(q^n,p_n)$. In the present section,
we assume the system to be regular, and thus to be free of any
further constraint on the phase space degrees of freedom.

Consequently, time evolution of any phase space observable $F(q^n,p_n;t)$
is determined from the Hamiltonian equation of motion
\begin{equation}
\dot{F}=\frac{dF}{dt}=\frac{\partial F}{\partial t}+\{F,H_0\}\ ,
\end{equation}
while in particular for the phase space degrees of freedom one obtains
the first-order Hamiltonian equations of motion
\begin{equation}
\dot{q}^n=\frac{\partial H_0}{\partial p_n}\ \ \ ,\ \ \ 
\dot{p}_n=-\frac{\partial H_0}{\partial q^n}\ .
\end{equation}
When solving for the conjugate momenta $p_n$ in terms of the velocities
$\dot{q}^n$ through the first set of these equations, and then substituting
these expressions for $p_n$ in the second set of equations, one recovers the
original Euler-Lagrange equations of motion of the Lagrangian formulation
of the same dynamical system. Note that the Hamiltonian equations
of motion also follow through the variational principle from a first-order
phase space action given by
\begin{equation}
S[q^n,p_n]=\int dt\left[\dot{q}^np_n-H_0(q^n,p_n)\right]\ ,
\label{eq:firstorderaction}
\end{equation}
or some other expression differing from the present one by a total
time derivative of an arbitrary function of phase space.

In the particular case of the simple one-dimensional harmonic oscillator 
described by the Lagrange function (\ref{eq:LHO}), one finds for the 
conjugate momentum
\begin{equation}
p=m\dot{q}\ ,
\end{equation}
with the canonical Poisson bracket structure
\begin{equation}
\{q(t_0),p(t_0)\}=1\ .
\end{equation}
The canonical Hamiltonian reads
\begin{equation}
H_0=\frac{p^2}{2m}+\frac{1}{2}m\omega^2q^2\ ,
\end{equation}
leading to the Hamiltonian equations of motion
\begin{equation}
\dot{q}=\frac{1}{m}p\ \ \ ,\ \ \ 
\dot{p}=-m\omega^2q\ .
\end{equation}
Their general solution may again be given in the general parametrization
used previously,
\begin{equation}
\begin{array}{r c l}
q(t)&=&\frac{1}{\sqrt{2m\omega}}
\left[\alpha\,e^{-i\omega(t-t_0)}+\alpha^*\,e^{i\omega(t-t_0)}\right]\ ,\\
 & & \\
p(t)&=&-\frac{im\omega}{\sqrt{2m\omega}}
\left[\alpha\,e^{-i\omega(t-t_0)}-\alpha^*\,e^{i\omega(t-t_0)}\right]\ .
\end{array}
\label{eq:phasespacesolution}
\end{equation}

From this particular example, we may also extend a general remark.
Clearly, there is a one-to-one correspondence between specific solutions
to the Hamiltonian equations of motion and the associated integration constants,
for instance in the present case between the solutions for $q(t)$ and $p(t)$
and the complex integration constant $\alpha$. Hence, rather than
viewing the Poisson bracket structure as being defined over phase space, 
namely the set of all functions $(q^n(t),p_n(t))$, one may equivalently view 
the Poisson bracket structure as being defined on the space of integration
constants for the associated equations of motion. This alternative but
equivalent point of view may in fact be quite relevant and physically
meaningfull. Indeed, as we shall see in the case of quantum field theories,
energy-momentum quanta that realize the quantum particle content of the
theory correspond precisely to the operators associated to the field
integration constants, whose commutation relations are nothing but the
operator realization of the corresponding classical Poisson bracket
structure defined on the space of integration constants. In the simpler
case of the harmonic oscillator, the same point of view implies the
following nontrivial Poisson bracket for the complex integration constants
$\alpha$ and $\alpha^*$,
\begin{equation}
\{\alpha,\alpha^*\}=-i\ ,
\end{equation}
while the canonical Hamiltonian then acquires the expression,
\begin{equation}
H_0=\frac{1}{2}\omega\left[\alpha^*\alpha+\alpha\alpha^*\right]\ ,
\end{equation}
an expression in which care has been exercised not to commute quantities
which at the quantum level will correspond to noncommuting operators.
Consequently, the Hamiltonian equations of motion for these integration
constants, or rather now variables, are
\begin{equation}
\dot{\alpha}=-i\omega\,\alpha\ \ \ ,\ \ \ 
\dot{\alpha}^*=i\omega\,\alpha^*\ ,
\end{equation}
with the general solution
\begin{equation}
\alpha(t)=\alpha_0e^{-i\omega(t-t_0)}\ \ \ ,\ \ \ 
\alpha^*(t)=\alpha_0^*e^{i\omega(t-t_0)}\ ,
\end{equation}
which thus corresponds to the following representation for the
general phase space solution in (\ref{eq:phasespacesolution}),
\begin{equation}
q(t)=\frac{1}{\sqrt{2m\omega}}\left[\alpha(t)+\alpha^*(t)\right]\ \ \ ,\ \ \ 
p(t)=-\frac{im\omega}{\sqrt{2m\omega}}\left[\alpha(t)-\alpha^*(t)\right]\ .
\label{eq:phasespacesolution2}
\end{equation}
Here, the initial values $\alpha_0$ and $\alpha^*_0$ at $t=t_0$ stand for the 
values $\alpha$ and $\alpha^*$ of the integration constants as they
appear in the general phase space solution (\ref{eq:phasespacesolution}).
At the quantum level, this correspondence between the solutions to the
equations of motion in phase space and the solutions on the space of 
integration constants will translate into the Schr\"odinger and
Heisenberg pictures for the quantized system, in which in the first
case states are time dependent while operators are time independent, 
and vice-versa in the second case.

\subsection{Canonical quantization}
\label{Subsect2.3}

Canonical quantization simply proceeds through the correspondence
principle. In the same manner that the classical Hamiltonian
formulation of any dynamical system is characterized by three structures,
namely the space of states or phase space, the kinematical
structure of that space as embodied algebraically through its Poisson 
bracket structure, and finally its dynamics generated by the Hamiltonian,
likewise any quantum system is cha\-rac\-te\-ri\-zed by three such structures,
namely the space of quantum states or Hilbert space, equipped with
a series of algebraic structures providing a linear re\-pre\-sen\-ta\-tion 
space of the algebraic structure of commutation relations which are in direct
correspondence with the classical Poisson brackets, and finally a quantum 
Hamiltonian which generates time evolution of states through the 
Schr\"odinger equation.

More specifically, the space of quantum states must be some complex
vector space equipped with an hermitean inner product denoted using
Dirac's bra-ket notation as $<\varphi|\psi>$ for two states
$|\varphi>$ and $|\psi>$. In the best of cases, this complex space
ought to be a Hilbert space in the strict mathematical sense, but
often it is difficult to meet that stringent requirement, and one
has to extend somewhat the relevant complex vector space. Such issues
are addressed in Prof.~S.T. Ali's lectures in these Proceedings.

Furthermore, this ``Hilbert" space must provide a linear representation
space of the quantum operator algebra of observables of the quantized
system. Ideally, what one would wish is that associated to any two
classical observables $A$ and $B$ defined over phase space, there should
exist two linear quantum operators $\hat{A}$ and $\hat{B}$ whose
commutator algebra is in correspondence with the Poisson bracket of the
classical observables through the rule
\begin{equation}
[\hat{A},\hat{B}]=i\hbar\,\hat{C}\ ,
\end{equation}
where $\hat{C}$ stands for the operator to be associated to
the result of the classical Poisson bracket $\{A,B\}=C$.
It is understood that this correspondence rule is to be considered at
a fixed reference time $t_0$, which has not been displayed in the
above relation. Furthermore, the choice of hermitean inner product
ought to be such that the operators $\hat{A}$ and $\hat{B}$ associated
to classical observables which are real under complex conjugation,
$A^*=A$ and $B^*=B$, are themselves self-adjoint operators,
$\hat{A}^\dagger=\hat{A}$ and $\hat{B}^\dagger=\hat{B}$.

These requirements, which actually define what is precisely
meant by ``quantization", are by no means easily met, as is discussed
in detail in Prof.~S.T. Ali's lectures. To understand part of the
difficulty, let us consider this requirement already for the
elementary phase space degrees of freedom $(q^n,p_n)$, which ought
thus to correspond to operators $(\hat{q}^n,\hat{p}_n)$ obeying the
Heisenberg algebra,
\begin{equation}[\,\hat{q}^n(t_0),\hat{p}_m(t_0)\,]=i\hbar\,\delta^n_m\ ,
\end{equation}
as well as the following hermiticity properties on the space of quantum states,
\begin{equation}
\left(\hat{q}^n(t_0)\right)^\dagger=\hat{q}^n(t_0)\ \ ,\ \ 
\left(\hat{p}_n(t_0)\right)^\dagger=\hat{p}_n(t_0)\ .
\end{equation}
Since all other observables of the system are defined as
composite operators built from these $\hat{q}^n$'s and $\hat{p}_n$'s,
clearly the fact that they no longer commute at the quantum level
implies that a specific choice has to be made as to the order in which
they are multiplied with one another. {\sl A priori\/} the only restriction
that should apply to this choice of ``operator ordering" is that a given
composite observable real under complex conjugation
should be associated to a self-adjoint operator. For some
particular observables related to symmetries that a given system may
possess, namely the associated Noether charges,
further restrictions would follow from the requirement that
these operators still generate the corresponding symmetry for the
quantized system. However, it is far from trivial that such a construction
is at all possible, and the fact of the matter is that it is indeed
problematic, leading to ``the problem of quantization" as discussed
in Prof. S.T. Ali's lectures. Not all classical observables may be
put into correspondence with a self-adjoint quantum operator.
When this is impossible for a symmetry generator while at the same time
preserving the symmetry algebra now at the quantum level, one says that 
the symmetry has become anomalous or that it suffers a quantum anomaly, 
even though a better term would be ``quantum symmetry breaking" given that 
the symmetry is explicitly broken at the quantum level, being no longer 
a symmetry of the quantized system. 

In practical terms, given a classical 
system, there may thus exist more than one unitarily inequivalent quantum 
system that corresponds to it, since there may exist more than one choice
of operator ordering consistent with the above correspondence rules
for a certain subset of all classical observables\cite{Klaud2}. 
It thus seems more
appropriate to advocate the point of view that one has to define some
quantum system, having in the back of one's mind some classical system,
in terms of some algebra of self-adjoint elementary and composite
operators, which in the limit $\hbar\rightarrow 0$ is in correspondence
with some classical system. After all, the physical world is quantum 
mechanical, rather than being that of classical mechanics whether 
relativistic or not.
The fact that a given classical system may correspond to more than one 
quantum system then becomes an issue to be resolved only through experiment. 
For instance, there exists a discrete infinity of rotationally invariant 
quantum systems labelled by the SU(2) spin value, which all correspond to 
the same classical system of a nonrelativistic point-particle invariant 
under spatial rotations. Only an experiment can determine which of these
quantum representations is in fact realized in a specific physical system, 
whether the electron or a heavy nucleus in some excited state, for example.

Finally, the last structure which defines the abstract canonical quantization
of a classical system is the Hamiltonian operator $\hat{H}_0$, which should
be in correspondence with the classical Hamiltonian $H_0$. Since in general
this observable is a composite quantity, the definition of the quantum
Hamiltonian $\hat{H}_0$ involves a specific choice of operator ordering
such that $\hat{H}_0$ be self-adjoint given the hermitean inner product
of which the space of quantum states is equipped. This is a necessary
requirement for the quantum unitarity, hence the physical consistency, of
the quantum system. Indeed, time evolution of quantum states $|\psi,t>$
is generated precisely through the quantum Hamiltonian and Schr\"odinger's
equation,
\begin{equation}
i\hbar\frac{d}{dt}|\psi,t>=\hat{H}_0\ |\psi,t>\ ,
\end{equation}
whose formal solution is of the form
\begin{equation}
|\psi,t>=\hat{U}(t,t_0)\,|\psi,t_0>\ ,
\end{equation}
where $|\psi,t_0>$ is some boundary state at the initial time $t_0$,
while the quantum evolution operator or ``propagator" of the system
is formally defined by the exponential operator
\begin{equation}
\hat{U}(t_2,t_1)=e^{-\frac{i}{\hbar}(t_2-t_1)\hat{H}_0}\ .
\label{eq:evolutionoperator1}
\end{equation}
Quantum unitarity of time evolution thus requires $\hat{U}(t_2,t_1)$
to be unitary and to obey the involution property
\begin{equation}
\hat{U}^\dagger(t_2,t_1)=\hat{U}^{-1}(t_2,t_1)=\hat{U}(t_1,t_2)\ \ ,\ \ 
\hat{U}(t_3,t_2)\,\hat{U}(t_2,t_1)=\hat{U}(t_3,t_1)\ ,
\label{eq:evolutionoperator2}
\end{equation}
which certainly requires $\hat{H}_0$ to be self-adjoint on the
space of quantum states. The choice of operator ordering should
therefore be consistent with this basic property for $\hat{H}_0$.

Note that the entire discussion concerning the definition of
the algebraic operator structure defining the quantization of
a given classical system is performed at a specific reference
time $t_0$, namely all operators are regarded as having been
defined at that reference time. Time dependency of the dynamical
quantum system is totally accounted for through the time dependency
of quantum states $|\psi,t>$ as solutions to the Schr\"odinger equation.
This formulation of a quantum system is referred to as ``the Schr\"odinger
picture", which is in correspondence with the classical phase space
description of the system in which the Poisson bracket structure is
carried by the time dependent degrees of freedom $(q^n(t),p_n(t))$.
The fact that the quantum operator algebra is defined at a fixed reference
time $t_0$ raises the question of whether the quantized system is
dependent on that choice. However, two quantizations defined at different
reference times in the Schr\"odinger picture
are unitarily equivalent through the evolution operator
$\hat{U}(t_2,t_1)$ associated to a specific choice of self-adjoint
quantum Hamiltonian $\hat{H}_0$. In other words, the quantum evolution
operator defines the isomorphism between all unitarily equivalent
representations of the algebraic structures associated to different
reference times. For this isomorphism to be a unitary one, it is
essential that the Hamiltonian $\hat{H}_0$ be self-adjoint.
Later on, we shall discuss briefly the ``Heisenberg picture" for
a quantum system, as an alternative to the Schr\"odinger one,
which will be seen to correspond at the classical level to the
Poisson bracket structure being defined on the space of integration
constants for the Hamiltonian equations of motion as discussed previously.

As a specific simple illustration of the above general discussion,
let us consider again the one-dimensional harmonic oscillator.
As a quantum system, its space of states should thus provide a
linear representation space of the Heisenberg algebra
\begin{equation}[\,\hat{q},\hat{p}\,]=i\hbar\ ,
\label{eq:Heisenberg1.1}
\end{equation}
equipped with an hermitean inner product for which these two operators
be self-adjoint,
\begin{equation}
\hat{q}^\dagger=\hat{q}\ \ \ ,\ \ \ 
\hat{p}^\dagger=\hat{p}\ .
\label{eq:Heisenberg1.2}
\end{equation}
Alternatively, given the physical parameters $m$ and $\omega$
carrying physical dimensions,
it is possible to represent in an equivalent way the same algebra in
terms of the following combinations,
\begin{equation}
\hat{q}=\sqrt{\frac{\hbar}{2m\omega}}\left[a+a^\dagger\right]\ \ ,\ \ 
\hat{p}=-i\sqrt{\frac{\hbar m\omega}{2}}\left[a-a^\dagger\right]\ ,
\end{equation}
which are seen to be in direct correspondence with the classical
solutions in (\ref{eq:phasespacesolution}), or equivalently,
\begin{equation}
a=\sqrt{\frac{m\omega}{2\hbar}}\left[\hat{q}+\frac{i}{m\omega}\hat{p}\right]
\ \ ,\ \ 
a^\dagger=\sqrt{\frac{m\omega}{2\hbar}}\left[\hat{q}
-\frac{i}{m\omega}\hat{p}\right]\ .
\label{eq:definitionFock}
\end{equation}
The operators $a$ and $a^\dagger$ are simply the
well known annihilation and creation operators of the harmonic
oscillator, obeying the Fock space algebra,
\begin{equation}
[a,a^\dagger]=\one\ \ \ ,\ \ \
\label{eq:Fock}
\end{equation}
with $a$ and $a^\dagger$ being the adjoint of one another.
In other words, given a physical parameter with the dimension
of $m\omega$, the Heisenberg and Fock space algebras are equivalent.
The quantum space of states should thus provide a representation
space of either algebra.

Finally, given the classical Hamiltonian $H_0=p^2/(2m)+m\omega^2q^2/2$,
the quantum Hamiltonian may be chosen to be
\begin{equation}
\hat{H}_0=\frac{\hat{p}^2}{2m}+\frac{1}{2}m\omega^2\hat{q}^2=
\frac{1}{2}\hbar\omega[a^\dagger a+ a a^\dagger]=\hbar\omega
\left[a^\dagger a+\frac{1}{2}\right]\ .
\end{equation}
Note that in terms of $\hat{q}$ and $\hat{p}$, $\hat{H}_0$ does
not require some operator ordering prescription, which would not be
the case had one chosen to consider its definition in terms of
the creation and annihilation operators. Of course in the present
case, the difference between the two possibilities amounts to a
constant shift in the energy eigenvalues without any physical
consequence, corresponding to the vacuum quantum fluctuation
contribution. For more general systems however, such a difference may not
be as innocuous.

\subsection{Representations of the Heisenberg algebra}
\label{Subsect2.4}

Canonical quantization thus raises generally the issue of the classification
of the representations of the Heisenberg algebra. Such representations
will be discussed here only in the simplest of cases, namely a set
of cartesian degrees of freedom $q^n$, restricting the presentation
to a single such degree of freedom, $N=1$, the general case being obtained
through a straightforward tensor pro\-duct over $n=1,2,\cdots,N$. 
A generalization to spaces of nontrivial geometry and/or topology,
for curvilinear coordinates and curved configurations spaces $q^n$ whose 
topology may be compact or not, is also possible, but shall not be addressed 
here.\cite{JG2,JG2bis} In the case of cartesian phase space canonical 
coordinates $(q,p)$,
it is well known that up to unitary equivalence, there exists only
one representation of the Heisenberg algebra defined in
(\ref{eq:Heisenberg1.1}) and (\ref{eq:Heisenberg1.2}). 
We refrain here from discussing how this conclusion may be reached
in terms familiar to most physicists,\cite{JG1,JG2,JG2bis} and only present 
the description of this representation in its different realizations,
namely the configuration space, the momentum space, the Fock space, 
and the coherent state representations.

\vspace{10pt}

\noindent\underline{The configuration space representation}

\vspace{10pt}

Given that the classical variable $q$ may take all real values, one
assumes that there exists a basis of position $\hat{q}$-eigenstates
$|q>$ with eigenvalues $q$ taking all possible real values, and
normalized to unity on the real line,
\begin{equation}
\hat{q}|q>=q|q>\ \ ,\ \ <q|q'>=\delta(q-q')\ \ ,\ \ 
\one=\int_{-\infty}^\infty dq\,|q><q|\ .
\end{equation}
This representation is known as ``the configuration space representation" of 
the Heisenberg algebra, leading to the configuration space wave function
re\-pre\-sen\-ta\-tion $\psi(q)$ of any state $|\psi>$ in the associated 
space of quantum states,
\begin{equation}
\psi(q)=<q|\psi>\ \ \ ,\ \ \ 
|\psi>=\int_{-\infty}^\infty dq\,|q>\psi(q)\ .
\end{equation}
This wave function $\psi(q)$ thus provides the components of the state 
$|\psi>$ in the configuration space basis $|q>$. In particular, inner 
products of states are simply given by
\begin{equation}
<\varphi|\psi>=\int_{-\infty}^\infty dq\,\varphi^*(q)\psi(q)\ ,
\end{equation}
as follows directly from the above spectral decomposition of the unit
ope\-ra\-tor $\one$ and the configuration space decomposition of the
states $|\varphi>$ and $|\psi>$. Given this construction, it then follows
that the abstract position and momentum operators $\hat{q}$ and $\hat{p}$
possess the following configuration space re\-pre\-sen\-tations,
\begin{equation}
<q|\hat{q}|\psi>=q\,\psi(q)\ \ \ ,\ \ \
<q|\hat{p}|\psi>=-i\hbar\frac{d}{dq}\psi(q)\ .
\end{equation}
This, of course, is nothing but the most familiar wave function quantization
of a single cartesian degree of freedom system.

\vspace{10pt}

\noindent\underline{The momentum space representation}

\vspace{10pt}

Likewise for the conjugate phase space degree of freedom, one has
the momentum space representation which is spanned by the momentum
eigenstate basis $|p>$, which in the case of the real line is associated
to all possible real momentum eigenvalues $p$, and is normalized to unity,
\begin{equation}
\hat{p}|p>=p|p>\ \ ,\ \ <p|p'>=\delta(p-p')\ \ ,\ \ 
\one=\int_{-\infty}^\infty dp\,|p><p|\ .
\end{equation}
Correspondingly, arbitrary states are decomposed in that basis in terms
of their momentum space wave function,
\begin{equation}
\psi(p)=<p|\psi>\ \ ,\ \ 
|\psi>=\int_{-\infty}^\infty dp\,|p>\psi(p)\ \ ,\ \ 
<\varphi|\psi>=\int_{-\infty}^\infty dp\,\varphi^*(p)\psi(p)\ .
\end{equation}
Finally, the elementary operators are represented as
\begin{equation}
<p|\hat{q}|\psi>=i\hbar\frac{d}{dp}\psi(p)\ \ \ ,\ \ \
<p|\hat{p}|\psi>=p\,\psi(p)\ .
\end{equation}
These results are of course most familiar.

\vspace{10pt}

\noindent\underline{The Fock space representation}

\vspace{5pt}

Given the fact that the Heisenberg algebra may alternatively be
represented through the Fock space algebra (\ref{eq:Fock}) (provided
a parameter with the physical dimension of $m\omega$ is available),
one may also consider the Fock space representation of the Heisenberg
algebra in the case of the harmonic oscillator. As is well known, a basis 
of this representation is constructed
as follows. There exists a vacuum state $|0>$ (not to be confused with
either of the states $|q=0>$ or $|p=0>$), normalized to unity, and
annihilated by the annihilation operator $a$,
\begin{equation}
a|0>=0\ \ \ ,\ \ \ <0|0>=1\ .
\end{equation}
Applying then in succession powers of the creation operator $a^\dagger$
onto the Fock vacuum $|0>$,
one obtains a discrete infinite set of linearly independent orthonormalized
states which span the representation space, defined by
\begin{equation}
|n>=\frac{1}{\sqrt{n!}}\left(a^\dagger\right)^n\,|0>\ \ \ ,\ \ \ 
<n|m>=\delta_{n,m}\ ,
\end{equation}
thus leading to the spectral representation of the unit operator
\begin{equation}
\one=\sum_{n=0}^\infty\,|n><n|\ .
\end{equation}
Correspondingly, arbitrary states may be decomposed in that basis
according to
\begin{equation}
\psi_n=<n|\psi>\ \ ,\ \ 
|\psi>=\sum_{n=0}^\infty |n>\,\psi_n\ \ ,\ \ 
<\varphi|\psi>=\sum_{n=0}^\infty\,\varphi^*_n\,\psi_n\ .
\end{equation}
From the definition of the states $|n>$, one readily establishes that
the annihilation and creation operators possess the following
representations,
\begin{equation}
a|n>=\sqrt{n}\,|n-1>\ \ ,\ \ 
a^\dagger|n>=\sqrt{n+1}\,|n>\ \ ,\ \ 
a^\dagger a|n>=n|n>\ ,
\end{equation}
from which the Heisenberg matrix representation for the position and
momentum operators $\hat{q}$ and $\hat{p}$ may be derived in terms of
semi-infinite matrices.

\vspace{10pt}

\noindent\underline{The coherent state representation}

\vspace{5pt}

Finally, an overcomplete set of basis vectors is provided by the
holomorphic or phase space coherent states defined by, 
respectively,\cite{Klaud3}
\begin{equation}
|z>=e^{-\frac{1}{2}|z|^2}\,e^{za^\dagger}\,|0>\ \ \ ,\ \ \ 
|q,p>=e^{-\frac{i}{\hbar}q\hat{p}}\,e^{\frac{i}{\hbar}p\hat{q}}\,|0>\ ,
\end{equation}
$z$ being an arbitrary complex number and $(q,p)$ arbitrary
real quantities in correspondence with the classical phase space
canonical degrees of freedom. These two sets of states are related as
\begin{equation}
|z>=e^{\frac{i}{2\hbar}qp}\,|q,p>\ ,
\end{equation}
with
\begin{equation}
z=\sqrt{\frac{m\omega}{2\hbar}}\left[q+\frac{i}{m\omega}p\right]\ ,
\end{equation}
to be compared to the definitions for $a$ and $a^\dagger$ in
(\ref{eq:definitionFock}). The spectral resolution of the unit
operator is then given by
\begin{equation}
\one=\int\frac{dz\,d\bar{z}}{\pi}\,|z><z|\ =\
\int_{(\infty)}\frac{dq\,dp}{2\pi\hbar}\,|q,p><q,p|\ ,
\end{equation}
while for matrix elements one finds for instance for the holomorphic
coherent states (a similar expression may easily be established for the
matrix element $<q_1,p_1|q_2,p_2>$)
\begin{equation}
<z_1|z_2>=e^{-\frac{1}{2}|z_1|^2}\,e^{-\frac{1}{2}|z_2|^2}\,
e^{\bar{z}_1z_2} \ .
\end{equation}
The interest of coherent states stems, among other reasons, from the fact
that they provide quantum states whose properties are the closest possible
to classical states. This is made manifest for example by the following
action of the annihilation operator on holomorphic coherent states,
\begin{equation}
a^n|z>=z^n|z>\ ,
\end{equation}
making these states an ideal choice of basis to compute matrix elements
of the creation and annihilation operators of the Fock space algebra 
representation for the Heisenberg algebra.

\vspace{10pt}

\noindent\underline{Changes of bases}

\vspace{5pt}

The fact that all these different realizations of the same basic
algebraic structure, namely that of the Heisenberg algebra, are unitarily
equivalent, may be made explicit in different ways, such as for example
by specifying the matrix elements for the different changes of bases
related to the different representations described above.
Thus, the relation between the configuration and momentum space
representations is provided by the matrix elements
\begin{equation}
<q|p>=\frac{1}{\sqrt{2\pi\hbar}}\,e^{\frac{i}{\hbar}qp}\ \ \ ,\ \ \ 
<p|q>=\frac{1}{\sqrt{2\pi\hbar}}\,e^{-\frac{i}{\hbar}qp}\ ,
\end{equation}
from which it should be clear that the relation between the configuration
and momentum wave functions $\psi(q)$ and $\psi(p)$ for a given state
$|\psi>$ is simply that of an ordinary Fourier transformation,
\begin{equation}
\begin{array}{r c l}
\psi(p)&=&<p|\psi>=\int_{-\infty}^\infty\,dq<p|q><q|\psi>=
\int_{-\infty}^\infty\,\frac{dq}{\sqrt{2\pi\hbar}}\,e^{-\frac{i}{\hbar}qp}\,
\psi(q)\ \ ,\ \ \\
 & & \\
\psi(q)&=&<q|\psi>=\int_{-\infty}^\infty\,dp<q|p><p|\psi>=
\int_{-\infty}^\infty\,\frac{dp}{\sqrt{2\pi\hbar}}\,
e^{\frac{i}{\hbar}qp}\,\psi(p)\ .
\end{array}
\end{equation}

Likewise, the change of basis between the configuration space
representation, say, and the Fock space one, is provided by the
following matrix elements,
\begin{equation}
<q|n>=\left(\frac{m\omega}{\pi\hbar}\right)^{1/4}\,\frac{1}{\sqrt{2^n\,n!}}\,
e^{-\frac{m\omega}{2\hbar}q^2}\,H_n\left(q\sqrt{\frac{m\omega}{\hbar}}\right)\ ,
\end{equation}
where $H_n(x)$ is the usual Hermite polynomial of order $n$. Given this
result, any quantity obtained within any one of these representations may be
transformed into its expression in any of the other representations.

Finally, the change of basis to coherent states is best given in terms
of the Fock space states, in which the relevant matrix elements reduce
to simple monomials of the complex variable $z$,
\begin{equation}
<n|z>=\frac{1}{\sqrt{n!}}\,z^n\,e^{-\frac{1}{2}|z|^2}\ .
\end{equation}
Given these different matrix elements expressing the changes of bases
between all four representations of the Heisenberg algebra, it is in
principle possible to establish some result using one representation
in which this is most convenient, and then convert to any other
representation in which to determine some physical quantity.

\vspace{10pt}

For instance, in the case of the one-dimensional harmonic oscillator,
the Hamiltonian $\hat{H}_0$ is most readily diagonalized in the Fock
space representation, since one has
\begin{equation}
\hat{H}_0=\hbar\omega\left[a^\dagger a+\frac{1}{2}\right]\ \ ,\ \ 
\hat{H}_0|n>=E_n|n>\ \ ,\ \ 
E_n=\hbar\omega(n+\frac{1}{2})\ ,\ n=0,1,2,\cdots\ .
\end{equation}
Consequently, orthonormalized configuration space wave functions for 
energy eigenstates are simply given as $\psi_n(q)=<q|n>$ whose expressions 
are known, while the configuration space propagator of the system, namely 
the matrix elements of the quantum evolution operator $\hat{U}(t_2,t_1)$ for
external position eigenstates $|q_1>$ and $|q_2>$, possesses the following 
simple representation,
\begin{equation}
<q_2|\hat{U}(t_2,t_1)|q_1>=\sum_{n=0}^\infty\,
\psi_n^*(q_2)\,e^{-\frac{i}{\hbar}(t_2-t_1)\hbar\omega(n+1/2)}\,
\psi_n(q_1)\ ,
\end{equation}
given the spectral resolution of the evolution operator
\begin{equation}
\hat{U}(t_2,t_1)=\sum_{n=0}^\infty\,|n>\,
e^{-\frac{i}{\hbar}(t_2-t_1)E_n}\,<n|\ .
\end{equation}

\vspace{10pt}

\noindent\underline{The Heisenberg picture}

\vspace{5pt}

Finally, let us reconsider the issue of the Schr\"odinger and Heisenberg
pictures for a quantum system. Given that in the former picture, the
matrix element of any operator $\hat{\cal O}(t_0)$ defined at the
reference time $t_0$ for arbitrary time dependent external states is
given by $<\varphi,t|\hat{\cal O}(t_0)|\psi,t>$,
using the evolution operator as expressed in general terms in
(\ref{eq:evolutionoperator1}), and obeying the properties 
(\ref{eq:evolutionoperator2}), the same matrix element may be expressed as
\begin{equation}
<\varphi,t|\hat{\cal O}(t_0)|\psi,t>=<\varphi,t_0|\hat{\cal O}(t)|\psi,t_0>\ ,
\end{equation}
where the time dependent operator $\hat{\cal O}(t)$ is simply defined by
\begin{equation}
\hat{\cal O}(t)=\hat{U}^\dagger(t,t_0)\,\hat{\cal O}(t_0)\,\
\hat{U}(t,t_0)\ .
\label{eq:Heisenbergpicture}
\end{equation}
In other words, within the so-called Heisenberg picture, the same physical
content as that of the Schr\"odinger picture is provided by time independent
quantum states $|\psi,t_0>$ defined at the reference time $t_0$, while
the whole time dependency of the system is now carried by the quantum
operators ${\cal O}(t)$ related to their definition in the Schr\"odinger
picture at the reference time $t_0$ by the above definition in terms
of the quantum evolution operator $\hat{U}(t_2,t_1)$. Thus, the Schr\"odinger
equation that governs time evolution of the system now applies to the
operators rather than to the states, and is given by 
\begin{equation}
i\hbar\frac{d}{dt}\hat{\cal O}(t)=\left[\hat{\cal O}(t),\hat{H}_0(t)\right]\ ,
\end{equation}
as follows directly from the definition of the Heisenberg
picture in (\ref{eq:Heisenbergpicture}). Note how this quantum operator 
equation of motion is again in direct correspondence with the associated
classical one in Hamiltonian form, namely
\begin{equation}
\dot{\cal O}=\{{\cal O},H_0\}\ .
\end{equation}
Furthermore, since the quantum Hamiltonian $\hat{H}_0(t)$ commutes
with itself, $[\hat{H}_0(t),\hat{H}_0(t)]=0$, in fact this operator
is totally time independent, hence $\hat{H}_0(t)=\hat{H}_0(t_0)$, expressing
a conservation law for the system, which is that of the energy
in the case that $t$ parametrizes physical time, and not some other
possible evolution parameter for the system's dynamics.

Once again turning to the one-dimensional harmonic oscillator in its Fock 
space representation, the Heisenberg picture is thus defined in terms of
time dependent creation and annihilation operators $a^\dagger (t)$ and
$a(t)$, whose solution to the ope\-rator Schr\"odinger equation 
$i\hbar\dot{a}(t)=[a(t),\hat{H}_0]=\hbar\omega a(t)$ is
\begin{equation}
a(t)=a(t_0)\,e^{-i\omega(t-t_0)}\ \ \ ,\ \ \ 
a^\dagger(t)=a^\dagger(t_0)\,e^{i\omega(t-t_0)}\ ,
\end{equation}
thus leading to the following representation for the position operator
$\hat{q}(t)$, say, in the Heisenberg picture,
\begin{equation}
\hat{q}(t)=\sqrt{\frac{\hbar}{2m\omega}}\left[a(t)+a^\dagger(t)\right]=
\sqrt{\frac{\hbar}{2m\omega}}\left[a(t_0)\,e^{-i\omega(t-t_0)}+
a^\dagger(t_0)\,e^{i\omega(t-t_0)}\right]\ ,
\end{equation}
an expression which ought to be compared to its classical counterparts
in (\ref{eq:phasespacesolution}) and (\ref{eq:phasespacesolution2}),
and thus justifying the comments made already at that stage of our
discussion with respect to the Schr\"odinger and Heisenberg pictures
for a quantum system and the corresponding characterization of the
Poisson bracket structure either on phase space or on the space of
integration constants for the classical Hamiltonian equations of motion.

\subsection{Path integral quantization}

It is well known that besides the canonical quantization path,
there is another royal avenue towards the quantization of a classical
system whose dynamics is defined through some action and the variational
principle, namely the so-called path integral or functional integral 
formulation of quantum mechanics.\cite{Feynman} Here we shall discuss how,
starting from the canonical quantization of any such system following
the approach outlined in the previous sections, it is possible to set up
integral representations for matrix elements of quantum operators, which
acquire the interpretation of functional integrals over phase space.
When reducing from these integrals the conjugate momentum degrees of
freedom, one recovers a functional integral over configuration space
in which the original classical action expressed in terms of the
Lagrange function plays again a central role. Further remarks as to
quantization directly through the functional integral are made
at the end of this discussion. It should already be clear
that these two approaches are complementary, each with its own
advantages and difficulties both with respect to an intuitive 
understanding of the physics that they both encode as well as to the
calculational advantages of one compared to the other. However, when properly 
implemented, they represent in complementary terms an identical
physical content.

The procedure to construct an integral representation for matrix
elements of operators, starting from canonical quantization, follows
essentially always the same avenue, based on the insertion of
complete sets of states in terms of which the unit operator possess
a spectral resolution. Here, we shall illustrate this feature for
the configuration space representation of the Heisenberg algebra,
even though more general cases may be envisaged as well, for instance
in terms of coherent states. Furthermore,
we shall consider configuration space matrix elements of the evolution
operator for a given quantum system, namely the propagator
$<q_f|\hat{U}(t_f,t_i)|q_i>$ of the system (in configuration space).
This operator thus writes as
\begin{equation}
\hat{U}(t_f,t_i)=e^{-\frac{i}{\hbar}(t_f-t_i)\hat{H}_0}=
\left[e^{-\frac{i}{\hbar}\epsilon\hat{H}_0}\right]^N=
\lim_{N\rightarrow\infty}\left[1-\frac{i}{\hbar}\epsilon\hat{H}_0\right]^N\ ,
\end{equation}
with $\epsilon=(t_f-t_i)/N$, while $N$ is some arbitrary positive
integer specifying an equally spaced slicing of the finite time
interval $(t_f-t_i)$. In what follows, the $n$ index for the degrees
of freedom $(q^n,p_n)$ is suppressed, to keep expressions as
transparent as possible. Given this time-sliced form of the evolution
operator, the idea now is to insert twice the spectral resolution of the
unit operator $\one$, once in terms of the position eigenstates,
and once in terms of the momentum eigenstates, and this in between each of
the $N$ factors that appear in the above $N$-factorized form for
$\hat{U}(t_f,t_i)$, as follows,
\begin{equation}
\one=\int_{-\infty}^\infty\,dp_\alpha\,\int_{-\infty}^\infty\,
dq_{\alpha+1}\,|q_{\alpha+1}><q_{\alpha+1}|p_\alpha><p_\alpha|
\ \ ,\ \ \alpha=0,1,2,\cdots,N-2\ .
\end{equation}
Setting then $q_f=q_{\alpha=N}$ and $q_i=q_{\alpha=0}$, a straightforward
substitution into the considered matrix element leads to the expression
(a substitution of the unit operator as
$\one=\int_{-\infty}^\infty dp|p><p|$ is also
performed to the right of the external final state $<q_f|$, leading
to one more integration over the $p_\alpha$'s than over the $q_\alpha$'s)
\begin{equation}
\begin{array}{r l}
&<q_f|\hat{U}(t_f,t_i)|q_i>= \\
 & \\
&=\int_{-\infty}^\infty\prod_{\alpha=1}^{N-1}
dq_\alpha\,\prod_{\alpha=0}^{N-1}dp_\alpha\,
\prod_{\alpha=0}^{N-1}\left[<q_{\alpha+1}|p_\alpha>
<p_\alpha|e^{-\frac{i}{\hbar}\epsilon\hat{H}_0}|q_\alpha>\right]\ .
\end{array}
\end{equation}
Using then the value for the matrix element $<q|p>$ given previously, 
this quantity finally reduces to
\begin{equation}
\begin{array}{r l}
&<q_f|\hat{U}(t_f,t_i)|q_i>= \\
 & \\
&=\lim_{N\rightarrow\infty}\int_{-\infty}^\infty\prod_{\alpha=1}^{N-1}
dq_\alpha\,\prod_{\alpha=0}^{N-1}\frac{dp_\alpha}{2\pi\hbar}\,
{\rm exp}\left\{\frac{i}{\hbar}\sum_{\alpha=0}^{N-1}\epsilon
\left[\frac{q_{\alpha+1}-q_\alpha}{\epsilon}p_\alpha-h_\alpha\right]\right\}\ ,
\end{array}
\label{eq:PI1}
\end{equation}
with the Hamiltonian matrix elements
\begin{equation}
h_\alpha=\frac{<p_\alpha|\hat{H}_0|q_\alpha>}
{<p_\alpha|q_\alpha>}\ .
\end{equation}
Clearly, the discretized integral representation (\ref{eq:PI1}) of
the configuration space propagator corresponds to a specific construction
of the otherwise formal expression for the phase space path integral
or functional integral corresponding to that quantity, namely
\begin{equation}
<q_f|\hat{U}(t_f,t_i)|q_i>=
\int_{q(t_i)=q_i}^{q(t_f)=q_f}
\left[{\cal D}q\frac{{\cal D}p}{2\pi\hbar}\right]\,
e^{\frac{i}{\hbar}S[q,p]}\ ,
\label{eq:formalPI1}
\end{equation}
in which the phase space action is that of the first-order Hamiltonian 
formulation of the system given in (\ref{eq:firstorderaction}), namely
\begin{equation}
S[q,p]=\int_{t_i}^{t_f}dt\left[\dot{q}p-H_0(q,p)\right]\ ,
\end{equation}
which is that associated to the choice of boundary conditions corresponding
to the configuration space propagator when imposing the variational
principle in a strong sense, namely with the induced boundary terms also
required to vanish through the boundary conditions $q(t_{i,f})=q_{i,f}$. 
Note that contrary to what the formal
expression (\ref{eq:formalPI1}) may lead one to believe, the integration
measure is not quite the phase space Liouville measure, since in fact
there is always one more $p_\alpha$ integration than the number of $q_\alpha$ 
integrations. One should always keep this remark in mind when developing
formal arguments based on the formal expression (\ref{eq:formalPI1})
of the functional integral.

Considering the momentum space matrix elements of the same
operator, a similar analysis leads to an analogous specific discretized
expression corresponding to the formal quantity
\begin{equation}
<p_f|\hat{U}(t_f,t_i)|p_i>=
\int_{p(t_i)=p_i}^{p(t_f)=p_f}
\left[\frac{{\cal D}q}{2\pi\hbar}{\cal D}p\right]\,
e^{\frac{i}{\hbar}S[q,p]}\ ,
\label{eq:formalPI2}
\end{equation}
where the appropriate Hamiltonian first-order action now reads
\begin{equation}
S[q,p]=\int_{t_i}^{t_f}\,dt\left[-q\dot{p}-H_0(q,p)\right]\ ,
\end{equation}
being this time associated to the choice of boundary conditions
$p(t_{i,f})=p_{i,f}$ as opposed to $q(t_{i,f})=q_{i,f}$ for the
propagator in configuration space. Note that the same remark as above
concerning the phase space Liouville measure applies here as well.

In the particular situation that the Hamiltonian is such that the
matrix elements $h_\alpha$ are quadratic in the momenta,
\begin{equation}
h_\alpha=\frac{p^2_\alpha}{2m}+V(q_\alpha)\ ,
\end{equation}
the integration over momentum space may be completed explicitly
in the above discretized expressions, thereby leading to the configuration
space functional integral representation,
\begin{equation}
\begin{array}{r l}
&<q_f|\hat{U}(t_f,t_i)|q_i>=\lim_{N\rightarrow\infty}
\left(\frac{m}{2i\pi\hbar\epsilon}\right)^{N/2}\,
\int_{-\infty}^\infty\prod_{\alpha=1}^{N-1}dq_\alpha\,\times \\
 & \\
&\times{\rm exp}\left\{\frac{i}{\hbar}\sum_{\alpha=0}^{N-1}\epsilon
\left[\frac{1}{2}m\left(\frac{q_{\alpha+1}-q_\alpha}{\epsilon}\right)^2
-V(q_\alpha)\right]\right\}\ ,
\end{array}
\label{eq:PI2}
\end{equation}
or at the formal level,
\begin{equation}
<q_f|\hat{U}(t_f,t_i)|q_i>
=\int_{q(t_i)=q_i}^{q(t_f)=q_f}
\left[{\cal D}q\right]\,e^{\frac{i}{\hbar}S[q]}\ ,
\label{eq:formalPI3}
\end{equation}
with
\begin{equation}
S[q]=\int_{t_i}^{t_f}\,dt\,L(q,\dot{q})\ \ \ ,\ \ \ 
L(q,\dot{q})=\frac{1}{2}m\dot{q}^2-V(q)\ .
\end{equation}
The above explicit discretized representation of this latter formal functional 
integral coincides exactly with the explicit construction performed by 
Feynman.\cite{Feynman}

Hence, we have come back full circle. Starting from the action principle
defined within the Lagrangian formulation of dynamics, the canonical
Hamiltonian formulation of the same dynamics on phase space has been
constructed, allowing for the canonical operator quantization of the
associated algebraic and geometric structures, for which operator
matrix elements may be given a functional integral representation
on phase space or configuration space, in which the classical Hamiltonian
or Lagrangian action functionals reappear on equal terms. The concept
which is central to this whole construction is that of the action,
through one of the many forms by which it contributes whether for the
classical or the quantum dynamics.

Having chosen to follow the operator quantization path, once a specific
choice of operator ordering has been made, in principle the functional
integral representations acquires a totally unambiguous and well defined
discretized expression, which defines in an exact manner otherwise
ill defined formal path integral expressions whose actual meaning always
still needs to be specified properly. Nonetheless, as we have indicated, 
difficulties lie at the operator level precisely in the choice of operator 
ordering in order to obtain a consistent unitary quantum theory.

Had one taken the functional integral path towards quantization,
whether from the Lagrangian or Hamiltonian classical actions,
the difficulty of a proper construction of the quantized system
then lies hidden in the necessity to give a precise definition and
meaning, through some discretization procedure or otherwise, to the
formal and thus ill defined functional integrals such as those in 
(\ref{eq:formalPI1}), (\ref{eq:formalPI2}) and (\ref{eq:formalPI3}). 
As a matter of fact, the arbitrariness which exists at this level in the 
choice of discretization procedure and functional integration measure
(whether over configuration, momentum or phase space) is in direct
correspondence with the arbitrariness which exists on the operator
side of this relationship in terms of the choice of operator ordering.
Taking either path towards quantization, for appropriate choices on
both sides which are in correspondence, the same dynamical quantum system 
is being represented in a complementary manner. It is extremely
fruitful to constantly keep in one's mind these equivalent
representations of a quantum dynamics when properly implemented, in particular
in a manner that should ensure its quantum unitarity.

\section{Relativistic Quantum Particles and Field Theories}
\label{Sect3}

Starting with this section, we shall explicitly work in four-dimensional
Minkowski spacetime with coordinates $x^\mu$ ($\mu=0,1,2,3)$ and
a metric $\eta_{\mu\nu}$ of signature $(+---)$. Furthermore as is
customary in quantum field theory, units such that $\hbar=1=c$ are also 
being used throughout, so that mass and energy on the one hand, as well
as time and space on the other, are each measured in the same units,
while energy and time, for instance, are of inverse dimensions. Hence,
any mechanical quantity may always be expressed in units of mass
to some power.

\subsection{Motivation}
\label{Subsect3.1}

It is an experimental fact that there exist particles in nature, which
behave both with relativistic and quantum properties, have definite
energy, momentum and thus invariant mass values, and may be created
or annihilated through different physical processes. Which type of
mathematical framework would be able to account for all these physical
properties all at once? 

As we have recalled above, the quantization of the harmonic oscillator
leads to such a framework. Indeed, the operators $a$ and $a^\dagger$,
which obey the Fock algebra $[a,a^\dagger]=\one$,
provide for the annihilation and creation of energy quanta, each
carrying an identical amount $\hbar\omega$ of energy. Furthermore,
we also know that associated to these operators, there exists some
configuration space operator $\hat{q}$ which in the Heisenberg picture
has a time dependency defined by (from now on, the choice of
reference time will be $t_0=0$)
\begin{equation}
\hat{q}(t)=\sqrt{\frac{\hbar}{2m\omega}}\left[a\,e^{-i\omega t}+
a^\dagger\,e^{i\omega t}\right]\ ,
\label{eq:HOsuperposition}
\end{equation}
which, in the classical limit, thus defines the entire real line
as the space of classical configurations of
the system. Hence, the configuration space quantum operator $\hat{q}(t)$
in the Heisenberg picture obeys the following equation
\begin{equation}
\left[\,\frac{d^2}{dt^2}\ +\,\omega^2\,\right]\,\hat{q}(t)=0\ ,
\end{equation}
which also coincides with the classical equation of motion for the
system, which derives from the Lagrangian action
\begin{equation}
S[q]=m\int\,dt\,\left[\frac{1}{2}\left(\frac{dq}{dt}\right)^2\,-\,
\frac{1}{2}\omega^2q^2\,\right]\ .
\end{equation}

Let us now try to extend this mathematical framework to spinless 
re\-la\-tivistic
quantum particles of definite energy-momentum $k^\mu=(k^0,\vec{k})$
and mass $m$ such that $k^0=(\vec{k}\,^2+m^2)^{1/2}=\omega(\vec{k}\,)$,
and which may be created or annihilated in specific physical processes. 
Thus, for each of the possible
momentum values $\vec{k}$, one should introduce a pair of creation
and annihilation operators $a^\dagger(\vec{k}\,)$ and $a(\vec{k}\,)$
obeying the Fock space algebra
\begin{equation}
\left[a(\vec{k}\,),a^\dagger(\vec{k}')\right]=
(2\pi)^3\,2\omega(\vec{k}\,)\,\delta^{3}(\vec{k}-\vec{k}')\ ,
\label{eq:Fockparticle}
\end{equation}
where, compared to the Fock algebra for the harmonic oscillator,
the normalization of the operators has been modified for a reason
to be specified presently. Thus in particular, 1-particle quantum states
are obtained from the normalized Fock vacuum $|0>$ as
\begin{equation}
|\vec{k}>=a^\dagger(\vec{k}\,)\,|0>\ \ \ ,\ \ \ 
<0|0>=1\ .
\end{equation}
Proceeding by analogy with the harmonic oscillator case, in order
to identify the configuration space for such a quantum system,
let us also consider superpositions of these operators such as in
(\ref{eq:HOsuperposition}). However, since we wish to develop a
formalism which is manifestly spacetime covariant under Lorentz
transformations, the product $\omega t$ appearing in the imaginary
exponentials that multiply the operators and which thus corresponds to
the product of the energy value of a quantum by the time interval,
must be extended into the Minkowski invariant product 
$\omega(\vec{k}\,)t-\vec{k}\cdot\vec{x}=k\cdot x$, where the last
expression denotes the inner product of four-vectors with the 
four-dimensional Minkowski metric. Furthermore, since in the present case
we have an infinity of quantum operators labelled by the vector values
$\vec{k}$ and which are all on an equal footing, one should consider a general
superposition of all such linear combinations of the creation and
annihilation operators with a $\vec{k}$-independent weight. Hence finally, 
one is led to consider the following operator, again in the Heisenberg picture,
as the relativistic invariant extension of (\ref{eq:HOsuperposition}),
\begin{equation}
\hat{\phi}(x^\mu)=\int_{(\infty)}\frac{d^3\vec{k}}{(2\pi)^32\omega(\vec{k}\,)}
\left[a(\vec{k}\,)e^{-ik\cdot x}+a^\dagger(\vec{k}\,)e^{ik\cdot x}\right]\ .
\label{eq:fieldparticle}
\end{equation}
Note that having rescaled the creation and annihilation operators by
a factor $(\omega(\vec{k}\,))^{1/2}$, the $d^3\vec{k}$ integration
measure includes the same dimensionful normalization factor as in 
(\ref{eq:HOsuperposition}) for the harmonic oscillator. The choice of 
numerical factor $(2\pi)^3$ is made
for later convenience. As a matter of fact, the reason for the specific
choice of normalization in (\ref{eq:Fockparticle}) is that the
integration measure in (\ref{eq:fieldparticle}), namely
$d^3\vec{k}/2\omega(\vec{k}\,)$, is invariant under Lorentz transformations,
as may easily be checked. In other words, this parametrization of the
operator $\hat{\phi}(x^\mu)$ is manifestly Lorentz covariant.

Hence, associated to the algebra (\ref{eq:Fockparticle}), one expects
that the actual con\-fi\-gu\-ra\-tions of the corresponding system is that
of a real scalar field in spacetime! Indeed, in the classical limit,
the combination (\ref{eq:fieldparticle}) defines a real number $\phi(x^\mu)$
attached at each spacetime point. In other words, an arbitrary collection 
of identical relativistic free quantum point-particles with causal and 
unitary propagation corresponds to quanta of a single relativistic quantum 
field in Minkowski spacetime. Furthermore, even though these particles 
display corpuscular properties by having definite energy-momentum values, 
their spacetime dynamical propagation also displays wavelike properties, 
since the field obeys the following equation of motion,
\begin{equation}
\left[\frac{\partial^2}{\partial t^2}-\vec{\nabla}^2+m^2\right]
\hat{\phi}(x^\mu)=0\ ,
\end{equation}
which is indeed a wave equation, known as the Klein-Gordon equation,
and is nothing but the straightforward relativistic invariant extension
of the equation of motion for the harmonic oscillator. Likewise, the
corresponding classical action principle thus reads, in a manifestly Lorentz
invariant form,
\begin{equation}
S[\phi]=\int\,dt\int_{(\infty)}d^3\vec{x}
\left[\frac{1}{2}\left(\frac{\partial}{\partial t}\phi\right)^2-
\frac{1}{2}\left(\vec{\nabla}\phi\right)^2-\frac{1}{2}m^2\phi^2\right]\ .
\end{equation}
From this point of view, the configuration space that has been
identified corresponds to an infinite set of harmonic oscillators sitting
all adjacent next to one another in the three dimensions of space, 
and while they each oscillate away from their equilibrium position, 
the gradient term $\vec{\nabla}\phi$ in the action or in the equation of 
motion induces a coupling between adjacent oscillators, thereby leading to a 
propagating wave behaviour of the system in space as a function of time. 
This term in $\vec{\nabla}\phi$ is required by Lorentz invariance
from the similar term in $\partial\phi/\partial t$ which is necessary
for the time dependent dynamics of the system.

In conclusion, having considered the possibility to describe an
arbitrary collection of identical relativistic free quantum spinless 
point-particles of de\-fi\-ni\-te energy-momentum and mass which
may be created and annihilated locally in Minkowski spacetime, we are
naturally led to consider a formulation which is that of a local real
relativistic scalar field in spacetime with its dy\-na\-mi\-cal wave 
properties, 
whose action is real under complex conjugation (which guarantees quantum 
unitarity), Poincar\'e invariant (necessary for causality, and also leading 
to states of definite energy-momentum and angular momentum, which are the 
conserved Noether charges for the Poincar\'e invariance group of Minkowski 
spacetime), and finally local in spacetime (thus gua\-ran\-tee\-ing spacetime 
causality and locality of particle propagation, and later on also for their 
interactions). At this stage, given the algebra
(\ref{eq:Fockparticle}), one is only describing interactionless
particles, since the complete space of energy eigenstates is the simple
tensor product over all $\vec{k}$ values of a Fock space representation,
without any nonvanishing matrix element of the Hamiltonian between
dif\-fe\-rent factors of this tensor product, which would otherwise indeed
represent energy-momentum exchange, namely interactions.

\subsection{The classical free relativistic real scalar field}
\label{Subsect3.2}

Let us thus consider as a classical system a real scalar field $\phi(x)$ 
over spacetime, whose dynamics is governed by the spacetime local action
\begin{equation}
S[\phi]=\int d^4x^\mu\,{\cal L}_0(\phi,\partial_\mu\phi)\ ,
\end{equation}
with the Lagrangian density
\begin{equation}
{\cal L}_0(\phi,\partial_\mu\phi)=
\frac{1}{2}\eta^{\mu\nu}\partial_\mu\phi\partial_\nu\phi-\frac{1}{2}m^2\phi^2=
\frac{1}{2}\left(\partial_\mu\phi\right)^2- \frac{1}{2}m^2\phi^2\ .
\end{equation}
We shall apply to this system exactly the same procedure of canonical
quantization as has been described in Section~\ref{Sect2}, and establish that
we have indeed a formulation of free relativistic quantum spinless
particles of mass $m$. The infinite number of degrees of freedom is 
parametrized by $\phi(x^0,\vec{x}\,)$, and is thus labelled by the values 
of the space vector $\vec{x}$. Note that there is an abuse
in our notation for the parameter $m$ in the above Lagrangian density.
At the classical level, only a length scale $\kappa$ may be introduced,
leading to a quadratic term of the form $\phi^2/\kappa^2$ rather than
$m^2\phi^2$ above. However, at the quantum level, it will found that
the field quanta possess an invariant mass given by $m=\hbar c/\kappa$,
which explains our abuse of notation at the classical level already.

In their manifestly Lorentz covariant form, the Euler-Lagrange equations read
\begin{equation}
\partial_\mu\frac{\partial{\cal L}_0}{\partial(\partial_\mu\phi)}-
\frac{\partial{\cal L}_0}{\partial\phi}=0\ ,
\end{equation}
or in the present case
\begin{equation}
\left[\partial_\mu\partial^\mu+m^2\right]\phi=0\ ,
\end{equation}
which is the Klein-Gordon equation. Through Fourier analysis, the general
solution is readily established, and may be expressed as
\begin{equation}
\phi(x^\mu)=\int_{(\infty)}\frac{d^3\vec{k}}{(2\pi)^32\omega(\vec{k}\,)}
\left[a(\vec{k}\,)e^{-ik\cdot x}+a^*(\vec{k}\,)e^{ik\cdot x}\right]\ ,
\label{eq:solutionphi}
\end{equation}
$a(\vec{k}\,)$ and $a^*(\vec{k}\,)$ being complex integration constants,
while in the plane wave contributions $e^{\mp ik\cdot x}$
the value $k^0=\omega(\vec{k}\,)$ is to be used.

In order to quantize the system, let us first consider its Hamiltonian
formulation. By definition, the momentum conjugate to the field
$\phi(x^0,\vec{x}\,)$ at each point $\vec{x}$ in space is
\begin{equation}
\pi(x^0,\vec{x}\,)=\frac{\partial{\cal L}_0}
{\partial(\partial_0\phi(x^0,\vec{x}\,))}=\partial_0\phi(x^0,\vec{x}\,)\ ,
\end{equation}
while the phase space degrees of freedom 
$(\phi(x^0,\vec{x}\,),\pi(x^0,\vec{x}\,))$ possess a Poisson bracket structure
defined by the canonical brackets at equal time $x^0$
\begin{equation}
\{\phi(x^0,\vec{x}\,),\pi(x^0,\vec{y}\,)\}=\delta^{(3)}(\vec{x}-\vec{y}\,)\ .
\end{equation}
The Hamiltonian density is
\begin{equation}
{\cal H}_0=\partial_0\phi\,\pi-{\cal L}_0=
\frac{1}{2}\pi^2+\frac{1}{2}\left(\vec{\nabla}\phi\right)^2+
\frac{1}{2}m^2\phi^2\ ,
\end{equation}
while the Hamiltonian equations of motion follow as usual from the
Hamiltonian $H_0=\int_{(\infty)}d^3\vec{x}\,{\cal H}_0$ (namely the sum
of ${\cal H}_0$ over all degrees of freedom labelled by $\vec{x}$)
through the Poisson brackets. For the elementary phase space degrees 
of freedom, one has,
\begin{equation}
\partial_0\phi=\pi\ \ \ ,\ \ \ 
\partial_0\pi=\left(\vec{\nabla}^2-m^2\right)\phi\ ,
\end{equation}
clearly leading back to the Klein-Gordon equation upon reduction of
the conjugate momentum $\pi$. Hence, given the solution (\ref{eq:solutionphi})
for the field $\phi(x^\mu)$, that for the conjugate momentum is
\begin{equation}
\pi(x^\mu)=\int_{(\infty)}\frac{d^3\vec{k}}{(2\pi)^32\omega(\vec{k}\,)}
\left(-i\omega(\vec{k})\,\right)
\left[a(\vec{k}\,)e^{-ik\cdot x}-a^*(\vec{k}\,)e^{ik\cdot x}\right]\ .
\label{eq:solutionpi}
\end{equation}

On basis of these expressions, it is possible to also determine the Poisson
bracket structure on the space of integration constants $a(\vec{k}\,)$
and $a^*(\vec{k}\,)$, rather than on the phase space 
$(\phi(x^0,\vec{x}\,),\pi(x^0,\vec{x}\,))$. A straightforward
calculation finds for the only nonvanishing bracket,
\begin{equation}
\{a(\vec{k}\,),a^*(\vec{k}')\}=-i(2\pi)^32\omega(\vec{k}\,)
\delta^{(3)}\left(\vec{k}-\vec{k}'\right)\ ,
\end{equation}
while the Hamiltonian then reads
\begin{equation}
H_0=\int_{(\infty)}\frac{d^3\vec{k}}{(2\pi)^32\omega(\vec{k}\,)}
\frac{1}{2}\omega(\vec{k}\,)\left[a^*(\vec{k}\,)a(\vec{k}\,)+
a(\vec{k}\,)a^*(\vec{k}\,)\right]\ ,
\end{equation}
hence leading to the Hamiltonian equations of motion
\begin{equation}
\dot{a}(\vec{k}\,)=-i\omega(\vec{k}\,)a(\vec{k}\,)\ \ \ ,\ \ \ 
\dot{a}^*(\vec{k}\,)=i\omega(\vec{k}\,)a^*(\vec{k}\,)\ ,
\end{equation}
whose solutions are of course consistent with the explicit expressions
already constructed above for $\phi(x^\mu)$ and $\pi(x^\mu)$.

\subsection{The quantum free relativistic real scalar field}
\label{Subsect3.3}

Canonical quantization of the system in the Schr\"odinger picture,
at the re\-fe\-rence time $t_0=x^0_0=0$, is straightforward. The space
of quantum states $|\psi>$, with hermitean inner product
$<\chi|\psi>$, provides a representation of the Heisenberg
algebra
\begin{equation}
\left[\hat{\phi}(\vec{x}\,),\hat{\pi}(\vec{y}\,)\right]=
i\delta^{(3)}\left(\vec{x}-\vec{y}\,\right)\ .
\label{eq:Heisenbergfield}
\end{equation}
In terms of the following representation for the quantum field
operators in the Schr\"odinger picture at $x^0_0=0$,
\begin{equation}
\begin{array}{r c l}
\hat{\phi}(\vec{x}\,)&=&
\int_{(\infty)}\frac{d^3\vec{k}}{(2\pi)^32\omega(\vec{k}\,)}
\left[a(\vec{k}\,)e^{i\vec{k}\cdot\vec{x}}+
a^*(\vec{k}\,)e^{-i\vec{k}\cdot\vec{x}}\right]\ ,\\
 & & \\
\hat{\pi}(\vec{x}\,)&=&
\int_{(\infty)}\frac{d^3\vec{k}}{(2\pi)^32\omega(\vec{k}\,)}
\left(-i\omega(\vec{k}\,)\right)
\left[a(\vec{k}\,)e^{i\vec{k}\cdot\vec{x}}-
a^*(\vec{k}\,)e^{-i\vec{k}\cdot\vec{x}}\right]\ ,
\end{array}
\end{equation}
alternatively one has the Fock space algebra
\begin{equation}
\left[a(\vec{k}\,),a^\dagger(\vec{k}')\right]=(2\pi)^32\omega(\vec{k})
\delta^{(3)}\left(\vec{k}-\vec{k}'\right)\ .
\label{eq:Fockfield}
\end{equation}

The Schr\"odinger equation for the time evolution of quantum states in the
Schr\"odinger picture also reads
\begin{equation}
i\hbar\frac{d}{dt}|\psi,t>=\hat{H}_0|\psi,t>\ ,
\end{equation}
with the quantum Hamiltonian given by
\begin{equation}
\hat{H}_0=\int_{(\infty)}d^3\vec{x}\left[
\frac{1}{2}\hat{\pi}^2+\frac{1}{2}\left(\vec{\nabla}\hat{\phi}\right)^2+
\frac{1}{2}m^2\hat{\phi}^2\right]\ .
\end{equation}
Note that this operator does not suffer any operator ordering ambiguity.
On the other hand, in terms of the Fock space operators, the same
quantum Hamiltonian reads
\begin{equation}
\hat{H}_0=\int_{(\infty)}\frac{d^3\vec{k}}{(2\pi)^32\omega(\vec{k}\,)}
\frac{1}{2}\omega(\vec{k}\,)\left[a^\dagger(\vec{k}\,)a(\vec{k}\,)+
a(\vec{k}\,)a^\dagger(\vec{k}\,)\right]\ ,
\end{equation}
which leads to finite matrix elements only after normal ordering
of the creation and annihilation operators, a procedure which is denoted by 
double dots on both sides of a quantity and is defined by
commuting all operators so that all creation operators are to the left
of all annihilation operators, such as for example
\begin{equation}
:a(\vec{k}\,)a^\dagger(\vec{\ell}\,):\ =a^\dagger(\vec{\ell}\,)a(\vec{k}\,)
\ \ \ ,\ \ \ 
:a^\dagger(\vec{k}\,)a(\vec{\ell}\,):\ =a^\dagger(\vec{k}\,)a(\vec{\ell}\,)\ .
\end{equation}
Applying this operator ordering prescription to the above expression
for $\hat{H}_0$, one thus finds in the Fock space representation the
normal ordered Hamiltonian
\begin{equation}
\hat{H}_0=\int_{(\infty)}\frac{d^3\vec{k}}{(2\pi)^32\omega(\vec{k}\,)}
\omega(\vec{k}\,)\,a^\dagger(\vec{k}\,)a(\vec{k}\,)\ ,
\label{eq:normalorderedH0}
\end{equation}
while an infinite normal ordering constant contribution is then
subtracted away, namely
\begin{equation}
\int_{(\infty)}\frac{d^3\vec{k}}{(2\pi)^32\omega(\vec{k}\,)}
\frac{1}{2}\omega(\vec{k}\,)(2\pi)^32\omega(\vec{0}\,)\delta^{(3)}(\vec{0})\ .
\end{equation}
This contribution corresponds to the sum of all vacuum quantum fluctuations
of all the $\vec{k}$-modes of the scalar field. Provided the system is
not coupled to gravity, such a renormalization of the energy eigenvalues
is without physical consequence. Nonetheless, it should imply that the
two representations of the quantized system, namely that achieved through
the Heisenberg algebra for the fields, or that achieved through the
Fock algebra for its modes, need no longer be unitarily equivalent
for such a system with an infinite set of degrees of freedom,\cite{CCR} in
contradistinction to the situation for a system with a finite number
of degrees of freedom such as the one-dimensional harmonic oscillator.

It thus appears that one might have available two possibly physically 
inequivalent approaches to the quantization of this system, the first based 
on the representations of the field Heisenberg algebra 
(\ref{eq:Heisenbergfield}), and the second based on the representations of 
the field Fock space algebra (\ref{eq:Fockfield}). Let us first consider 
the Heisenberg algebra realization, say in its configuration space 
representation. In the Schr\"odinger picture, the basis of states is then 
spanned by states $|\phi>$ which are associated to specific classical field 
configurations $\phi(\vec{x}\,)$ defined over space at the reference time 
$x^0_0=0$, and which are eigenstates of the quantum field operator 
$\hat{\phi}(\vec{x}\,)$,
\begin{equation}
\hat{\phi}(\vec{x}\,)\,|\phi>=\phi(\vec{x}\,)\,|\phi>\ .
\end{equation}
The values for the vector $\vec{x}$ being the label for degrees of freedom,
at least formally one has the following normalization of these states,
together with the associated spectral resolution of the unit operator,
\begin{equation}
<\phi|\phi'>=\prod_{\vec{x}}\delta\left(\phi(\vec{x}\,)-\phi'(\vec{x}\,)\right)
\ \ \ ,\ \ \ 
\one=\int_{-\infty}^\infty\prod_{\vec{x}}d\phi(\vec{x}\,)\,|\phi><\phi|\ ,
\end{equation}
in direct analogy with the situation for a system with a finite number
of degrees of freedom. Hence, arbitrary quantum states $|\psi>$ possess now 
a configuration space wave functional representation $\Psi[\phi]$ defined by
\begin{equation}
\Psi[\phi]=<\phi|\psi>\ \ ,\ \ 
|\psi>=\int_{-\infty}^\infty\prod_{\vec{x}}d\phi(\vec{x}\,)\,|\phi>\,\Psi[\phi]
\ ,
\end{equation}
which thus represents the probability amplitude for observing the given
quantum state $|\psi>$ in the classical field configuration $\phi(\vec{x}\,)$,
again in direct analogy with the meaning of the configuration space
wave function for a finite dimensional system.

Furthermore, since the field operators $\hat{\phi}(\vec{x}\,)$ and
$\hat{\pi}(\vec{x}\,)$ possess the following configuration space
representations,
\begin{equation}
<\phi|\hat{\phi}(\vec{x}\,)|\psi>=\phi(\vec{x}\,)\,\Psi[\phi]\ \ \ ,\ \ \ 
<\phi|\hat{\pi}(\vec{x}\,)|\psi>=-i\hbar\frac{\delta}{\delta\phi(\vec{x}\,)}\,
\Psi[\phi]\ ,
\end{equation}
the action of the quantum Hamiltonian on quantum states in their configuration
space wave functional representation is
\begin{equation}
<\phi|\hat{H}_0|\psi>=\int_{(\infty)}d^3\vec{x}\,
\frac{1}{2}\left[-\hbar^2\left(\frac{\delta}{\delta\phi(\vec{x}\,)}\right)^2+
\left(\vec{\nabla}\phi(\vec{x}\,)\right)^2+m^2\phi^2(\vec{x}\,)\right]\,
\Psi[\phi]\ .
\end{equation}

This Schr\"odinger functional representation of a quantum field
theory could prove to be an appropriate framework in which to attempt 
a nonperturbative quantization. Even though it may well be that for a
noninteracting field, which is the above situation, this approach
would be unitarily equivalent to the Fock space one to
be discussed presently, it is far from clear that such an equivalence
should survive the introduction of nonlinear interactions. Given the
wide success of the perturbative treatment of particle interactions,
based on the Fock space quantization of a field theory briefly described
hereafter, such nonperturbative functional quantizations have not been
developed to the same extent, making this issue a worthwhile topic of 
further investigation,\cite{CCR} especially when it comes to nonlinear
field theories whose space of classical solutions includes topological
configurations such as solitons and higher dimensional monopole-like
configurations.

Turning now to the field Fock space algebra (\ref{eq:Fockfield}) and
its representations, it is clear that the space of states is spanned
by all possible $n$-particle states $(n=0,1,2,\cdots)$ of arbitrary
momentum values $\vec{k}_i$ $(i=1,2,\cdots,n)$, which are built through the
action of the creation operators $a^\dagger(\vec{k}\,)$ from
the normalized Fock vacuum $|0>$, itself annihilated by the $a(\vec{k}\,)$
operators, $a(\vec{k}\,)|0>=0$, 
\begin{equation}
|\vec{k}_1,\vec{k}_2,\cdots,\vec{k}_n>=
N(\vec{k}_1,\vec{k}_2,\cdots,\vec{k}_n)\,
a^\dagger(\vec{k}_1)\,a^\dagger(\vec{k}_2)\cdots a^\dagger(\vec{k}_n)|0>\ ,
\end{equation}
where $N(\vec{k}_1,\vec{k}_2,\cdots,\vec{k}_n)$ denotes some normalization
factor. In particular, the 1-particle quantum states correspond to
\begin{equation}
|\vec{k}>=a^\dagger(\vec{k}\,)|0>\ \ \ ,\ \ \ 
<\vec{k}|\vec{k}'>=
(2\pi)^32\omega(\vec{k}\,)\,\delta^{(3)}\left(\vec{k}-\vec{k}'\right)\ .
\end{equation}
In addition, given the manifest spacetime invariance of the system under
the Poincar\'e group, the quantum operators $\hat{P}^\mu$ and 
$\hat{M}^{\mu\nu}$ associated to the conserved Poincar\'e Noether charges
generate the Poincar\'e algebra on the space of quantum states, the latter
thus getting organized into irreducible representations of that symmetry.
The eigenstates of these operators, thus of de\-fi\-ni\-te energy-momentum, 
angular-momentum and invariant mass, define the 1-particle states
of the quantized field. Clearly, these eigenstates must correspond to
the 1-particle quantum states $|\vec{k}>$ constructed above, which is
indeed the case. For instance, the energy-momentum operator in Fock space
is given by
\begin{equation}
\hat{P}^\mu=\int_{(\infty)}\frac{d^3\vec{k}}{(2\pi)^32\omega(\vec{k}\,)}\,
k^\mu\,a^\dagger(\vec{k}\,)a(\vec{k}\,)\ ,
\end{equation}
so that the 1-particle states $|\vec{k}>$ are eigenstates of
this operator, namely $\hat{P}^\mu|\vec{k}>=k^\mu|\vec{k}>$, with the
eigenvalues
\begin{equation}
\hat{P}^0\ :\ \ k^0=\omega(\vec{k}\,)\ \ \ ;\ \ \
\hat{\vec{P}}\ :\ \ \vec{k}\ .
\end{equation}
In particular, the relativistic invariant mass eigenvalue of
these states is $m^2$, showing that the parameter $m$ indeed
measures the mass of the quanta of the quantized field.
Likewise for the generalized angular-momentum operator $\hat{M}^{\mu\nu}$,
the 1-particle states $|\vec{k}>$ possess an eigenvalue which measures
their orbital angular-momentum, thus expressing the fact that the
quanta associated to the scalar field $\phi(x^\mu)$ are indeed
spinless particles. In order to obtain 1-particle states with a
nontrivial spin value, one has to use fields which transform
nontrivially under the Lorentz group SO(3,1), such as a vector field
lea\-ding then to particles of unit spin or helicity (the latter in the
massless case), or a spinor field (whether a Weyl, a Dirac or a Majorana
spinor) leading to particles of $1/2$ spin or helicity values (Grassmann odd 
variables must be used to parametrize spinor field degrees of freedom, 
leading, at the classical level, to Grassmann graded Poisson bracket
structures and, at the quantum level, to anticommutation rather than
commutation rules for fermionic quantum operators).

Hence, as expected on basis of the heuristic construction of 
Section~\ref{Subsect3.1}, the Fock space representation of a
relativistic quantum field theory (whose action is quadratic
in the field) shows that the physical content of such a system
is that of an arbitrary ensemble of identical free relativistic quantum 
point-particles of definite mass, energy- and angular-momentum. 
The interpretation
of the field quanta as being such relativistic particles is made consistent 
by the manifest Poincar\'e invariance of the action principle. 

The above Fock space construction of the quantized field is performed
within the Schr\"odinger picture at the reference time $x^0_0=0$. Within the
cor\-res\-pon\-ding Heisenberg picture, states are time independent
whilst the quantum operators, among which the basic field
$\hat{\phi}(\vec{x}\,)$, are rather now explicitly time dependent and
carry the whole dynamics of the system. Given the quantum Hamiltonian
(\ref{eq:normalorderedH0}), it is straightforward to show, based exactly
on the definition (\ref{eq:Heisenbergpicture}), that in the Heisenberg
picture the relativistic quantum scalar field is given precisely
by the expression (\ref{eq:fieldparticle}) which was constructed
heuristically in Section~\ref{Subsect3.1}. Hence, it is precisely the ordinary
rules of canonical quantization, and only these, which, when applied to the
classical system describing the dynamics of a relativistic field theory,
lead to a framework which rea\-dily accounts for all the observed physical
spacetime properties of relativistic quantum particles including the
possibility of their creation and annihilation, which is possible
only within a formalism which includes both special relativity and
quantum mechanics.

In particular, acting with the quantum field $\hat{\phi}(x^\mu)$ in the
Heisenberg picture on the Fock vacuum, one obtains a plane wave superposition
of 1-particle states of definite momentum,
\begin{equation}
\hat{\phi}(x^\mu)|0>=
\int_{(\infty)}\frac{d^3\vec{k}}{(2\pi)^32\omega(\vec{k}\,)}\,
e^{ik\cdot x}\,|\vec{k}\,>\ .
\end{equation}
Such a state may thus be viewed as the quantum configuration of the field
such that one particle has been created exactly at the spacetime point 
$x^\mu$, which, as a consequence of Heisenberg's uncertainty principle,
thus possesses a totally undertermined energy-momentum value with its
characteristic plane wave probability amplitude. More generally,
this interpretation also enables one to construct the probability amplitude
for the process in which one particle is created at a given initial spacetime
point $x^\mu_i$ and then annihilated at the final point $x_f$, while it
propagates in a causal manner between these two positions. This quantity
is thus defined by the time-ordered two-point function of the field
operator,
\begin{equation}
\begin{array}{r l}
<0|T\left(\hat{\phi}(x_f)\hat{\phi}(x_i)\right)|0>=&
\theta(x^0_f-x^0_i)\,<0|\hat{\phi}(x_f)\hat{\phi}(x_i)|0>\\
 & \\
+&\theta(x^0_i-x^0_f)\,<0|\hat{\phi}(x_i)\hat{\phi}(x_f)|0>\ ,
\end{array}
\end{equation}
($\theta(x)$ is the usual step function such that $\theta(x>0)=1$
and $\theta(x<0)=0$) and corresponds to what is called the Feynman propagator 
for single field quanta. Using the explicit expansion (\ref{eq:fieldparticle})
of the field operator in the Heisenberg
picture in terms of the creation and annihilation operators, it is
straightforward to establish that the Feynman propagator is given by the
manifestly spacetime invariant expression
\begin{equation}
<0|T\left(\hat{\phi}(x_f)\hat{\phi}(x_i)\right)|0>=
\int_{(\infty)}\frac{d^4k^\mu}{(2\pi)^4}\,e^{-ik\cdot(x_f-x_i)}\,
\frac{i}{k^2-m^2+i\epsilon}\ ,
\label{eq:Feynmanpropagator}
\end{equation}
where the infinitesimal parameter $\epsilon>0$ is introduced in order
to specify the contour integration in the complex plane for the
energy contribution $k^0$, so that the correct causal structure of this 
propagator is recovered. This quantity is also one of the Green
functions for the Klein-Gordon operator $[\partial_\mu\partial^\mu+m^2]$.

\vspace{5pt}

Hence, the marriage of special relativity and of quantum me\-cha\-nics, 
namely of the constants $c$ and $\hbar$, leads in a most natural way
to a fundamental convergence and unification of concepts: relativistic 
quantum particles
are nothing but the quanta of relativistic quantum fields, displaying
at the same time the corpuscular properties of particles and the
wavelike properties of the spacetime dynamics of fields. This is indeed
a most powerful and all encompassing outcome of the unification of
relativity and quantum me\-cha\-nics. Among other consequences, it explains
at once why identical particles are necessarily indistinguishable,
since they simply correspond to actual physical quantum fluctuations
of a single physical entity filling all of spacetime, namely the
corresponding relativistic quantum field, and which may be excited or
absorbed, namely created or annihilated, by acting on the system
through some interaction with another field. In fact, and as shall
become clear in Section~\ref{Sect4}, even interactions, namely changes
in the total energy-momentum content of given quantum field states,
are understood in terms of exchanges of such 1-particle quanta between
given quantum states. The notion of a force acting on a re\-la\-ti\-vistic
particle, or of a potential energy contributing to the Hamiltonian
of a quantum system, is also superseded by that of fields filling all
of spacetime, and interacting with one another through local spacetime
couplings, thereby leading to the exchanges of 1-particle quanta.
Other profound consequences of the relativistic quantum field picture
of physical reality are the spin-statistics connection (namely the fact
that integer spin particles obey the Bose-Einstein statistics while
half-integer spin particles the Fermi-Dirac statistics), the invariance
of any relativistic quantum field theory under the combined product of
the parity, time reversal and charge conjugation transformations (the 
so-called CPT theorem), and the particle/antiparticle duality (only this 
latter point is discussed explicitly hereafter).

It is clear that the Fock space quantization of field theory is
ideally suited for a perturbative description of interactions, namely
by starting with a situation with only free quanta, corresponding to
an action which is quadratic in the fields, and then adding as
perturbations to be summed through a series expansion
further corrections involving locally in spacetime higher order
pro\-ducts of the fields and their couplings, thus leading to
successive perturbative corrections to quantum matrix elements of specific
observables which may be viewed in terms of specific 1-particle exchanges
among quantum states. This procedure will briefly be outlined in 
Section~\ref{Sect4}. On the other hand, the Schr\"odinger functional 
quantization of a field theory is from the outset nonperturbative in 
character, and may thus be better suited to study nonperturbative issues 
in quantum field theory, in ways that have not been explored to the same 
extent as the perturbative picture of quantum field theory.

A final remark may also be in order concerning some vocabulary. Note that
exactly the same methods of canonical quantization are applied whether
for a finite or an infinite dimensional dynamical system. Often in the
literature, one finds written that the first situation is that of
``first quantization", while the second that of ``second quantization".
Furthermore, there is also quite often mention of ``negative energy
states" and ``negative probabilities", which must be circumvented through
``second quantization". The fact of the matter is that this vocabulary
is due to an historical accident. Initially, one wished to develop a
relativistic extension of the nonrelativistic Schr\"odinger equation
for, say, the harmonic oscillator and its configuration space wave function. 
Doing so, one unavoidably encounters diverse
problems of negative energy and/or probability states, which defy a consistent
physical interpretation. Considering then that the 
``relativistic wave function" itself needs to be quantized, one
discovered that these issues are evaded altogether, leading in fact to the
quantum field theory representations that were described above.
In other words, the correct physical point of view is that, rather than
quantizing some relativistic wave function, from the outset one is in fact 
(first!) quan\-ti\-zing a classical field theory which obeys some relativistic 
invariant wave equation, and at no point whatsoever do issues of ``negative 
energy or probability states" arise.\cite{Wein1,Wein2,PS} In the same way 
that quantum mechanics,
whether relativistic or not, is the quantization of finite dimensional
systems whose configurations represent as a function of time, say, 
the positions in space of a finite collection of particles, quantum field 
theory is the quantization of infinite dimensional systems whose 
configurations are, say, the values taken by a finite collection of
fields in space as a function of time, all in a spacetime
invariant manner in the case of a relativistic field theory.

\section{Interactions and the Gauge Symmetry Principle}
\label{Sect4}

Having understood how the dynamics of a relativistic quantum field
whose Lagrangian density is quadratic in the field in fact describes
a system whose quantum states correspond to an arbitrary number of
identical free relativistic quantum particles of definite energy-momentum,
spin and invariant mass, it becomes possible to envisage an extension of this
formalism in order to account for interactions among such particles,
namely the exchange of energy and momentum between such quantum states
through the creation and annihilation of the associated quanta. Clearly,
such a formulation is perturbative in character, since the free particle
picture provides the starting point for a perturbative expansion in
which an increasing number of interaction points are included for a given
physical process. The purpose of the present section is to briefly
outline how this point of view, which has proved to be so powerful and
relevant to high energy particle physics and their fundamental interactions
except for the gravitational one, has led, on the one hand, to the local 
gauge symmetry principle as an essential requirement for any theory of the 
fundamental interactions, and on the other hand, to the Feynman diagrammatic 
representation of physical processes through a perturbative expansion of 
the associated probability amplitudes order by order in the exchanges of 
interacting particles.

\subsection{Field coupling and interactions}
\label{Subsect4.1}

For definitiness, the discussion to be presented uses the simplest of
examples, namely that of an interacting real scalar field $\phi(x^\mu)$
whose Lagrangian density now includes also a quartic term in the
potential contribution, in addition to the quadratic contribution
considered so far,
\begin{equation}
{\cal L}(\phi,\partial_\mu\phi)=\frac{1}{2}\left(\partial_\mu\phi\right)^2-
\frac{1}{2}m^2\phi^2-\frac{1}{4}\lambda\phi^4\ ,
\end{equation}
$\lambda>0$ being a real positive parameter which turns out to correspond
to a coupling constant measuring the strength of a spacetime local interaction 
in which four quanta of the field $\phi(x)$ are involved in a perturbative 
expansion. Compared to the free field case, we thus have
\begin{equation}
{\cal L}={\cal L}_0\ +\ {\cal L}_{\rm int}\ \ ,\ \ 
{\cal L}_{\rm int}=-\frac{1}{4}\lambda\phi^4\ ,
\end{equation}
${\cal L}_0$ being the free field Lagrangien density whose quantization
has been discussed above, while ${\cal L}_{\rm int}$ corresponds
to an additional contribution associated to some specific interaction.
The canonical quantization of such a system follows the same rules as those
applied in the free field case, with in particular the fundamental
Poisson brackets
\begin{equation}
\{\phi(x^0,\vec{x}\,),\pi(x^0,\vec{y}\,)\}=
\delta^{(3)}\left(\vec{x}-\vec{y}\,\right)\ ,
\end{equation}
which remain those of the free field case. Note that the conjugate momentum
is still given by the relation $\pi(x)=\partial_0\phi(x)$ (had
the interacting Lagrangian ${\cal L}_{\rm int}$ included some derivative 
coupling of the field $\phi$, the conjugate momentum would have been 
different). However, the canonical Hamiltonian density acquires an 
additional contribution directly related to and determined by 
${\cal L}_{\rm int}$, namely
\begin{equation}
\begin{array}{r c l}
{\cal H}&=&\frac{1}{2}\pi^2+\frac{1}{2}\left(\vec{\nabla}\phi\right)^2
+\frac{1}{2}m^2\phi^2+\frac{1}{4}\lambda\phi^4\\
 & & \\
&=&{\cal H}_0\ + \ {\cal H}_{\rm int}\ \ ,\ \ 
{\cal H}_{\rm int}=-{\cal L}_{\rm int}=\frac{1}{4}\lambda\phi^4\ .
\end{array}
\end{equation}
The restriction on the coupling constant $\lambda>0$ stems from the
requirement that the energy spectrum of the system be bounded below,
since otherwise no stable ground state may exist. The same requirement
also explains why a purely $\phi^3$ term, without the quartic contribution
in ${\cal L}_{\rm int}$, is not considered in the above discussion,
even though the perturbative expansion to be described presently is then 
somewhat simpler to implement in actual calculations.

Consequently, the canonical quantization of the system, even in the 
pre\-sen\-ce of the interaction contribution, may still be performed, say, 
in the Fock space representation in terms of the creation and annihilation
operators of free particle quanta, with a specific definition of a
self-adjoint Hamiltonian operator $\hat{H}=\int_{(\infty)}d^3\vec{x}\,
\hat{\cal H}$ through normal ordering in these operators. However, what
then becomes a nontrivial issue is the actual diagonalization of
this Hamiltonian, namely the identification of the actual spectrum
of the quantized interacting field theory. A perturbative approach in
the parameter $\lambda$ enables an order by order identification of
the quantum physical content of such a system and of its physical properties,
starting from the free field quanta.

\vspace{15pt}

\noindent\underline{The scattering matrix}

\vspace{10pt}

In practical terms, an extremely important method for the experimental 
investigation of the quantum relativistic properties of physical systems is 
that of scattering measurements. Different components of a given system are
prepared in a given initial configuration in causally separated regions of
space, and are then made to scatter within a given local neighbourdhood of
an interaction point, from where interaction products emerge whose properties
are then measured and analyzed, in order to infer the specific characteristics
of the interactions at work and responsible for the observed process.
In other words, all the physical information related to these interactions
is encoded into the corresponding scattering probability amplitude.

Given such a general scheme, the basic implicit idea is that the interaction
takes place over a region of space whose extent is so small that for all
practical purposes the interactions are only short-ranged, so that beyond
that interaction region the separated components of the system are free
from interactions. In a classical picture, such components may be viewed as
independent free particles each following asymptotically a straight 
trajectory. When the interactions are ``turned off", these trajectories are 
not modified as they pass one another, and are thus not scattered. 
However, when the interaction is ``turned on", the more the particles 
approach one another, the more their trajectories deviate from a straight 
path, leading in the asymptotic final state to a scattered configuration of 
straight trajectories as the final state components which emerge from the 
spatial interaction region. In other words, the cha\-rac\-te\-ri\-za\-tion of a 
nontrivial scattering process proceeds by extrapolating to both the infinite 
past and the infinite future the time dependent dynamics of a given 
configuration of the system, and by comparing the asymptotic states to what 
they would have been had there not been any interaction.

Clearly, the same heuristic understanding of the characterization of the
scattering process applies at the quantum level, by comparing the time
dependency of given in- and out-states in the presence or absence of some
given interaction, provided the initial asymptotic states are identical. The 
characterization of the scattering process, and of the interaction 
responsible for it, is then obtained by identifying the operator in Hilbert 
space which leads to this transition between the in- and out-asymptotic 
states. This is the scattering operator $S$ whose matrix elements are thus 
the quantities of interest, which represent the probability amplitude for 
a given physical scattering process to occur.

Let us translate this reasoning in mathematical terms. Concentrating first on 
the initial state, let us represent the free Hamiltonian by $\hat{H}_0$, 
the total Hamiltonian including interactions by $\hat{H}$, and assume to be
working in the Schr\"odinger picture at some reference time $t_0$.
A given state $|\psi_{\rm in},t_0>$ of the free theory is then evolved
backwards in time into the asymptotic in-state
\begin{equation}
|\psi_{\rm in},-\infty>=
\lim_{t\rightarrow -\infty}e^{-i(t-t_0)\hat{H}_0}\,|\psi_{\rm in},t_0>=
\lim_{t\rightarrow -\infty}|\psi_{\rm in},t>\ ,
\end{equation}
while a given state $|\psi,t_0>$ of the interacting theory is likewise
propagated back in the infinite past according to
\begin{equation}
|\psi,-\infty>=
\lim_{t\rightarrow -\infty}e^{-i(t-t_0)\hat{H}}\,|\psi,t_0>=
\lim_{t\rightarrow -\infty}|\psi,t>\ .
\end{equation}
However, these two asymptotic states should correspond to an identical
asymptotic quantum in-state, so that the asymptotic correspondence is
defined by the relation
\begin{equation}
|\psi,-\infty>=|\psi_{\rm in},-\infty>\ .
\end{equation}
Likewise for the asymptotic quantum out-state, one has the identification
\begin{equation}
|\chi,+\infty>=|\chi_{\rm out},+\infty>\ ,
\end{equation}
where
\begin{equation}
|\chi_{\rm out},+\infty>=
\lim_{t\rightarrow +\infty}e^{-i(t-t_0)\hat{H}_0}\,|\chi_{\rm out},t_0>=
\lim_{t\rightarrow +\infty}|\chi_{\rm out},t>\ ,
\end{equation}
\begin{equation}
|\chi,+\infty>=
\lim_{t\rightarrow +\infty}e^{-i(t-t_0)\hat{H}}\,|\chi,t_0>=
\lim_{t\rightarrow +\infty}|\chi,t>\ .
\end{equation}
Note that behind this construction lies the fact that the quantum theories
based on $\hat{H}_0$ and $\hat{H}$ share a common space of quantum states,
namely an identical representation space of a common algebraic structure
of commutation relations for the fundamental degrees of freedom. The scattering
operator, whose matrix elements we are about to characterize, is thus an
operator acting withing this common space of quantum states, which must 
reduce to the identity operator in the absence of any interaction, 
$\hat{H}=\hat{H}_0$.

Given the above formulation, it is clear that the transition probability
amplitude between the asymptotic in- and out-states of the interacting
theory is simply given by
\begin{equation}
<\chi,t|\psi,t>=<\chi,t_0|\psi,t_0>\ ,
\end{equation}
the value of this matrix element being independent of the time $t$ at which 
it is evaluated since the evolution operator $e^{-i(t-t_0)\hat{H}}$ for the
interacting theory defines a unitary isomorphism between all Schr\"odinger 
pictures for all values of $t$. However, this matrix element may also be 
expressed in terms of the in- and out-states of the free theory, since the 
asymptotic in-states for either theory are identical. A direct substitution 
of the above relations then finds
\begin{equation}
<\chi,t|\psi,t>=<\chi,t_0|\psi,t_0>=<\chi_{\rm out},t_0|S|\psi_{\rm in},t_0>\ ,
\label{eq:Svalue}
\end{equation}
where the scattering operator $S$ is defined by the asymptotic limits
\begin{equation}
S=\lim_{t_\mp\rightarrow\mp\infty}\,M(t_+,t_0)\,M^\dagger(t_-,t_0)\ ,
\end{equation}
with
\begin{equation}
M(t,t_0)=e^{i(t-t_0)\hat{H}_0}\,e^{-i(t-t_0)\hat{H}}\ .
\end{equation}
Note that in the absence of any interaction, $\hat{H}=\hat{H}_0$,
the scattering operator $S$ indeeds reduces to the identity operator.
Since the operator $M(t,t_0)$ plays such a central role in the construction
of the scattering operator $S$, it is important to obtain alternative
expressions for it. In particular, one readily establishes
the differential equation
\begin{equation}
\begin{array}{r c l}
i\partial_t\,M(t,t_0)&=&e^{i(t-t_0)\hat{H}_0}\,\left[\hat{H}-\hat{H}_0\right]\,
e^{-i(t-t_0)\hat{H}}\\
 & & \\
&=&e^{i(t-t_0)\hat{H}_0}\,\hat{H}_{\rm int}(t_0)\,e^{-i(t-t_0)\hat{H}_0}\,
M(t,t_0)\\
 & & \\
&=&\hat{H}^{(I)}_{\rm int}(t)\,M(t,t_0)\ ,
\end{array}
\label{eq:Moller}
\end{equation}
having introduced
\begin{equation}
\hat{H}^{(I)}_{\rm int}(t)=e^{i(t-t_0)\hat{H}_0}\,\hat{H}_{\rm int}(t_0)\,
e^{-i(t-t_0)\hat{H}_0}\ \ ,\ \ \hat{H}_{\rm int}(t_0)=\hat{H}-\hat{H}_0\ .
\end{equation}
Note that this latter definition coincides with that of the Heisenberg
picture associated to the free Hamiltonian $\hat{H}_0$. Since in the
interacting theory the Heisenberg picture should be defined
in a likewise manner but in terms of the full Hamiltonian $\hat{H}$ rather
than the free Hamiltonian $\hat{H}_0$, one refers to the ``interaction 
picture" as being associated to the general definition of time dependent 
operators ${\cal O}_{(I)}$ given by
\begin{equation}
{\cal O}_{(I)}(t)=e^{i(t-t_0)\hat{H}_0}\,{\cal O}(t_0)\,
e^{-i(t-t_0)\hat{H}_0}\ ,
\end{equation}
where ${\cal O}(t_0)$ is the operator as constructed through canonical 
quantization of the interacting theory in its Schr\"odinger picture. 

In other words, in the interaction picture, quantum states as well as 
ope\-ra\-tors carry a split time dependency, such that the one carried by the 
quantum states is solely induced by the interactions and the interacting 
Hamiltonian $\hat{H}_{\rm int}$, while the one carried by the quantum 
operators is solely induced by the time dependency related to the free 
field dynamics and the free Hamiltonian $\hat{H}_0$. In the interaction 
picture, any time dependency in the quantum states is totally ascribed to 
the interactions only.

Returning to the equation (\ref{eq:Moller}) characterizing the operator 
$M(t,t_0)$, one sees that its solution may also be expressed in the form
\begin{equation}
M(t,t_0)=T\,e^{-i\int_{t_0}^t\,dt'\,\hat{H}^{(I)}_{\rm int}(t')}\ ,
\end{equation}
where the symbol $T$ in front of the exponential in the r.h.s. of this
expression stands for the time-ordered product and exponential in which 
products of time-dependent operators are integrated from left to right in
decreasing order of their time arguments (this is indeed required given
that the operator $M(t,t_0)$ is to the right of $\hat{H}^{(I)}_{\rm int}(t)$
in (\ref{eq:Moller})). 

Hence, using this solution for the operator
$M(t,t_0)$, the scattering ope\-ra\-tor acquires the expression
\begin{equation}
S=T\,e^{-i\int_{t_0}^\infty\,dt\hat{H}^{(I)}_{\rm int}(t)}\,
T\,e^{-i\int_{-\infty}^{t_0}\,dt\hat{H}^{(I)}_{\rm int}(t)}=
T\,e^{-i\int_{-\infty}^\infty dt\int_{(\infty)}d^3\vec{x}\,
\hat{\cal H}^{(I)}_{\rm int}}\ ,
\label{eq:ST}
\end{equation}
which, in the absence of any derivative coupling in the interacting Lagrangian
density, so that ${\cal L}_{\rm int}=-{\cal H}_{\rm int}$, finally reduces to
\begin{equation}
S=T\,e^{-i\int_{(\infty)}d^4x^\mu\,\hat{\cal H}^{(I)}_{\rm int}}\ =\
T\,e^{i\int_{(\infty)}d^4x^\mu\,\hat{\cal L}^{(I)}_{\rm int}}\ .
\end{equation}

\vspace{5pt}

In this form, it should be clear why this formulation of any scattering
process is ideally suited for a perturbative treatment. Since scattering
matrix elements are given by matrix elements of the operator $S$ for
free field external states, see (\ref{eq:Svalue}), it suffices to consider 
the creation and annihilation mode expansions of the field and its conjugate 
momentum in the interaction picture, and substitute these in the expressions 
for the interacting Lagrangian and Hamiltonian densities in the interaction 
picture. In particular, these fields in the interaction picture retain their 
expressions valid for the Heisenberg picture of the free field theory. One has
\begin{equation}
\begin{array}{r c l}
\hat{\phi}_{(I)}(t,\vec{x}\,)&=&e^{i(t-t_0)\hat{H}_0}\,
\hat{\phi}(t_0,\vec{x}\,)\,e^{-i(t-t_0)\hat{H}_0}\ ,\\
 & & \\
\hat{\pi}_{(I)}(t,\vec{x}\,)&=&e^{i(t-t_0)\hat{H}_0}\,
\hat{\pi}(t_0,\vec{x}\,)\,e^{-i(t-t_0)\hat{H}_0}\ ,
\end{array}
\end{equation}
with the mode expansions
\begin{equation}
\hat{\phi}_{(I)}(x)=\int_{(\infty)}\frac{d^3\vec{k}}
{(2\pi)^32\omega(\vec{k}\,)}\left[a(\vec{k}\,)e^{-ik\cdot x}\,+\,
a^\dagger(\vec{k}\,)e^{ik\cdot x}\right]\ ,
\end{equation}
\begin{equation}
\hat{\pi}_{(I)}(x)=\int_{(\infty)}\frac{d^3\vec{k}}
{(2\pi)^32\omega(\vec{k}\,)}\left(-i\omega(\vec{k}\,)\right)
\left[a(\vec{k}\,)e^{-ik\cdot x}\,-\,
a^\dagger(\vec{k}\,)e^{ik\cdot x}\right]\ ,
\end{equation}
while the creation and annihilation operators still obey the usual
algebra
\begin{equation}
\left[a(\vec{k}\,),a^\dagger(\vec{\ell}\,)\right]=(2\pi)^32\omega(\vec{k}\,)
\delta^{(3)}\left(\vec{k}-\vec{\ell}\,\right)\ ,
\end{equation}
since canonical quantization in the Schr\"odinger picture of the interacting
theory still requires the commutation relations
\begin{equation}
\left[\hat{\phi}(t_0,\vec{x}\,),\hat{\pi}(t_0,\vec{y}\,)\right]=
i\delta^{(3)}\left(\vec{x}-\vec{y}\,\right)\ .
\end{equation}
Furthermore, once such a substitution has been effected, a straightforward
expansion of the time-ordered exponential (\ref{eq:ST}) defining the 
scattering ope\-ra\-tor in terms of the interacting Hamiltonian in the 
interaction picture leads to an expansion in powers of the coupling 
coefficient $\lambda$, namely a perturbative representation of the 
probability amplitude associated to a given set of external states in terms 
of successive exchanges of free particle quanta being created and 
annihilated through the interaction couplings of the fields as they contribute 
to the interacting Hamiltonian.

In particular, it should be clear that successive contractions of these
creation and annihilation operators as they are commuted past one another
in the evaluation of the matrix elements, all in a manner consistent with 
the causal time ordering implied by the solution (\ref{eq:ST}), always lead 
precisely to the time-ordered two-point function of the field operator in the
interaction picture, namely the Feynman propagator computed previously
for the free field theory,
\begin{equation}
<0|T\left(\hat{\phi}_{(I)}(x)\hat{\phi}_{(I)}(y)\right)|0>=
\int_{(\infty)}\frac{d^4k^\mu}{(2\pi)^4}\,\frac{i}{k^2-m^2+i\epsilon}\,
e^{-ik\cdot(x-y)}\ ,
\end{equation}
where $|0>$ still denotes the perturbative normalized Fock vacuum annihilated 
by the operators $a(\vec{k}\,)$, $a(\vec{k}\,)|0>=0$.

Even though we cannot consider here a discussion of perturbation
theory in any detail whatsoever, once put within such a framework,
it takes little effort of imagination to understand how a systematic
set of rules for such a perturbative expansion and evaluation of
scattering matrix elements may be identified, thus providing an
efficient approach towards the determination
of scattering cross sections of direct relevance to experimental
results. Such a discussion would consist in a whole set of lectures
on their own, which is not the purpose of the present notes and may be found
exposed in great detail in any quantum field theory 
textbook.\cite{Wein2,PS,IZ,Ramond}
Nonetheless, from the above description, it should be clear that
Fock space quantization of relativistic quantum field theory is ideally
suited for a perturbative representation of interacting relativistic quantum
particles, and that this perturbation theory approach is directly based
on the interacting Hamiltonian and Lagrangian contribution to the total
Lagrangian density, namely all those contributions which are not
purely quadratic in the fields, the latter on their own being relevant to the
description of free relativistic quantum particles.

\vspace{10pt}

\noindent\underline{Perturbation theory}

\vspace{10pt}

In spite of the fact that this is not the place for a detailed presentation
of perturbative quantum field theory, let us nevertheless highlight some
points of relevance to the discussion hereafter, particularizing again to
the simplest ${\cal L}_{\rm int}=-\lambda\phi^4/4$ interacting Lagrangian.
As far as scattering processes are concerned, all possible results are
encoded into the scattering operator
\begin{equation}
S=\one\ +\ \sum_{n=1}^\infty\,\frac{1}{n!}\,
T\left(-i\int_{(\infty)}d^4x^\mu\,\hat{\cal H}^{(I)}_{\rm int}(x)\right)^n\ ,
\end{equation}
where now the interacting Hamiltonian density in the interaction picture
is defined according to the usual normal ordering prescription for the
creation and annihilation operators $a^\dagger(\vec{k}\,)$ and $a(\vec{k}\,)$,
\begin{equation}
\hat{\cal H}^{(I)}_{\rm int}(x)=\frac{1}{4}\lambda\,:\hat{\phi}^4_{(I)}(x):\ .
\end{equation}
Clearly, when considering the scattering operator in this series expanded
form and the evaluation of its matrix elements for external states associated
to de\-fi\-ni\-te numbers of incoming and outgoing particles, time ordering
of operator products commuted with one another implies the contribution
of the Feynman propagator which, in momentum space, simply leads to the
following contribution for any internal propagating line connecting two
interaction vertices at which particle quanta are created or annihilated,
\begin{equation}
\frac{i}{k^2-m^2+i\epsilon}\ .
\end{equation}
Likewise, whenever the operator $\hat{\cal H}^{(I)}_{\rm int}(x)$ contributes
at a given order of the perturbative expansion, it implies a spacetime
local interaction in which four particle quanta are either created or
annihilated, with an amplitude given by the factor
\begin{equation}
-\frac{1}{4}i\lambda\ ,
\end{equation}
up to some combinatorics factor depending on the topology of the associated
diagram.

In other words, it is possible to translate the mathematical expression for 
the relevant matrix element evaluation into a diagrammatic representation
in which internal lines are connected to interaction vertices, and for which
the above contributions are then multiplied with one another, and integrated
over internal momenta in a manner such as to obey the rules of energy-momentum
conservation at each vertex, in order to determine the associated probablity
amplitude. These rules relating such Feynman diagrams to the required
ma\-the\-ma\-ti\-cal quantity are the Feynman rules of perturbative quantum 
field theory. In the specific case of the $\lambda\phi^4/4$ scalar field theory,
the above discussion thus establishes that these rules consist only of
the single interaction vertex accompanying the scalar Feynman
propagator. In principle, given such rules, any scattering amplitude for
whatever physical process may be computed to an arbitrary order in the
perturbative expansion in the coupling constant $\lambda$.

\vspace{5pt}

As far as we are concerned, the main conclusion to be drawn from the above
is that once relativistic quantum fields are coupled to one another
through local spacetime couplings, such as 
${\cal L}_{\rm int}=-\lambda\phi^4/4$, one in facts has made available
within a perturbative picture a formalism in which local and causal quantum 
interactions are directly understood in terms of exchanges of quantum 
particles free to propagate between interaction vertices that occur
locally in spacetime but at arbitrary positions which are integrated over
when they are not observed. The marriage of $\hbar$ and $c$ leading
to quantum field theory as the natural framework for the description
of relativistic quantum point-particles also implies a physical understanding
of the physical origin of forces and interactions simply as following
from the spacetime local couplings of fields, which also translate in
the dual corpuscular picture into a process in which particles are being 
created, annihilated and exchanged, thereby lea\-ding to 
changes in their energy-momentum, hence to their interactions.
The mysterious action at a distance of classical mechanics is forever gone,
superseded by relativistic quantum fields which provide a natural framework
not only for a unified description both of the corpuscular properties of matter
and of the wavelike properties of their spacetime dynamics, but also a
unified understanding of the fundamental quantum interactions in terms of
both spacetime local couplings of fields and causal exchanges of particle 
quanta, all in a manner consistent with the principles of special relativity,
of unitary quantum mechanics, and of causality.

\vspace{5pt}

However, this amazing convergence of physical concepts based on a few
general basic principles comes with a price. When considering the perturbative
expansion of scattering matrix elements, one soon comes across loop diagram
contributions in which one must integrate over the internal momenta running
around closed loops. For instance when considering the propagation of
a single particle quantum, the first order correction to the propagator
is obtained by inserting into it the four-point vertex $\lambda\phi^4/4$
and then contracting two of its four external lines with one another, 
leading to a 1-loop contribution with the factor
\begin{equation}
\left(-\frac{1}{4}i\lambda\right)\int_{(\infty)}\frac{d^4p^\mu}{(2\pi)^4}\,
\left(\frac{i}{p^2-m^2+i\epsilon}\right)\ ,
\end{equation}
the origin of each of the factors in parentheses being obvious, while the 
closed loop propagator must be integrated over the associated energy-momentum.
Likewise, when considering a $2\rightarrow 2$ scattering process with
two initial and two final particles, beyond the nonscattering and
one interaction vertex contributions, there appears a 1-loop correction
in which two 4-point vertices are inserted with two lines of each being
contracted in pairs with two lines of the other. The corresponding 
contribution is given by
\begin{equation}
\left(-\frac{1}{4}i\lambda\right)^2\int_{(\infty)}\frac{d^4p^\mu}{(2\pi)^4}\,
\left(\frac{i}{p^2-m^2+i\epsilon}\right)\,
\left(\frac{i}{(p+k)^2-m^2+i\epsilon}\right)\ ,
\end{equation}
where $p^\mu$ is again the energy-momentum running around the closed loop
(say, that running through one of the two internal contracted lines), while 
$k^\mu$ is the total external energy-momentum of the two initial or final 
particles ($k+p$ being then the energy-momentum running through the other 
internal line).

The characteristic feature of such contributions, which arise whenever
closed loops appear in a diagram, is their divergence for large values 
of the internal momentum, namely in the ultra-violet at small distances.
The fundamental reason for this feature is that interactions occur locally
in spacetime at given points where the fields are multiplied with one another.
In order to perform calculations nonetheless, one has to introduce some
regularization procedure to tame such divergencies, and hope that at the
very end, when all contributions are summed up again, all the divergent
contributions would combine is such a manner that physical observables remain
nevertheless finite, even if affected by finite renormalization. Many different
regularization procedures have been developed, and this is not the place
to discuss such issues.\cite{Wein2,PS,IZ,Ramond} 
The most straightforward one is to introduce an
upper cut-off value $\Lambda_c$ in the momentum integration,
to keep track of the different types of divergencies that may arise.
For instance, the 1-loop correction to the scalar field propagator given
above leads to a quadratic divergence proportional to $\Lambda^2_c$,
while that to the $2\rightarrow 2$ scattering process is only
logarithmically divergent and proportional to $\ln\Lambda_c$,
as may easily be seen through simple power counting and dimensional analysis
of the relevant expressions.

The crucial issue thus arises as to which are the interacting quantum
field theories which, in a perturbative quantization, lead to physically
meaningful and thus finite predictions for scattering processes, in spite of
the existence of these ultraviolet short distance higher-loop divergencies.
In practical terms, and to put it into just a few words, here is how the
procedure works. Given any specific regularization procedure, order after
order in perturbation theory, one needs to add further and further
corrections (``counterterms") to the initial Lagrangian density,
in order to introduce additional contributions to scat\-te\-ring amplitudes
such that the perturbative series summed up to the given order remains
finite when the regulator is removed, thereby leading to a finite physical
result, even though the basic quantities appearing separately in the
renormalized Lagrangian may be divergent. However, if the number of the
required countertems grows with the order in perturbation theory, no
specific prediction remains possible, since each new counterterm requires
the specification of a new coupling constant whose value may be inferred
only from experiment. Hence, a quantum field theory possesses any predictive
power provided only a finite number of counterterms is required to render
the renormalized scattering amplitudes to whatever order of perturbation
theory finite and thus physical. Field theories for which this programme is 
feasable are called renormalizable. In fact, all such renormalizable field 
theories are such that all counterterms belong to a finite class of local
quantum operators such that the renormalization of the
theory amounts to a redefinition of the field normalizations, masses and 
couplings (the ``bare" quantities of the classical Lagrangian density) in 
terms of renormalized and finite physical observables directly related to 
the physical external states, their masses and couplings. The ``bare" 
quantities are obtained in terms of the re\-nor\-ma\-li\-zed 
ones through factors multiplying the latter, these factors
being given as power series expansions in the coupling constants whose
coefficient are divergent as the regulator is removed. Theories for
which finite renormalization is achieved in this manner are called
``multiplicatively renormalizable". These are the only perturbative
quantum field theories of possible relevance to relativistic quantum 
particle physics and their fundamental quantum interactions. Under such
circumstances, one thus obtains a predictive framework for the
representation and evaluation of these processes.

The above $\lambda\phi^4/4$ scalar field theory is the simplest example
of such a renormalizable quantum field theory. All the required counterterms
to all orders of perturbation theory simply amount to a redefinition, through 
a multiplicative factor, of the field normalization, its mass $m^2$ and its
self-coupling $\lambda$, each of these renormalization factors being given
as power series expansions in the coupling $\lambda$ whose coefficients
include both finite and infinite contributions as the regulator is
removed. Nevertheless, all physical quantities remain finite in that same 
limit, and may be predicted in terms only of the renormalized mass and 
coupling of asymptotic quanta.

\vspace{10pt}

\noindent\underline{Renormalizable relativistic quantum field theories}

\vspace{5pt}

Among all possible Lagrangian densities for collections of fields of
a variety of spin values, how does one characterize those that define
a renormalizable quantum field theory? Through power counting and
dimensional analysis of loop amplitudes, a necessary
condition, though not a sufficient one, for renormalizability may be
established. Namely, when working in units such that $\hbar=1=c$
so that all dimensionful quantities may be measured in units of mass,
whenever the Lagrangian density contains a specific contribution
whose coupling coefficient, say $\lambda$, has a mass dimension to some 
strictly negative power, $\lambda=\alpha_0/\Lambda^\kappa$ with $\alpha_0$
dimensionless, $\Lambda$ some mass scale and $\kappa>0$, then the associated 
interactions are not renormalizable.

For example, let us consider a real scalar field $\phi$ whose dynamics
derives from the Lagrangian density
\begin{equation}
{\cal L}=\frac{1}{2}\left(\partial_\mu\phi\right)^2-\frac{1}{2}m^2\phi^2-
V(\phi)\ .
\end{equation}
Since in units such that $\hbar=1=c$ the quantum action must be dimensionless,
in a four-dimensional spacetime the scalar field must have a mass dimension
of unity, as well as the mass parameter $m$. Consequently, any trilinear
coupling $g\phi^3$ contribution to the potential density $V(\phi)$ must
have a coupling strength $g$ of mass dimension unity, while a quartic
interaction $\lambda\phi^4$ a dimensionless strength coupling $\lambda$. 
In other words, in four-dimensional spacetime, any quartic potential $V(\phi)$
leads to a renormalizable quantum scalar field theory (in the absence of
a quartic coupling, a cubic coupling is excluded on physical grounds, since
otherwise the energy is not bounded below). However, any coupling of
higher order, $\lambda\phi^n$ with $n>4$, requires a strength coupling of mass
dimension $[\lambda]=4-n$, and thus represents a nonrenormalizable interaction
in a four-dimensional spacetime.

A similar analysis may be developed for any other field theory
of higher spin content. Incidentally, in the case of general relativity,
the fact that Newton's constant, which then defines the coupling strength for
gravity, has a strictly negative mass dimension is one of the reasons why
the perturbative quantization of that classical metric field theory of
spacetime geometry is nonrenormalizable.

Historically, the requirement of renormalizability was viewed as de\-fi\-ning,
albeit for physical arguments not thoroughly convincing, a basic restriction 
on the construction of realistic quantum field theories for the fundamental
interactions of the elementary particles. Nowadays, this point of view
has considerably shifted, and renormalizable quantum field theories are
rather considered to define effective low energy approximations to some
more fundamental underlying description of the basic physical phenomena,
which need not be given even in terms of a quantum field theory.\cite{Wein1} 
By integrating
out from a given theory its high energy modes above its characteristic energy
scale $\Lambda$, one recovers a low energy effective description in terms of 
a field theory in which the effects of the underlying theory relevant to the 
higher energy scales contribute only through nonrenormalizable effective 
coupling coefficients of the form $\lambda=\alpha_0/\Lambda^\kappa$. 
Hence, as the energy scale of the underlying theory becomes arbitrary
large, only renormalizable couplings survive in its low energy effective
field theory approximation, thereby leading to a decoupling of energy scales
as one passes from one level of effective description to the next.
From that point of view, the principle of renormalizability for the
construction of physical quantum field theories is nothing but a principle
for the decoupling of energy scales when formulating a theory capable of
describing phenomena up to some characteristic energy scale, without the
knowledge and independently of the physics lying beyond that energy scale.
The procedure of renormalization described above is then also seen to 
correspond to a renormalization of the low energy observables through the
resummation of all the known contributions up to some cut-off energy scale, 
beyond which there may lie some unknown territory, and then at the same time
make sure that the low energy observables remain independent of this unknown 
physics, and thus remain finite as well, as this cut-off scale possibly 
characteristic of some unknown interactions and particles is pushed to 
arbitrary large values. In effect, this indeed corresponds to a decoupling 
of scales for the effective low energy approximate quantum field theory 
description.

Nonetheless, this rationale for the decoupling of scales translated into
the requirement of renormalizability still leaves us with the general
issue of the construction of such theories. The necessary condition
mentioned above in terms of the mass dimension of interaction coupling
constants, even when met, is not sufficient to ensure renormalizability
of the corresponding coupling. The answer to this issue has been given above
in the case of scalar fields, but not yet for spinor nor vector fields
in interactions, which are certainly required for a description for the
fundamental interactions of quarks and leptons. It turns out this is far
from a trivial matter, and throughout the 1960's and early 1970's, it has 
been established\cite{HV} that the only renormalizable interactions of 
vector fields, 
massive or not, with matter are those governed by the ge\-ne\-ral gauge 
symmetry principle of Yang-Mills theories based on some internal symmetry 
whose algebraic group is a compact Lie group. The stringent and elegant 
symmetry constraints brought about by the local gauge symmetry principle on 
the structure of such interactions are just powerful enough to guarantee 
renormalizability.

Hence, in conclusion, the general principles of special relativity,
quantum mechanics and decoupling of scales for effective field theory
descriptions of the fundamental interactions and particles has led
to the general gauge symmetry principle, and its actual realization
in terms of internal symmetries, as the guiding principle for the
construction of renormalizable interacting relativistic quantum
unitary local field theories as the appropriate framework for the description
of the causal interactions of relativistic quantum point-particles and their
wavelike spacetime dynamics. Quite an achievement for the marriage of
$\hbar$ and $c$, the genuine third conceptual revolution of XX$^{\rm th}$ 
century physics following general relativity and quantum mechanics!

\subsection{Global internal symmetries}
\label{Subsect4.2}

Hence, it is time now to turn to the meaning of internal symmetries, namely
symmetries acting on a system but which are not associated to transformations
in spacetime. In technical terms, a symmetry is a transformation of a
system such that it leaves its equations of motion form invariant. Or in
other words, a symmetry transforms a given solution to the dynamics of
a system into another solution to the same dynamics. Note that a symmetry
is not necessarily an invariance property of configurations of the system,
but rather it is an invariance property of the set of its dynamical
configurations. In particular, it may be that even for the lowest
energy configuration of a system, this solution may or may not be invariant
under the action of a symmetry of the equations of motion. As we shall see,
this possibility has profound consequences in the context of field theory,
especially when it comes to symmetries that are realized locally at each
point in spacetime, so-called local gauge symmetries.

Given the character of these notes, only the simplest examples of these
different issues are presented here. However, the reader should be aware
that many generalizations have been developed, that are available in the
literature as well as standard quantum field theory textbooks.

\vspace{10pt}

\noindent\underline{The simplest example}

\vspace{5pt}

So far, we have considered only the case of a single real scalar field
of mass $m$. Let us now extend the discussion to a system composed of two 
such fields $\phi_1(x)$ and $\phi_2(x)$ sharing identical masses $m$ and 
interaction couplings. Consequently, such a system possesses a continuous 
symmetry whose transformations mix these two fields by an arbitrary amount 
while preserving their normalization, namely a rotation of arbitrary angle in
the two-dimensional space $(\phi_1,\phi_2)$ in which they take their values.
Specifically, combining the two fields into a single complex valued
scalar field,
\begin{equation}
\phi(x)=\frac{1}{\sqrt{2}}\left[\phi_1(x)+i\phi_2(x)\right]\ ,
\end{equation}
the corresponding total Lagrangian density, which then reads
\begin{equation}
\begin{array}{r c l}
{\cal L}&=&\partial_\mu\phi^\dagger\partial^\mu\phi-m^2\phi^\dagger\phi-
V(|\phi|)\\
 & & \\
&=&\frac{1}{2}(\partial_\mu\phi_1)^2-\frac{1}{2}m^2\phi^2_1+
\frac{1}{2}(\partial_\mu\phi_2)^2-\frac{1}{2}m^2\phi^2_2-
V\left(\sqrt{\phi^2_1+\phi^2_2}\right)\ ,
\end{array}
\label{eq:Lphi}
\end{equation}
$V(|\phi|)$ being an arbitrary renormalizable interaction potential,
is clearly invariant under the class of continuous transformations
\begin{equation}
\phi'(x)=e^{i\alpha}\,\phi(x)\ ,
\label{eq:symU(1)}
\end{equation}
$\alpha$ being an arbitrary constant real parameter representing the
rotation angle of this SO(2)=U(1) symmetry.

This symmetry, which leaves the action and thus also the equations of motion
invariant, is a global symmetry, since it acts in an identical fashion on
the field $\phi(x)$ irrespective of the spacetime point labelled by $x^\mu$.
The symmetry shifts the phase of the complex field by an identical amount
globally throughout the whole of spacetime, namely not only instantaneously
through all of space but also identically throughout the whole time history
of the system. Furthermore, the action of the symmetry is not on the spacetime
points at which the field is evaluated, but rather within the ``internal 
two-dimensional space" in which the complex field takes its values. From
that point of view, these values for $\phi(x)$ define a two-dimensional
space associated to each of the spacetime points, the ``internal" space of 
the system. Consequently, one says that the symmetry is a global internal one.

By virtue of Noether's theorem, associated to such a continuous symmetry,
there exists a current and its charge which are locally conserved for
solutions to the equations of motion. In the present instance, these conserved
Noether current and charge are given by
\begin{equation}
J^\mu=-i\left[\phi^\dagger\partial^\mu\phi-
\partial^\mu\phi^\dagger\phi\right]\ \ ,\ \ 
Q=\int_{(\infty)}d^3\vec{x}\,J^0(x^0,\vec{x}\,)\ ,
\end{equation}
while for solutions to the dynamics of the system, these quantities
obey the conservation conditions,
\begin{equation}
\partial_\mu J^\mu=0\ \ \ ,\ \ \ \frac{dQ}{dt}=0\ .
\end{equation}
These Noether current and charge thus characterize the specific properties 
of the system that follow from its U(1) continuous global internal symmetry. 
In particular, in its Hamiltonian formulation, the charge $Q$ generates 
the algebra of the symmetry group, in the present case that of the abelian 
group U(1), through the Poisson bracket structure. Acting on phase space 
through these brackets, the Noether charge also generates, in linearized 
form, the associated symmetry transformations of the phase space degrees of 
freedom. Through the correspondence principle, the same properties should 
remain valid at the quantum level in terms of commutation relations. However, 
because of possible operator ordering ambiguities for composite quantities 
such as Noether charges and currents, it may be that the quantum consistency 
requirements for the definition of quantum physical observables clash with the
symmetry properties, namely that the symmetry algebra is no longer
realized in terms of the commutation relations of the Noether charges.
In such a case, the symmetry is said to be anomalous, by which is meant in
fact that the symmetry is explicitly broken for the quantized system.

In the present case, it may be checked by straightforward construction that
the U(1) internal symmetry is not anomalous. Associated to the creation and
annihilation mode expansions of the real fields $\phi_1$ and $\phi_2$, the
complex field $\phi(x)$ acquires of course also such an expansion, but
in terms of creation and annihilation operators which are superpositions
of those of the initial fields. Having initially two independent fields,
one still obtains two independent sets of creation and annihilations
operators, given by
\begin{equation}
\begin{array}{r c l c r c l}
a(\vec{k}\,)&=&\frac{1}{\sqrt{2}}\left[a_1(\vec{k}\,)+ia_2(\vec{k}\,)\right]\
&,&\
b^\dagger(\vec{k}\,)&=&\frac{1}{\sqrt{2}}\left[a^\dagger_1(\vec{k}\,)+
ia^\dagger_2(\vec{k}\,)\right]\ ,\\
 & & & & \\
b(\vec{k}\,)&=&\frac{1}{\sqrt{2}}\left[a_1(\vec{k}\,)-ia_2(\vec{k}\,)\right]\
&,&\
a^\dagger(\vec{k}\,)&=&\frac{1}{\sqrt{2}}\left[a^\dagger_1(\vec{k}\,)-
ia^\dagger_2(\vec{k}\,)\right]\ ,
\end{array}
\end{equation}
and obeying the appropriate Fock space algebras
\begin{equation}
[a(\vec{k}\,),a^\dagger(\vec{\ell}\,)]=(2\pi)^32\omega(\vec{k}\,)
\delta^{(3)}\left(\vec{k}-\vec{\ell}\,\right)=
[b(\vec{k}\,),b^\dagger(\vec{\ell}\,)]\ .
\end{equation}
The mode expansion of the complex field in the interacting picture is then
\begin{equation}
\hat{\phi}_{(I)}(x)=\int_{(\infty)}\frac{d^3\vec{k}}
{(2\pi)^32\omega(\vec{k}\,)}\left[a(\vec{k}\,)e^{-ik\cdot x}+
b^\dagger(\vec{k}\,)e^{ik\cdot x}\right]\ ,
\end{equation}
while a direct substitution in the normal ordered expression for the quantum
Noether charge $\hat{Q}$ finds
\begin{equation}
\hat{Q}=-\int_{(\infty)}\frac{d^3\vec{k}}{(2\pi)^32\omega(\vec{k}\,)}\,
\left[a^\dagger(\vec{k}\,)a(\vec{k}\,)\,-\,
b^\dagger(\vec{k}\,)b(\vec{k}\,)\right]\ .
\end{equation}
In comparison with the mode expansion for a real scalar field, one notices
a common structure with, however, the role played by the creation operator
component now taken over by that of the independent mode $b^\dagger(\vec{k}\,)$
rather than $a^\dagger(\vec{k}\,)$, since the field need no longer be real
under complex conjugation. Furthermore, it is precisely this complex
character of the field which makes possible the existence of the U(1)
symmetry, whose Noether charge should thus distinguish the two types of
modes present in the system. Indeed, a direct calculation finds, for
instance for the creation operators,
\begin{equation}
[\hat{Q},a^\dagger(\vec{k}\,)]=-a^\dagger(\vec{k}\,)\ \ ,\ \ 
[\hat{Q},b^\dagger(\vec{k}\,)]=+b^\dagger(\vec{k}\,)\ ,
\end{equation}
with in particular
\begin{equation}
\hat{Q}\,a^\dagger(\vec{k}\,)\,|0>=-a^\dagger(\vec{k}\,)\,|0>\ \ ,\ \ 
\hat{Q}\,b^\dagger(\vec{k}\,)\,|0>=+b^\dagger(\vec{k}\,)\,|0>\ .
\end{equation}
In other words, the conserved quantum number $Q$ associated to this Noether
quantum charge which generates the U(1) symmetry of the system, is an additive
quantum number for quantum states, and takes opposite values for the
field quanta created by either the operators $a^\dagger(\vec{k}\,)$
or $b^\dagger(\vec{k}\,)$. To put it still differently, these two
types of field quanta are distinguished by an opposite U(1) charge
under the U(1) global internal symmetry. Fields neutral under complex
conjugation are associated to neutral particles under some given continuous
symmetry, while fields complex under complex conjugation lead to charged
particles for the associated U(1) global internal symmetry. Hence, these
two types of quanta correspond to particles and their antiparticles, since
except for the opposite values for the U(1) conserved charge, they otherwise
share identical physical properties under the spacetime Lorentz symmetry,
namely their mass and spin values.

Consequently, this is yet one more
outcome of the marriage of $\hbar$ and $c$: the existence of particles and
antiparticles of identical mass and spin, but opposite charge under internal
continuous symmetries, such as their electric charge. Even for electrically
neutral particles, it could be that the particle and antiparticle species
are still distinct due to some other conserved quantum number than the
electric charge taking opposite values. Of course, a particle which coincides
with its antiparticle, and whose field is thus necessarily real under
complex conjugation, is necessarily electrically neutral.

The Noether charge operator $\hat{Q}$ being the generator of the U(1)
global symmetry, finite transformations of parameter $\alpha$ are induced 
through the exponentiated form
\begin{equation}
e^{i\alpha\hat{Q}}
\end{equation}
acting on the space of quantum states of the system. In particular, note
that the perturbative vacuum $|0>$ carries a vanishing U(1) charge,
$\hat{Q}|0>=0$, hence is also invariant under the action of the symmetry
group,
\begin{equation}
e^{i\alpha\hat{Q}}\,|0>\ =\ |0>\ .
\end{equation}
When the ground state or vacuum of the system is left invariant under the
action of the symmetry, one says that the symmetry is realized in its
Wigner mode.

It is straightforward to extend the above considerations to any internal
compact Lie symmetry group. Assume that a given system of fields is invariant
under a continuous group $G$ whose algebra is spanned by a set of generators
$T^a$ such that
\begin{equation}
[T^a,T^b]=if^{abc}\,T^c\ ,
\end{equation}
$f^{abc}$ being its structure constants, and for which the collection of
fields spans some linear representation of that algebra. Hence, if
$\phi(x)$ denotes this collection of fields (with the representation index
suppressed), and $T^a$ now stand for the $G$-generators in that specific
representation, the action of the symmetry on the fields may be represented
as
\begin{equation}
\phi'(x)=e^{i\theta^aT^a}\,\phi(x)\ ,
\end{equation}
$\theta^a$ being arbitrary constant but continuous parameters for 
$G$-transformations. These quantities being constant and acting independently
of the value of $x^\mu$, such transformations define a global internal
symmetry, assuming of course that the Lagrangian density 
${\cal L}(\phi,\partial_\mu\phi)$ is invariant under these transformations.
Consequently, because of Noether's theorem, there exists conserved
currents $J^a_\mu(x)$ and charges $Q^a=\int_{(\infty)}d^3\vec{x}\,J^{a0}$
generating the symmetry algebra and its transformations on the space
of classical as well as quantum states of the system. In particular,
if the ground state of the system is invariant under all $G$-transformations,
namely if the symmetry is realized in the Wigner mode,
the quantum space of states gets organized into irreducible representations 
of $G$, with in particular the one-particle states falling into the same 
$G$-representations as the original fields $\phi(x)$, since the creation 
and annihilation operators also carry that same representation index. 
All the latter properties are clearly met in the simple U(1) example above,
and it should be straightforward to understand why they should remain valid
for an arbitrary nonabelian symmetry group as well.

\vspace{10pt}

\noindent\underline{Spontaneous global symmetry breaking}

\vspace{5pt}

The above results still leave open the case of a symmetry which is not
realized in the Wigner mode, namely when the vacuum or ground state of the
system is not invariant under the action of the symmetry. It is well known 
that specific physical systems may possess such a property, as is the case
for instance for spontaneous magnetization in a ferromagnetic material
below the transition temperature. Let us recall the point made already
previously, namely that what is meant by a symmetry is not the invariance of
any of its configurations in particular, but rather the invariance of its 
equations of motion, hence also of the set of its configurations solving 
these equations viewed as a whole. If a given solution is not invariant, 
the existence of the continuous symmetry simply implies that there exists 
an infinite degeneracy of distinct solutions of identical energy all related 
through the action of the symmetry transformations. For example, imagine a 
simple linear stick standing along the vertical direction, onto which a 
certain pressure is applied along that axis. This system is obviously 
invariant for all rotations around the vertical axis. As long as the applied 
pressure is mild enough, the stick does not bend, and the lowest energy 
configuration of the system is indeed invariant under the axial symmetry. 
However, as soon as the applied pressure exceeds a specific critical value, 
the stick does bend until it reaches some equilibrium con\-fi\-gu\-ra\-tion. 
The horizontal direction in which this bending occurs is arbitrary, but it 
clearly spontaneously breaks the axial symmetry. Nevertheless, all the 
configurations of the system associated to all possible horizontal bending 
directions are degenerate in energy, and are related to one another precisely 
by the action of the axial symmetry group. The specific solution to the 
equations of motion singled out by the bending process is no longer 
invariant, but the set of all these solutions remains invariant, all the 
degenerate solutions being related through the axial symmetry group. When 
a symmetry is realized in such a manner, namely when the ground state of 
the system is not invariant under the symmetry, one says that the symmetry 
is spontaneously broken, or that it is realized in the Goldstone mode.

Whether a symmetry is realized in the Wigner or in the Goldstone mode
is governed by the details of the dynamics of the system, whether in
a perturbative or a nonperturbative regime. Once again for the purpose of
simplicity, here we only discuss the simplest example, namely that
of the spontaneous symmetry breaking already at the level of the classical
theory of a single complex scalar field $\phi(x)$ possessing the U(1) 
global symmetry
\begin{equation}
\phi'(x)=e^{i\alpha}\,\phi(x)\ ,
\end{equation}
with the real constant angular parameter $\alpha$.

Let us consider again the Lagrangian density
\begin{equation}
{\cal L}\left(\phi,\partial_\mu\phi\right)=
|\partial_\mu\phi|^2-V(|\phi|)\ ,
\end{equation}
where the potential contribution is given by
\begin{equation}
V(|\phi|)=\mu^2|\phi|^2+\lambda|\phi|^4\ 
\end{equation}
with $\lambda>0$. In our previous considerations, the quantity $\mu^2$
was taken to be positive, in which case it defined the mass-squared of the
particle quanta associated to the field, describing the quantum
excitations of this field above its ground state, namely the perturbative
vacuum $|0>$ associated to the classical value $\phi=0$ up to the vacuum
quantum fluctuations subtracted away through normal ordering, which is
invariant under the U(1) symmetry. 

Presently however, we shall consider the situation when $\mu^2<0$,
cor\-res\-pon\-ding to the so-called mexican hat potential, which very much 
looks like the bottom of a wine bottle. In such a case, the configuration 
$\phi=0$ no longer defines the lowest energy configuration of the system, 
since the potential $V(|\phi|)$ now reaches its lowest value for
\begin{equation}
|\phi(x)|=\frac{1}{\sqrt{2}}v\ \ \ ,\ \ \ 
v=\sqrt{\frac{-\mu^2}{\lambda}}\ .
\end{equation}
Such a configuration also defines the lowest energy state of the field,
since all field gradient contributions to the energy then vanish
identically, the field being constant throughout spacetime. Such a 
configuration however, is no longer invariant under the U(1) symmetry,
which is thus realized in the Goldstone mode. What are then the physical
consequences of this spontaneous symmetry breaking in the vacuum?

In order to properly identify the physical quanta of the field, it is
necessary to consider the field fluctuations about its vacuum configuration.
Note that the two independent degrees of freedom per spacetime point
defined by the complex scalar field may also be represented through
a polar decomposition around a given choice of vacuum configuration,
\begin{equation}
\phi(x)=\frac{1}{\sqrt{2}}e^{i\xi(x)/v}\,
\left[\rho(x)+v\right]\ ,
\label{eq:fluct}
\end{equation}
where $\xi(x)$ and $\rho(x)$ are two real scalar fields with a mass dimension
of unity. Note that the vacuum about which this expansion is performed is
\begin{equation}
\phi_0=\frac{1}{\sqrt{2}}\,v\ ,
\end{equation}
but that choice may easily be modified by adding to the mode $\xi(x)$
an arbitrary real constant quantity. This remark also shows that
the U(1) symmetry now leaves the radial field $\rho(x)$ invariant,
while it simply shifts the field $\xi(x)$ by the product $\alpha v$.
All the minimal energy configurations correspond the constant field $\phi$ 
lying at the bottom of the potential, with the norm $|\phi|=v/\sqrt{2}$ but an
arbitrary phase. The U(1) symmetry simply induces a transformation of
any such vacuum into any another such vacuum, the difference in their phases
being set by the value of the U(1) angle $\alpha$ (note the perfect analogy
with the above example of a bent stick). Hence, one should expect that
the fluctations associated to the field $\xi(x)$ are massless, since they
may be excited at zero-momentum at no extra energy cost. On the other hand,
the radial fluctuation $\rho(x)$, moving the field out from its lowest
energy configuration, must correspond to massive quanta of the field.
Furthermore, this physical conclusion does not depend on the choice of
complex phase for the reference constant vacuum configuration $\phi_0$, since
this amounts to a simple constant shift in the massless field $\xi(x)$.

More explicitly, a direct substitution of the mode expansion (\ref{eq:fluct})
gives
\begin{equation}
{\cal L}=\frac{1}{2}\partial_\mu\rho\partial^\mu\rho+
\frac{1}{2}\left(1+\frac{1}{v}\rho\right)^2\partial_\mu\xi\partial^\mu\xi
-\frac{1}{2}\mu^2(\rho+v)^2-\frac{1}{4}\lambda(\rho+v)^4\ .
\end{equation}
Isolating then the terms quadratic in $\xi(x)$ and $\rho(x)$ indeed confirms
that the mode $\xi(x)$ is massless, while the $\rho(x)$ field is massive, 
with the values
\begin{equation}
m^2_\rho=-2\mu^2>0\ \ \ ,\ \ \ 
m^2_\xi=0\ .
\end{equation}
Hence, we reach the conclusion that since the vacuum is not invariant
under the action of transformations which nevertheless define a symmetry
of the system and its equations of motion, necessarily in the Goldstone
mode rea\-li\-zation of the symmetry there exist massless modes, namely 
massless quanta for a quantized field, which in the zero momentum limit 
correspond to the excitation of one vacuum state into another one, all 
these vacuum states being degenerate in energy and infinite in number. 
Hence, rather than being explicitly realized in the space of states as is 
the case for the Wigner mode, the symmetry is now hidden through the 
existence of Golstone bosons. Nonetheless, the symmetry is still active 
within the system, even though it is no longer realized in a linear fashion. 
Indeed, within the field basis which diagonalizes its fluctuations, 
the symmetry acts as
\begin{equation}
\rho'(x)=\rho(x)\ \ \ ,\ \ \ \xi'(x)=\xi(x)+\alpha\,v\ ,
\end{equation}
which, among other consequences, implies that the Goldstone modes may
only possess derivative or gradient couplings with other fields.
The symmetry thus restricts to some extent the form of interactions
of Goldstone fields.

In fact, it should be quite clear that this is a conclusion valid in full
generality, which is known as Goldstone's theorem. Whenever a continuous
global symmetry is spontaneously broken in the vacuum, associated to each
of its broken generators, there exist massless quanta carrying the
cor\-res\-pon\-ding quantum numbers, known as the Goldstone bosons of the
symmetry. This conclusion is valid whether the spontaneous symmetry
breaking mechanism is perturbative or nonperturbative, and whether the
symmetry is abelian or nonabelian. The only specific requirement is that
the symmetry be a con\-ti\-nuous one (in the case of a fermionic or spacetime
symmetry, the Goldstone mode need not be bosonic, as is the case for instance
for spontaneous supersymmetry breaking leading to a spin 1/2 goldstino
massless mode).

\subsection{Local or gauged internal symmetries}
\label{Subsect4.3}

So far, we have briefly discussed the meaning of a global internal
symmetry, and described some of its physical consequences, whether
in the Wigner or the Goldstone mode. However, the
existence of a global symmetry is not very appealing, at least from
some theoretical aesthetic point of view. Indeed, any global internal
symmetry defines transformations on the set of fields which act
in an identical manner irrespective of the spacetime point at which the field
values are being considered. For instance in the case of the U(1) symmetry
associated to the electric charge and the electromagnetic interactions,
this would mean that in order to render the transformation unobservable,
one is required to change the phases of all the electrons of the Universe
by exactly the same amount instantaneously throughout all of infinite
space and troughout the whole of spacetime history! Although there is
no technical or mathematical inconsistency that arises with such a
relativistic quantum field theory, certainly it is a property of such
symmetry transformations which runs counter to our belief that causality
ought to be a stringent requirement on the construction of any physical
theory.

Hence, one should rather prefer to develop a formalism in which internal
symmetries are still possible, but such that now transformations may
be realized locally in spacetime, though in a continuous fashion as
to their spacetime dependency, while they would remain nevertheless
unobservable to any conceivable experiment. Namely, is it possible to
locally change the quantum phase of some electron while not at the same
time by the same amount that of all the other electrons of the Universe,
and nevertheless keep such a change hidden from any experimentalist?
Clearly, this would require some information to be sent to all the
other electrons in the Universe to tell them how to adjust their quantum
phases accordingly, and this at the speed of light so that no experimentalist
may catch up with this signal and measure the phase of some electron before 
it would have had the opportunity to adjust itself to the action of the
symmetry transformation. In other words, by making the symmetry local, or 
by gauging the symmetry, one must introduce some additional propagating 
field coupling with equal strength to all other matter carrying the same 
symmetry charge, and whose quanta are necessarily massless.

This is the heuristic idea of the local gauge symmetry principle.
As we shall explicitly see through the simplest examples, such a principle
in fact provides a unifying principle for the existence of fundamental
interactions, whose quantum carriers are massless and couple with identical
strength to all other quanta with which they interact. These gauge bosons 
are necessarily vector fields for internal symmetries,
and as stated previously, such Yang-Mills gauge theories based on
compact Lie groups are the only possible re\-nor\-ma\-li\-za\-ble field 
theories including spin 0 and 1/2 matter fields interacting with vector fields.

\vspace{10pt}

\noindent\underline{The simplest example}

\vspace{5pt}

As the simplest illustration of the above description, let us consider
once again the theory of a single complex scalar field $\phi$ whose
Lagrangian density is U(1) invariant under global phase transformations
of the field, see (\ref{eq:Lphi}) and (\ref{eq:symU(1)}). Clearly,
if one wishes to gauge this symmetry, namely to construct a
system which remains invariant under the local phase transformations
\begin{equation}
\phi'(x)=e^{i\alpha(x)}\,\phi(x)\ ,
\end{equation}
$\alpha(x)$ now being an arbitrary spacetime dependent parameter rather
than a constant angle as in the case of a global symmetry, a problem
arises with the original Lagrangian. Indeed, this Lagrangian is no longer
invariant, since the gradient contribution does not transform in
the same covariant manner as the original field does,
\begin{equation}
\partial_\mu\phi'(x)=e^{i\alpha(x)}\left[\partial_\mu\phi(x)+
i\partial_\mu\alpha(x)\,\phi(x)\right]\ .
\end{equation}
However, this expression suggests a modification of the ordinary derivative
or gradient of the field of the form
\begin{equation}
\partial_\mu\longrightarrow D_\mu(x)=\partial_\mu+igA_\mu(x)\ ,
\end{equation}
where $g$ is some dimensionless real quantity, which turns out to represent
the coupling strength of the U(1) gauge interaction, and $A_\mu(x)$ the vector
field for the gauge boson associated to the gauging of the U(1) symmetry.
Indeed, it now suffices to assume that this vector field transforms under
the local U(1) symmetry according to
\begin{equation}
A'_\mu(x)=A_\mu(x)-\frac{1}{g}\partial_\mu\,\alpha(x)\ ,
\end{equation}
to check that the modified gradient does possess the same covariant 
transformation as the field does under the symmetry,
\begin{equation}
D'_\mu(x)\phi'(x)=
\left[\partial_\mu+igA'_\mu(x)\right]e^{i\alpha(x)}\phi(x)=
e^{i\alpha(x)}\,D_\mu(x)\phi(x)\ ,
\end{equation}
hence the name ``covariant derivative" for the differential operator 
$D_\mu(x)$. Clearly, a simple substitution of the ordinary derivative
by the covariant one in the original Lagrangian density invariant under
the global U(1) symmetry leads to an expression invariant now under any
local U(1) symmetry transformation. The U(1) symmetry has been gauged.

However, we still need to provide the vector field $A_\mu(x)$ with some
dynamics, which is done by adding the pure gauge Lagrangian density to
that of the matter field,
\begin{equation}
{\cal L}_A=-\frac{1}{4}F_{\mu\nu}F^{\mu\nu}\ \ ,\ \ 
F_{\mu\nu}=\partial_\mu A_\nu-\partial_\nu A_\mu\ ,
\end{equation}
$F_{\mu\nu}$ being the gauge field strength, indeed the sole gauge 
invariant quantity that may be constructed out of the gauge field $A_\mu$ and
its first-order gradients, in order to obtain a Lagrangian density
which is of second-order in spacetime gradients, and thus represents
a causal propagation of the gauge field (for the same reason, the absolute 
sign and normalization of this Lagrangian density are fixed as given). 
This field being real under complex conjugation, its mode expansion is of 
the form, in the interacting picture,
\begin{equation}
A_\mu(x)=\int_{(\infty)}\frac{d^3\vec{k}}{(2\pi)^32|\vec{k}\,|}
\sum_{\lambda=\pm}
\left[e^{-ik\cdot x}\epsilon_\mu(\vec{k},\lambda)a(\vec{k},\lambda)\ +\ 
e^{ik\cdot x}\epsilon^*_\mu(\vec{k},\lambda)a^\dagger(\vec{k},\lambda)\right]
\ ,
\end{equation}
$a(\vec{k},\lambda)$ and $a^\dagger(\vec{k},\lambda)$ being annihilation
and creation operators with the Fock space algebra normalized in the
usual manner for massless quanta, and $\lambda$ denotes the different
polarization states possible associated to the polarization vectors
$\epsilon_\mu(\vec{k},\lambda)$. These polarization tensors are subjected
to some restrictions which stem from the gauge invariance properties of
the field, and shall not be discussed here (even though the issue of the
quantization of gauge invariant systems is discussed hereafter, but
not explicitly for such abelian and nonabelian Yang-Mills theories).
Note that the mass dimension of the gauge field indeed needs to be unity,
hence leading to a dimensionless gauge coupling constant $g$.

In conclusion, the gauging of the simplest U(1) invariant scalar field
theory is defined by the total Lagrangian density
\begin{equation}
{\cal L}_{\rm total}={\cal L}_A+{\cal L}_\phi\ ,
\end{equation}
with the pure gauge Lagrangian ${\cal L}_A$ given above, and
the matter one by
\begin{equation}
\begin{array}{r c l}
{\cal L}_\phi&=&{\cal L}\left(\phi,D_\mu\phi\right)=
|(\partial_\mu+igA_\mu)\phi|^2-m^2|\phi|^2-V(|\phi|)\\
 & & \\
 &=&|\partial_\mu\phi|^2-m^2|\phi|^2-V(|\phi|)
-igA_\mu\left[\phi^\dagger\partial^\mu\phi-\partial^\mu\phi^\dagger\phi\right]
+g^2A_\mu A^\mu\ .
\end{array}
\end{equation}
In the case that the U(1) symmetry is that associated to the 
elec\-tro\-ma\-gne\-tic
interaction, this system is simply that of scalar electrodynamics, namely
that describing the interactions of a massive charged spin 0 particle 
with the photon.

From the latter expression, we immediately read off the different interaction
terms coupling the matter and gauge fields. The term linear in $A_\mu$ is
in fact $gA_\mu J^\mu$, namely the coupling of gauge field to the U(1)
Noether current, and represents the coupling of one gauge quantum to
two scalar field quanta of opposite U(1) charges. Such a feature is generic
for all Yang-Mills theories: gauge fields always couple linearly to the
associated Noether currents. The term quadratic in $A_\mu$ describes the 
coupling of two gauge quanta to two scalar quanta, also of opposite
U(1) charges, in order for the total U(1) charge to be conserved in the
interactions. Note that the single gauge boson interaction is proportional
to $ig$, while the quadratic interaction is proportional to $ig^2$.
In other words, the gauge symmetry principle not only explains, on the basis
of a given internal symmetry, the appearance of local interactions, but
it also sets specific restrictions on the properties of these interactions
by predicting particular relations between the coupling strengths of
different interactions, such restrictions being a consequence of the symmetry.

Among the interactions, the gauge boson $A_\mu(x)$ does not couple to itself,
but only to the charged matter field with the universal coupling strength $g$.
The reason for the fact that the gauge boson lacks such a self-coupling 
is that it is neutral under the U(1) symmetry, and does not carry any 
U(1) charge. Indeed, under a global symmetry transformation 
$\alpha(x)=\alpha$, we simply have for the transformed field $A'_\mu=A_\mu$. 
Furthermore, it is also the U(1) symmetry, but this time in its gauged 
embodiement, which explains why the gauge boson quanta are massless 
particles. Indeed, any mass term of the form $M^2_AA_\mu A^\mu$ is clearly 
not gauge invariant under the local gauge transformations of the vector 
field. Hence, it is the local gauge symmetry which protects the gauge 
boson from acquiring any mass. In particular, this implies that physical 
(gauge invariant) quanta of that field may possess only two transverse 
polarization states, such that $k^\mu\epsilon_\mu(\vec{k},\lambda)=0$, 
$\lambda=\pm$, a fact related to the issue of the quantization of such 
Yang-Mills fields.

All the above considerations are readily extended to other matter fields,
including fermionic ones not addressed in these notes. Furthermore,
even though our discussion concentrates on the abelian U(1) case,
the same de\-ve\-lop\-ments apply to a nonabelian internal symmetry group $G$,
leading then to Yang-Mills gauges theories. In such a case, for a
collection of fields trans\-for\-ming in a $G$-representation whose generators
are $T^a$, the covariant derivative, which now is Lie-algebra valued, reads
\begin{equation}
D_\mu=\partial_\mu+igA^a_\mu T^a\ ,
\end{equation}
$g$ being the real gauge coupling constant, and $A^a_\mu$ the real gauge
vector fields, which, for infinitesimal local gauge transformations of
parameters $\theta^a(x)$, transform according to
\begin{equation}
{A'}^a_\mu=A^a_\mu-\frac{1}{g}\partial_\mu\theta^a-
f^{abc}\theta^bA^c_\mu\ ,
\end{equation}
$f^{abc}$ being the structure constants of the Lie algebra of $G$
(it is also straightforward to establish the transformations of the gauge
bosons for finite gauge transformations). 
The total Lagrangian of such a system is again given by the sum of the
original $G$-invariant Lagrangian of the matter fields in which the ordinary 
derivative is substituted by the covariant derivative $D_\mu$, to which one 
simply adds the pure Yang-Mills Lagrangian density
\begin{equation}
{\cal L}_A=-\frac{1}{4}F^a_{\mu\nu}F^{a\mu\nu}\ \ ,\ \ 
F^a_{\mu\nu}=\partial_\mu A^a_\nu-\partial_\nu A^a_\mu-
gf^{abc}A^b_\mu A^c_\nu\ .
\end{equation}
Once again, any mass term for the gauge bosons $A^a_\mu$ is forbidden by
local gauge invariance, while gauged matter interactions are directly
read off from the matter Lagrangian, leading again to linear and quadratic
interactions of scalar fields with the gauge bosons. However, for a nonabelian
symmetry, given the nonvanishing structure constants $f^{abc}$, the gauge
bosons themselves possess now $G$-charges, actually those of the adjoint
representation as may be seen from their gauge transformations for constant
parameters $\theta^a(x)=\theta^a$. Consequently, from the expansion of
the pure Yang-Mills Lagrangien, one identifies cubic and quartic terms
representing gauge boson trilinear and quadrilinear couplings, whose
strengths are directly proportional to $g$ and $g^2$, respectively.
Hence once again, the symmetry governs the details of all the gauge
interactions, namely their strengths and their symmetry properties as well.
Such predictions are specific to Yang-Mills theories, and provide important
signatures for high energy experiments as to the relevance of the gauge 
symmetry principle for the physics of the fundamental interactions and the
elementary particles. Note also that it is precisely these nonlinear
gauge boson self-couplings which must be, in ways still to be thoroughly
understood, at the origin of the specific nonperturbative phenomena of 
nonabelian theories, such as the property of confinement for the theory of 
the strong interactions among quarks, namely quantum chromodynamics (QCD) 
based on the local gauge symmetry SU(3)$_C$ for colour degree of freedom 
of quarks.

\vspace{10pt}

\noindent\underline{Spontaneous gauge symmetry breaking}

\vspace{5pt}

The above discussion of the construction of abelian and nonabelian internal
gauge symmetries implicitly assumed the symmetry to be realized in the Wigner
mode. Hence, it is also important to consider the situation when the
symmetry is rather realized in the Goldstone mode. For the purpose
of illustration in the simplest case, let consider once again the
U(1) gauged single scalar field theory, but this time with a potential
leading to spontaneous symmetry breaking. The associated Lagrangian
density is thus
\begin{equation}
{\cal L}=-\frac{1}{4}F_{\mu\nu}F^{\mu\nu}+|(\partial_\mu+igA_\mu)\phi|^2-
V(|\phi|)\ ,
\end{equation}
with
\begin{equation}
V(|\phi|)=\mu^2|\phi|^2+\lambda|\phi|^4\ \ \ ,\ \ \ 
\mu^2<0\ \ ,\ \ \lambda>0\ .
\end{equation}
This time however, because of the U(1) local symmetry transformation 
pro\-per\-ties of the fields,
\begin{equation}
\phi'(x)=e^{i\alpha(x)}\,\phi(x)\ \ \ ,\ \ \ 
A'_\mu(x)=A_\mu(x)-\frac{1}{g}\partial_\mu\alpha(x)\ ,
\end{equation}
when expanding any scalar field configuration about one of its vacuum
configurations,
\begin{equation}
\phi(x)=\frac{1}{\sqrt{2}}e^{i\xi(x)/v}\,\left[\rho(x)+v\right]\ \ \ ,\ \ \ 
\phi_0(x)=\frac{1}{\sqrt{2}}v\ \ \ ,\ \ \ 
v=\sqrt{\frac{-\mu^2}{\lambda}}\ ,
\end{equation}
it is always possible to effect a local U(1) gauge transformation, with
pa\-ra\-me\-ter
\begin{equation}
\alpha(x)=-\frac{1}{v}\xi(x)\ ,
\end{equation}
(note that since in general $\xi(x)$ is spacetime dependent, such a
procedure is possible only when the internal symmetry is gauged),
such that the Goldstone mode is completely gauged away from the scalar field,
but lies hidden now in the transformed gauge field $A'_\mu$,
\begin{equation}
\phi'(x)=\frac{1}{\sqrt{2}}\left[\rho(x)+v\right]\ \ \ ,\ \ \ 
A'_\mu(x)=A_\mu(x)+\frac{1}{gv}\partial_\mu\xi(x)\ .
\label{eq:Higgs}
\end{equation}
Upon substitution of the transformed fields in the Lagrangian density, 
which is physically equivalent to the original expression for the Lagrangian 
on account of local gauge invariance, one then finds
\begin{equation}
{\cal L}=-\frac{1}{4}F'_{\mu\nu}{F'}^{\mu\nu}+
\frac{1}{2}\left[\partial_\mu\rho+igA'_\mu(\rho+v)\right]^2-
\frac{1}{2}\mu^2(\rho+v)^2-\frac{1}{4}\lambda(\rho+v)^4\ .
\end{equation}
Isolating now all quadratic terms in the fields, one immediately notices
that the radial field $\rho$ still possesses the mass $m^2_\rho=-2\mu^2>0$,
but that in place of a massless Goldstone mode $\xi(x)$ which no longer
appears in this Lagrangian by having been gauged away, there now appears
an explicit mass term for the gauge boson field, with value
\begin{equation}
m^2_A=g^2v^2\ .
\end{equation}
Hence, even though a local symmetry when realized in the Wigner mode
forbids any mass for its gauge bosons, when spontaneously broken in
the vacuum and realized in the Goldstone mode, gauge bosons do acquire a mass!
Nevertheless, their mass is then not just any parameter in the Lagrangian,
but is in fact governed by the symmetry properties and takes a very
specific value proportional both to the gauge coupling constant $g$ and the
scalar field vacuum expectation value $v$ which spontaneously breaks
those symmetry generators whose gauge bosons are massive. The counting
of degrees of freedom is also in order. In the Wigner phase, one has
two real scalar modes (one massless and one massive, the Goldstone and the 
radial ones, $\xi$ and $\rho$) and two massless gauge modes (the two 
transverse modes of the gauge field). In the Goldstone phase, one has one real 
massive scalar mode (the radial field $\rho$) and three massive gauge boson
polarization modes. Note that the longitudinal massive gauge boson component 
is nothing but the would-be Goldstone mode $\xi$ which has been gauged away and
turned into the longitudinal component of the gauge field $A'_\mu$, see
(\ref{eq:Higgs}).

These general features of the spontaneous symmetry breaking of a local gauge
symmetry remain valid in general, and characterize the so-called Higgs 
mechanism. Whenever a local internal symmetry is spontaneously broken
in the vacuum, those gauge bosons associated to the generators which do not
leave invariant the vacuum acquire a mass proportional to the product of
the gauge coupling and the scalar vacuum expectation value. Moreover,
the Goldstone modes in the case of a global symmetry then provide
the longitudinal polarization states of the massive gauge bosons, leaving 
over the massive scalar modes, referred to as higgs scalars, as the only 
remnants of the spontaneously broken scalar matter sector.
The gauge transformation which gauges away the Goldstone modes from the scalar
sector to hide them in the gauge fields is known as the unitary gauge.
It is in the unitary gauge that the physical content of such a theory
is most readily identified. In the simplest example above, we thus conclude
that the physical field content is that of a neutral massive spin 0 particle 
of mass $\sqrt{-2\mu^2}$, the higgs particle, interacting with itself
and with a neutral massive spin 1 particle of mass $|gv|$.

\vspace{10pt}

\noindent\underline{Remarks}

\vspace{5pt}

As already mentioned, it turns out that the gauge symmetry principle
uniquely singles out among all possible quantum field theories of
interacting spin 0, 1/2 and 1 particles, all those that are
renormalizable, whether the gauge bosons are massive or not, provided
however that in the former case their mass arises through the Higgs
mechanism.\cite{HV} This is quite a re\-mar\-ka\-ble result, since such a local 
internal symmetry principle also implies the existence of specific interactions
between matter particles and gauge bosons, whose detailed properties are
totally governed by the underlying symmetry, whether abelian or nonabelian.
In other words, all the relativistic and quantum dynamics of fundamental
interactions among elementary point-particles, through the marriage of 
$\hbar$ and $c$, appears to follow simply from the very elegant and powerful
idea of a fundamental symmetry based on a compact Lie group.

Thus in order to describe all the known strong, electromagnetic and weak
interactions observed to act between all known quarks and leptons,
a gauge group as simple as SU(3)$_c\times$SU(2)$_L\times$U(1)$_Y$
suffices, with a specific choice of representations for the quark
and leptons fermionic fields, as well as for the scalar sector
required for the Higgs mechanism leading to massive electroweak gauge
bosons but nonetheless a massless photon. If not yet totally unified
within this Standard Model of these interactions, at least all these
interactions are brought within the unified framework of relativistic
quantum Yang-Mills theories, leading to predictions whose precision is
without precedent and which are confirmed through remarkable particle
physics experiments. Nevertheless, this raises the issue of the
rationale behind such a principle, as well as for the choice of internal
symmetry and matter content.

From another perspective, with such Yang-Mills theories we are encountering
dynamical theories whose quantization requires an approach more ge\-ne\-ral
than that which was briefly reviewed in Sect.\ref{Sect2}. Indeed,
considering the issue for example from within the Hamiltonian approach,
when identifying the momentum conjugate to the U(1) gauge field $A_\mu$
coupled to the single scalar field through the Lagrangian density discussed 
above, one finds
\begin{equation}
\pi^\mu=\frac{\partial{\cal L}_{\rm total}}{\partial(\partial_0A_\mu)}=
-F^{0\mu}\ ,
\end{equation}
thus leading to the following constraint for its time component
\begin{equation}
\pi^0=0\ .
\end{equation}
In other words, all phase space degrees of freedom of the system are
not independent. Some are in fact constrained, and as we
shall see in the forthcoming section, this is a generic feature for
any system possessing a local symmetry whose parameters are not constant.
How is one then to quantize such systems, since their physical dynamics
is not contained within all of phase space, but only within some subspace
of it? Clearly, gauge invariance implies that all degrees of freedom are
not physical and relevant to the dynamics. How does one then account
consistenly for such redundant features of a gauge invariant system
in its quantization? In the above example, it would be possible to
solve for these gauge degrees of freedom, but at the cost of loosing
a manifestly spacetime covariant description of such systems, which is
also not welcome in itself. Hence, it is time now to turn to the
discussion of the quantization of constrained dynamics.

\section{Dirac's Quantization of Constrained Dynamics}
\label{Sect5}

\subsection{Classical Hamiltonian formulation of singular systems}
\label{Subsect5.1}

\noindent\underline{The system of constraints}

\vspace{5pt}

The Hamiltonian approach towards canonical quantization discussed in
Sect.\ref{Sect2} explicitly assumed that the Lagrange function be
``regular", see (\ref{eq:regular}). Now, we have to develop the
Hamiltonian formulation associated to a ``singular" Lagrangian,\cite{Dirac,JG1}
namely one for which the Hessian possesses some local zero modes,
\begin{equation}
{\rm det}\,\frac{\partial^2 L}{\partial\dot{q}^{n_1}\partial\dot{q}^{n_2}}=0\ .
\label{eq:singular}
\end{equation}
When considered in terms of the conjugate momenta, 
$p_n=\partial L/\partial q^n$, for which the canonical Poisson bracket
structure is still in effect,
\begin{equation}
\{q^{n_1}(t),p_{n_2}(t)\}=\delta^{n_1}_{n_2}\ ,
\end{equation}
since the condition (\ref{eq:singular}) also writes as
\begin{equation}
{\rm det}\,\frac{\partial p_{n_1}}{\partial\dot{q}^{n_2}}=0\ ,
\end{equation}
it follows that singular systems are characterized by the existence of
a series of primary constraints on phase space, of the form
\begin{equation}
\phi_m(q^n,p_n)=0\ ,
\end{equation}
where in this section, the index $m$ will be reserved for the primary
constraints. To be precise, a technical restriction on the choice of
expression for these primary constraints, known as the regularity condition,
is assumed, such that the phase space gradient of these constraints is a 
matrix of constant maximal rank on the constraint hypersurface (as the 
discussion proceeds, further constraints beyond the primary ones appear; 
the subset of phase space defined by the entire set of constraints is known 
as the constraint hypersurface). Indeed, whether a given constraint 
$\phi=0$ is expressed as $\phi^2=0$ or $\phi^{1/2}=0$ may 
seem {\sl a priori\/} equally acceptable, but further developments in fact 
require this re\-gu\-la\-ri\-ty condition for the constraints. In practice, 
it may be established that under the regularity condition, any quantity that
vanishes on the constraint hypersurface may be written as a local
linear combination of the constraints.

The canonical Hamiltonian
\begin{equation}
H_0(q^n,p_n)=\dot{q}^np_n-L(q^n,\dot{q}^n)\ 
\end{equation}
is still defined in the usual way, in spite of the existence of constraints.
It may be shown that whether the Lagrange function is regular
or singular, this function $H_0(q^n,p_n)$ is indeed always a function
defined over phase space $(q^n,p_n)$, even in the presence of primary 
constraints. However, in contradistinction to the regular situation,
time evolution of the system may {\sl a priori\/} be generated by a 
Hamiltonian more general than the canonical one, since one may add to the
latter some combination of the primary constraints. Hence, let us
consider the primary Hamiltonian
\begin{equation}
H_*(q^n,p_n;t)=H_0(q^n,p_n)+{\cal U}^m(q^n,p_n;t)\phi_m(q^n,p_n)\ ,
\end{equation}
${\cal U}^m(q^n,p_n;t)$ being {\sl a priori\/} some arbitrary functions
of phase space, possibly time dependent as well. These functions parametrize
the arbitrariness which exists as to the time dependent dynamics of the
system restricted to the constraint hypersurface. Some of these functions
may be restricted or be totally determined by consistency conditions on this
time development, as we shall discuss, whereas others may remain totally
undetermined, a possibility that should be expected in the case of
gauge theories since the solutions to such systems always depend on
some arbitrary time dependent functions related to the gauge degrees
of freedom.

The functions ${\cal U}^m(q^n,p_n;t)$ could indeed be restricted, since
the primary Hamiltonian, which should generate a consistent time evolution,
must be such that given initial data within the constraint hypersurface, the
evolved con\-fi\-gu\-ra\-tion must always belong to that same hypersurface. 
In other words, whenever time evolution pulls some initial configuration away 
from the constraint surface, some projection mechanism must push it back onto 
it. This requirement is certainly met if time evolution of the primary 
constraints is such that their Hamiltonian equations of motion vanish
identically on the constraint hypersurface, namely
\begin{equation}
\dot{\phi}_m=\left\{\phi_m,H_*\right\}_{|_{\phi_m=0}}=0\ ,
\end{equation}
a condition also expressed as
\begin{equation}
\dot{\phi}_m\approx 0\ ,
\end{equation}
where the {\sl weak equality\/} sign ``$\approx$" stands for a relation
which is valid when restricted to the constraint hypersurface. Since the
Hamiltonian equation of motion for an arbitrary phase space observable
$f(q^n,p_n;t)$ is
\begin{equation}
\begin{array}{r c l}
\frac{df}{dt}&=&\frac{\partial f}{\partial t}+
\{f,H_0\}+{\cal U}^m\{f,\phi_m\}+\{f,{\cal U}^m\}\phi_m\\
 & & \\
&\approx&\frac{\partial f}{\partial t}+\{f,H_0\}+{\cal U}^m\{f,\phi_m\}\ ,
\end{array}
\end{equation}
it follows that a consistent time evolution of the constraints requires
\begin{equation}
\{\phi_m,H_0\}+{\cal U}^{m'}\{\phi_m,\phi_{m'}\}\approx 0\ .
\end{equation}
This set of equations for each of the values for the index $m$ labelling the
primary constraints either implies a trivial identity, $0=0$, or else
a nontrivial phase space restriction $\chi(q^n,p_n)=0$ independent of
the functions ${\cal U}^m$, or finally a genuine linear equation for
the functions ${\cal U}^m$. Whenever restrictions $\chi(q^n,p_n)=0$
appear at this stage, they in fact represent new or secondary constraints
on phase space, for which conditions of a consistent time evolution
have again to be considered. Consequently, a whole hierarchy of constraints
may appear generation after generation of such secondary constraints, 
through the requirement of a consistent time evolution, until the whole set 
of constraints is exhausted. Let us denote the whole set of constraints as 
$\phi_j(q^n,p_n)=0$, the index $j$ being reserved for that purpose. Note 
that whenever a new constraint $\phi_j=0$ is uncovered, there is no reason 
not to include it as well in the Hamiltonian through some combination 
${\cal U}^j(q^n,p_n)\phi_j$ to be added to the linear combination of 
constraints, hence leading {\sl in fine\/} to the total Hamiltonian
\begin{equation}
H_T=H_0+{\cal U}^j\phi_j\ .
\end{equation}
Hence finally, a consistent time evolution generated by this Hamiltonian
requires that all constraints $\phi_j(q^n,p_n)=0$ be preserved through
the dynamical equations of motion, $\dot{\phi}_j\approx 0$, leading to the 
set of equations
\begin{equation}
\{\phi_j,H_0\}+{\cal U}^{j'}\{\phi_j,\phi_{j'}\}\approx 0\ .
\end{equation}
The general solution to these equations is
\begin{equation}
{\cal U}^j(q^n,p_n)=U^j(q^n,p_n)+\lambda^\alpha(t)V^j_\alpha(q^n,p_n)\ ,
\end{equation}
where $U^j(q^n,p_n)$ are a specific set of solutions to the inhomogeneous 
linear equations
\begin{equation}
\{\phi_j,\phi_{j'}\}U^{j'}=-\{\phi_j,H_0\}\ ,
\end{equation}
while $V^j_\alpha(q^n,p_n)$ provide a basis for the space of solutions to
the homogeneous equations
\begin{equation}
\{\phi_j,\phi_{j'}\}V^{j'}_\alpha=0\ .
\end{equation}
Note that these solutions induce the space of zero modes of the matrix
$\{\phi_j,\phi_{j'}\}$. Finally, the quantities $\lambda^\alpha(t)$ are
arbitrary time dependent functions which define arbitrary linear combinations
of the zero mode solutions $V^j_\alpha$. {\sl A priori\/}, these functions
could also depend on the phase space variables, $\lambda^\alpha(q^n,p_n;t)$,
with no added advantage to the description of any such constrained
system however. As a matter of fact, it suffices to consider them to be
solely functions of time. Indeed, as discussed hereafter, the freedom in the
choice of these functions is directly related to the existence of local
gauge symmetries, and as is well known, gauge symmetries imply the
appearance of arbitrary (space)time dependent functions in the general
solutions to the equations of motion. The quantities $\lambda^\alpha(t)$
are related to nothing but these arbitrary (space)time dependent functions 
defining general solutions.

In terms of the above results, the dynamics of a constrained system is
generated by a total Hamiltonian of the form
\begin{equation}
H_T=H+\lambda^\alpha(t)\phi_\alpha\ ,
\end{equation}
with
\begin{equation}
H=H_0+U^j\phi_j\ \ \ ,\ \ \ 
\phi_\alpha=V^j_\alpha\phi_j\ .
\end{equation}
As we shall now discuss, the role and meaning of the quantities $U^j$,
$H$, $\phi_\alpha$ and $\lambda^\alpha$ are essential to the understanding
of constrained dynamics.

\vspace{10pt}

\noindent\underline{First- and second-class quantities and constraints}

\vspace{5pt}

As a matter of fact, constrained dynamical systems may possess physically
equivalent but nonetheless different Lagrangian formulations. Whether a
given constraint then appears as a primary or a secondary constraint is
then function of the chosen Lagrangian formulation used as the starting
point for the Hamiltonian analysis of constraints. However, there exists
a characterization of constraints which is not dependent on that choice,
in a way which is already suggested by the above conclusions regarding the
quantities $H$ and $\phi_\alpha$. This characterization leads to a 
classification of constraints which is due to Dirac.\cite{Dirac}

Consider any phase space quantity $R(q^n,p_n)$. By definition,
this quantity is said to be a {\sl first-class\/} quantity if and only if
its Poisson brackets with all the constraints $\phi_j$ vanish weakly,
\begin{equation}
\{R,\phi_j\}\approx 0\ \ \Longleftrightarrow\ \ 
\{R,\phi_j\}={R_j}^{j'}\phi_{j'}\ ,
\end{equation}
the equivalence being valid provided the constraints are regular.
Otherwise, if at least one of these Poisson brackets does not vanish
on the constraint hypersurface, the quantity $R$ is said to be
{\sl second-class\/}.

From the Jacobi identity obeyed by Poisson brackets, it follows
that the Poisson bracket $\{R_1,R_2\}$ of any two first-class
quantities $R_1$ and $R_2$ is itself a first-class quantity. Furthermore,
these definitions are not void of content. Indeed, from the definition
of the coefficients $U^j$ and the quantity $H$, it follows that
\begin{equation}
\{H,\phi_j\}\approx 0\ ,
\end{equation}
showing that the Hamiltonian $H$ is in fact a first-class quantity.
Likewise, from the definition of the coefficients $V^j_\alpha$, it follows
that
\begin{equation}
\{\phi_\alpha,\phi_j\}\approx 0\ ,
\end{equation}
showing also that all the linear combinations of constraints $\phi_\alpha$
are themselves first-class quantities. Consequently, the total Hamiltonian
is itself a first-class quantity, being given by the first-class Hamiltonian
$H$ summed with an arbitrary linear combination of the first-class
constraints $\phi_\alpha$.

This characterization as being first- or second-class quantities
applies in particular to the constraints $\phi_j$. This classification
of constraints in terms of first- and second-class constraints is in fact
an invariant one, independent of the starting Lagrangian used in the analysis
of constraints, in contradistinction to their primary or secondary character.
Hence in the following, we shall assume that through appropriate
linear combinations, a subset as large as possible of the original constraints
$\phi_j$ is brought into the first-class subset, while the remaining
linearly independent set of constraints is such that none of its linear
combinations could be first-class. Given the above definitions of the
coefficients $U^j$ and $V^j_\alpha$, the whole set of first-class
constraints corresponds to the constraints $\phi_\alpha$, while the
remaining linearly independent
second-class constraints shall be denoted $\chi_s$. Note that
whenever the whole set of constraints $\phi_j$ is reducible, namely
not linearly independent, the set of reducible constraints is necessarily
included among the first-class one.

\vspace{10pt}

\noindent\underline{Second-class constraints and Dirac brackets}

\vspace{5pt}

Having identified this invariant characterization of constraints,
let us now address its intrinsic meaning, first in the case
of second-class constraints. The simplest example of such a situation
is provided by the constraints
\begin{equation}
q^1\approx 0\ \ \ ,\ \ \ p_1\approx 0\ ,
\end{equation}
since $\{q^1,p_1\}=1$. Clearly, the meaning of such constraints is that
the degree of freedom $n=1$ does not partake in no manner whatsoever
in the dynamics of the system. One might as well remove that sector of
the system altogether from its inception, namely subtract away from the
definition of Poisson bra\-ckets all those contributions stemming for the
degree of freedom $n=1$, without any consequence as to the actual and genuine
dynamics of the system.

From this simple example, it thus appears that the meaning of second-class
constraints is that they are associated to the appearance of degrees of
freedom which are totally redundant and irrelevant to the dynamics of the
system, and may thus be suppressed altogether from the dynamics. This is
achieved through a redefinition of the bracket structure on phase space,
leading to so-called Dirac brackets in the general case, which in effect
subtracts away any second-class constraint contributions to the dynamics.

For this purpose, let us consider the matrix of Poisson brackets of the
second-class constraints $\chi_s$,
\begin{equation}
\Delta_{ss'}=\{\chi_s,\chi_{s'}\}\ .
\end{equation}
Given the property that none of the combinations of the second-class
constraints $\chi_s$ is first-class, it follows necessarily that this
matrix is regular, even on the constraint hypersurface,
\begin{equation}
{\rm det}\,\Delta_{ss'}\not{\!\!\approx}\ 0\ .
\end{equation}
Indeed, there would exist otherwise a nontrivial combination $\chi_sC^s$
of the constraints $\chi_s$ which would be first-class,
\begin{equation}
{\rm det}\,\Delta_{ss'}\approx 0\ \ 
\Longleftrightarrow\ \ \{\chi_s,\chi_{s'}C^{s'}\}\approx 0\ ,
\end{equation}
$C^s$ being a zero mode of the matrix $\Delta_{ss'}$
(note that this result also establishes that the combination $U^j\phi_j$
appearing in the first-class Hamiltonian $H$ belongs to the set of
combinations of the second-class constraints $\chi_s$).
Hence, the matrix of Poisson brackets of all second-class constraints
$\chi_s$ is invertible, even on the constraint hypersurface, leading
to the definition of the Dirac bracket $\{f,g\}_D$ of any two quantities
$f(q^n,p_n)$ and $g(q^n,p_n)$ on phase space,
\begin{equation}
\{f,g\}_D=\{f,g\}-\{f,\chi_s\}\left(\Delta^{-1}\right)^{ss'}
\{\chi_{s'},g\}\ .
\end{equation}
This definition is such that the Dirac bracket of any phase space quantity $f$
with any of the second-class constraints $\chi_s$ vanishes exactly as an
identity valid throughout all of phase space,
\begin{equation}
\{f,\chi_s\}_D=\{f,\chi_s\}-\{f,\chi_{s'}\}\left(\Delta^{-1}\right)^{s's''}
\{\chi_{s''},\chi_s\}=0\ .
\end{equation}
Consequently, provided one uses Dirac rather than Poisson brackets,
the second-class constraints $\chi_s=0$ may be imposed even before
any such calculation, which is certainly not the case for Poisson brackets
in which case it is essential that constraints never be enforced before any
bracket evaluation. Furthermore, when considering the Hamiltonian equations
of system, their expression in terms of Dirac rather than Poisson brackets
does not modifiy the dynamics on the constraint hypersurface either, since
one has for an arbitrary phase space quantity $f(q^n,p_n;t)$,
\begin{equation}
\begin{array}{r c l}
\frac{df}{dt}&=&\frac{\partial f}{\partial t}+\{f,H_T\}\\
 & & \\
&=&\frac{\partial f}{\partial t}+\{f,H_T\}_D+
\{f,\chi_s\}\left(\Delta^{-1}\right)^{ss'}\{\chi_{s'},H_T\}\approx
\frac{\partial f}{\partial t}+\{f,H_T\}_D\ .
\end{array}
\end{equation}

Consequently, through the use of Dirac rather than Poisson brackets,
it becomes possible to subtract away all those contributions of the
redundant degrees of freedom which are irrelevant to the time evolution
of the system because of the second-class constraints $\chi_s=0$.
One may then as well explicitly solve for these constraints, leading
to a set of coordinates $z_A$ parametrizing the associated reduced phase 
space equipped with the bracket structure induced by the Dirac brackets,
\begin{equation}
\{z_A,z_B\}_D=C_{AB}(z_A)\ .
\end{equation}
The total Hamiltonian is then still given by an expression of the form
\begin{equation}
H_T=H+\lambda^\alpha(t)\phi_\alpha\ .
\end{equation}
Hence, the actual and genuine time dependent dynamics of the constrained
system is now reduced into a constrained dynamics characterized by the sole
first-class constraints $\phi_\alpha$ restricted to the reduced phase space 
$\{z_A\}$. Henceforth, we shall assume to have effected such a reduction
of the second-class constraints through Dirac brackets, and no longer display
the ``$D$" subindex on bracket evaluations.

It should be remarked though, that generally such a reduction of second-class
constraints often entails a loss of manifest spacetime Poincar\'e invariance
in the case of relativistic field theories, not a welcome feature.
Furthermore, quite often the Dirac bracket structure that is obtained is
not canonical, with in particular phase space dependent bracket values
$C_{AB}(z_A)$ which sometimes are not even spacetime local functions
in the case of field theories. Such circumstances then render
canonical quantization of Dirac brackets problematic. In principle, by
Darboux's theorem, it is always possible to locally bring the phase space
coordinate system to canonical form, but again in practice this is often no
small feat. However such issues may only be addressed on a case by case
basis. Let us only point out here that the physical projector approach
to be discussed in Sect.\ref{Subsect5.5} readily circumvents all these issues.

As an example, the reader is invited to consider the following Lagrange
function,
\begin{equation}
L(q^n,\dot{q}^n)=\dot{q}^nK_n(q^n)-V(q^n)\ ,
\end{equation}
where $K_n(q^n)$ and $V(q^n)$ are arbitrary functions such that the
matrix
\begin{equation}K_{nm}=
\frac{\partial K_m}{\partial q^n}-\frac{\partial K_n}{\partial q^m}
\end{equation}
is regular. Clearly, this system is already in its Hamiltonian 
form,\cite{JG1,FJ}
with phase space degrees of freedom $q^n$, Hamiltonian $H(q^n)=V(q^n)$ and
a Poisson bracket structure $\{q^n,q^m\}$ encoded in the functions 
$K_n(q^n)$. To identify the latter, it suffices to consider the Euler-Lagrange 
equations following from the above first-order Lagrangian,
\begin{equation}
K_{nm}\dot{q}^m=\frac{\partial V}{\partial q^n}\ ,
\end{equation}
and require these equations to be equivalent to the Hamiltonian ones,
\begin{equation}
\dot{q}^n=\{q^n,q^m\}\frac{\partial H}{\partial q^m}\ .
\end{equation}
Hence, we must have for the Poisson brackets of the fundamental phase space
degrees of freedom
\begin{equation}
\{q^n,q^m\}=\left(K^{-1}\right)^{nm}\ .
\end{equation}
Of course, all this follows provided one notices that the above Lagrange
function is already that of the Hamiltonian formulation of the system,
since it is linear in the first-order time derivatives of the degrees of
freedom. However, in practical instances such a feature is not necessarily
so obvious, in which case one would embark onto the constraint analysis
path following the ge\-ne\-ral discussion of the present section. Indeed, one
immediately notices the primary constraints for the conjugate momenta
\begin{equation}
p_n=\frac{\partial L}{\partial\dot{q}^n}=K_n(q^n)\ \ \ ,\ \ \ 
\phi_n(q^n,p_n)=p_n-K_n(q^n)\ ,
\end{equation}
whose brackets $\{q^n,p_m\}=\delta^n_m$ are now canonical. The reader
is thus invited to pursue the constraint analysis of this system, to
conclude that these primary constraints $\phi_n=0$ already exhaust all
the constraints of the system, that these constraints are all second-class
and are solved precisely by reducing the conjugate momenta $p_n=K_n(q^n)$, 
and that finally the reduced description based on the relevant Dirac brackets 
is nothing else than the Hamiltonian formulation identified above in terms of
the phase space $\{q^n\}$, the Hamiltonian $H(q^n)=V(q^n)$ and the
brackets $\{q^n,q^m\}=(K^{-1})^{nm}$. Note that in the case of bosonic degrees
of freedom $q^n$, the regularity of the antisymmetric tensor $K_{nm}$
requires an even number of coordinates $q^n$, as befits indeed any bosonic
phase space.

Among possible examples of such systems of great interest, the most
obvious one is certainly the Dirac Lagrangian density for a Dirac spinor
in whatever spacetime dimension,
\begin{equation}
{\cal L}=\bar{\psi}\left(i\gamma^\mu\partial_\mu-m\right)\psi\ ,
\end{equation}
$\psi$ being the Dirac spinor with $\bar{\psi}=\psi^\dagger\gamma^0$,
$m$ the mass of its quanta, and $\gamma^\mu$ the usual Dirac
matrices obeying the Clifford-Dirac algebra 
$\{\gamma^\mu,\gamma^\nu\}=2\eta^{\mu\nu}$. This system is thus already
in its Hamiltonian form. Note however that its degrees of freedom $\psi(x^\mu)$
are now Grassmann odd quantities, hence leading to a Grassmann odd
graded bracket structure, whose canonical quantization requires now
anticommutation relations rather commutation ones.

\vspace{10pt}

\noindent\underline{First-class constraints and gauge invariance}

\vspace{5pt}

After explicit resolution of any second-class constraints, the constrained
Hamiltonian dynamics is characterized in terms of a phase space of coordinates
$z_A$ whose bracket structure is generally of the form
\begin{equation}
\{z_A,z_B\}=C_{AB}(z_A)\ ,
\end{equation}
and whose time evolution is generated by a total Hamiltonian
\begin{equation}
H_T=H+\lambda^\alpha\phi_\alpha\ ,
\end{equation}
where the first-class Hamiltonian $H$ and constraints $\phi_\alpha$ thus obey
the bracket algebra
\begin{equation}
\{H,\phi_\alpha\}={C_\alpha}^\beta\phi_\beta\ \ \ ,\ \ \ 
\{\phi_\alpha,\phi_\beta\}={C_{\alpha\beta}}^\gamma\phi_\gamma\ ,
\label{eq:gaugealgebra}
\end{equation}
${C_\alpha}^\beta$ and ${C_{\alpha\beta}}^\gamma$ being specific quantities
which, in a general situation, may even be functions of phase space.
The functions $\lambda^\alpha(t)$ are totally arbitrary,
and thus parametrize an intrinsic freedom active within the system and
directly related to the existence of the first-class constraints.

Given the above simple example of second-class constraints suggested as
an exercise, it should
be clear now that this whole information may also be encoded into the
specification of a Hamiltonian variational principle based on the following
first-order action
\begin{equation}
S[z_A;\lambda^\alpha]=\int dt\left[\dot{z}_AK^A(z_A)-H(z_A)-
\lambda^\alpha(t)\phi_\alpha(z_A)\right]\ ,
\label{eq:HS}
\end{equation}
where the functions $K^A(z_A)$ are such that
\begin{equation}
\frac{\partial K^B}{\partial z_A}-\frac{\partial K^A}{\partial z_B}=
\left(C^{-1}\right)^{AB}\ .
\end{equation}
Hence, from that point of view, the functions $\lambda^\alpha(t)$ are nothing
but Lagrange multipliers for the first-class constraints $\phi_\alpha=0$.
But then what is the meaning of the existence of these first-class constraints?

We shall now argue to establish that first-class constraints are the
ge\-ne\-ra\-tors of local gauge symmetries of such a system, namely 
transformations
of the phase space degrees of freedom $z_A$ and the functions $\lambda^\alpha$
leaving the equations of motion invariant and whose parameters are local
functions of time (or spacetime in the case of local field theories).
The most direct way to establish this fact is by considering the following
infinitesimal variations generated by the constraints
\begin{equation}
\delta_\zeta z_A=\{z_A,\zeta^\alpha\phi_\alpha(z_A)\}\ \ \ ,\ \ \ 
\delta_\zeta\lambda^\alpha=\dot{\lambda}^\alpha+
\lambda^\gamma\zeta^\beta{C_{\beta\gamma}}^\alpha-
\zeta^\beta{C_\beta}^\alpha\ ,
\label{eq:gaugetransf}
\end{equation}
where $\zeta^\alpha(t)$ are arbitrary time dependent infinitesimal parameters.
By direct substitution into the first-order action (\ref{eq:HS}), one then
finds that indeed this action is invariant up to a surface term,
\begin{equation}
\delta_\zeta S=\int dt\,\frac{d}{dt}\,\left[
\zeta^\alpha\left(K^AC_{AB}\frac{\partial\phi_\alpha}{\partial z_B}-
\phi_\alpha\right)\right]\ ,
\end{equation}
(which may vanish for an appropriate choice of boundary conditions, though
this is by no means a necessary requirement), so that the Hamiltonian
equations of motion are invariant. Hence indeed, the transformations
(\ref{eq:gaugetransf}) do define local gauge symmetries of the system.

Alternatively, let us consider a set of initial data lying within the
constraint hypersurface and let them evolve in time given two different
choices $\lambda^\alpha_1(t)$ and $\lambda^\alpha_2(t)$ for the Lagrange 
multipliers. Accordingly, the change in the phase space variables $z_A$
associated to an infinitesimal time interval $\delta t$ for each choice is
\begin{equation}
\delta_1z_A=\{z_A,H+\lambda^\alpha_1\phi_\alpha\}\delta t\ \ ,\ \ 
\delta_2z_A=\{z_A,H+\lambda^\alpha_2\phi_\alpha\}\delta t\ ,
\end{equation}
so that the corresponding phase space trajectories differ in such a way that
\begin{equation}
\delta_2z_A-\delta_1z_A=
\{z_A,(\lambda^\alpha_2-\lambda^\alpha_1)\delta t\phi_\alpha\}=
(\lambda^\alpha_2-\lambda^\alpha_1)\delta t\,\{z_A,\phi_\alpha\}\ .
\label{eq:gaugetraject}
\end{equation}
However, since the choice for the Lagrange multipliers $\lambda^\alpha(t)$,
which directly partake in the time dependency of the system dynamics, is
totally arbitrary, the physical content and interpretation of the
associated description should be equivalent irrespective of that choice.
In other words, different choices of $\lambda^\alpha(t)$ are to be viewed
as defining transformations between different phase space trajectories within
the constraint hypersurface which describe one and the same physical
configuration of the system. Namely, the freedom related to the choice
in $\lambda^\alpha$ is nothing but a local gauge symmetry freedom.
As the transformations (\ref{eq:gaugetransf}) establish, this gauge
freedom is also the one which is generated by all first-class constraints
within the Hamiltonian formulation of the system. In particular, it thus
appears, from (\ref{eq:gaugetraject}), that the freedom in the choice of
Lagrange multipliers $\lambda^\alpha(t)$ is nothing but the freedom that
the system affords to include within its time evolution the possibility
to also effect arbitrary gauge transformations as the system proceeds
along one of its physically equivalent phase space trajectories within
the constraint hypersurface. The Lagrange multipliers simply parametrize
the freedom in local gauge transformations available throughout the
time evolution of the system.

Having established the consistency of the gauge symmetry interpretation
of the action of the first-class constraints $\phi_\alpha$, it now appears
that the algebra (\ref{eq:gaugealgebra}) of these quantities is nothing
but the local Hamiltonian gauge symmetry algebra, with structure
coefficients ${C_{\alpha\beta}}^\gamma$,
\begin{equation}
\{\phi_\alpha,\phi_\beta\}={C_{\alpha\beta}}^\gamma\,\phi_\gamma\ .
\end{equation}
In case these coefficients are in fact constant, one says that the
algebra is closed, whereas otherwise it is open. In the latter case,
this implies that one is in fact dealing with an algebraic structure in a 
strict sense provided only gauge transformed quantities are restricted onto 
the constraint hypersurface. Indeed, given gauge transformations associated
to two independent sets of gauge parameters $\zeta^\alpha_1(t)$ and 
$\zeta^\alpha_2(t)$, the commutator of the bracket induced transformations
of any phase space quantity $f$ is such that, using Jacobi's identity,
\begin{equation}
\left[\delta_{\zeta_1},\delta_{\zeta_2}\right]f=
\{f,\{\zeta^\alpha_1\phi_\alpha,\zeta^\beta_2\phi_\beta\}\}=
\{f,\zeta^\alpha_1\zeta^\beta_2{C_{\alpha\beta}}^\gamma\phi_\gamma\}
\approx\zeta^\alpha_1\zeta^\beta_2{C_{\alpha\beta}}^\gamma
\{f,\phi_\gamma\}\ .
\end{equation}

Finally, not only do phase space configurations and trajectories of the
system fall into gauge equivalence classes in this manner, each such class
being associated to a distinct physical configuration of the system,
but a likewise classification of physical phase space observables as
the gauge equivalence classes of first-class quantities is also relevant.
Indeed, given any first-class quantity $f$ such that
\begin{equation}
\{f,\phi_\alpha\}={f_\alpha}^\beta\,\phi_\beta\ ,
\end{equation}
clearly this property is preserved through time evolution since the
total Hamiltonian $H_T=H+\lambda^\alpha\phi_\alpha$ is first-class,
while it also implies that all its gauge equivalent representations are
all of the form $f+\chi^\alpha\phi_\alpha$ for some coefficient functions
$\chi^\alpha(z_A)$. Hence, gauge invariant physical observables of the
system are nothing but the gauge equivalence classes of first-class
quantities, whose time evolution is well defined and independent of the
choice of Lagrange multipliers $\lambda^\alpha(t)$, as it should since the
latter parametrize gauge transformations throughout the time history of
the system. Obvious examples of such gauge invariant physical observables are
the first-class Hamiltonian $H$, the first-class constraints $\phi_\alpha$
which must vanish for physical configurations, and thus also the total
Hamiltonian $H_T$, which takes values independent of $\lambda^\alpha(t)$
for physical configurations.

A few final remarks are in order.\cite{JG1} First, it should be stressed that
even though there is in general a correspondence between Lagrangian and
Hamiltonian local gauge invariances, there is by no means any necessity that
the corresponding algebraic structures should be identical. The example of
the scalar relativistic particle to be discussed hereafter provides an
illustration. Furthermore, it may be that given the complete Hamiltonian
formulation of the system, a specific choice for some of the Lagrange
multipliers is implicitly made before the reduction of some conjugate 
momenta is effected in order to obtain a particular Lagrangian formulation 
of the same dynamics. In such a case, this Lagrangian shares only part of the
original gauge freedom of the Hamiltonian formulation, with a specific
correspondence between these gauge symmetries, rather than an identity, 
given the effected partial reduction of phase space. Likewise, if one fails
to notice that some of the original configuration space degrees of freedom
are in fact Lagrange multipliers for some constraints, and thus applies
an analysis of constraints for the associated tri\-vial conjugate momenta,
one obtains further first-class constraints expressing the Lagrange multiplier 
character of that sector of the system. As a matter of fact, such
redundant features may be gauged away without compromising the
genuine dynamics of the system, leading {\sl in fine\/} to the 
{\sl fundamental\/} or {\sl basic\/} Hamiltonian formulation\cite{JG1} of a 
gauge invariant system. The above con\-si\-de\-ra\-tions also indicate
how a same gauge invariant system may in fact possess quite a number
of distinct Hamiltonian and Lagrangian formulations. Only a complete
analysis of its constraints can uncover its actual basic Hamiltonian
formulation.

Finally, let us also stress that those gauge symmetries generated by
the first-class constraints are {\sl small\/} Hamiltonian gauge transformations,
namely are transformations which belong to the same homotopy class as
the identity transformation, being continuously connected to the latter.
Systems may also be invariant under {\sl large\/} gauge transformations,
namely gauge symmetries whose parameters are (space)time dependent functions
but such that nonetheless the associated transformations are not
continuously connected to the identity and thus belong to a homotopy
class of the gauge symmetry group different from the identity class.
In considering the Hamiltonian formulation of gauge invariant systems,
whereas its small gauge transformations are directly accounted for through
the first-class constraints, invariance under large gauge transformations, 
if relevant, has to be enforced separately.

To conclude this section, let us invite the reader to develop the
analysis of constraints of a pure nonabelian Yang-Mills theory, whose
Lagrangian density is
\begin{equation}
{\cal L}=-\frac{1}{4}F^a_{\mu\nu}F^{a\mu\nu}\ \ ,\ \ 
F^a_{\mu\nu}=\partial_\mu A^a_\nu-\partial_\nu A^a_\mu-
gf^{abc}A^b_\mu A^c_\nu\ ,
\label{eq:LYM}
\end{equation}
$A^a_\mu(x^\mu)$ being the gauge field vector associated to the nonabelian
compact Lie algebra of structure constants $f^{abc}$ such that
$[T^a,T^b]=if^{abc}T^c$, and $g$ the gauge coupling constant. Such an
analysis is quite instructive, for instance in what concerns the
Lagrange multiplier status of the time component $A^a_0$ of the gauge vector
potentials and the identification of the basic Hamiltonian formulation of
such a system.

As a particularization of pure Yang-Mills theory, it is also instructive
to consider its dimensional reduction to 0+1 dimensions,\cite{JG3} namely to
a gauge invariant mechanical system, leading to a Lagrangian of the
form
\begin{equation}
L(q^a_i,\dot{q}^a_i)=\frac{1}{2g^2}\left[\dot{q}^a_i+
f^{abc}\lambda^bq^c_i\right]^2-V(q^a_i)\ \ ,\ \ 
V(q^a_i)=\frac{1}{2}\omega^2\left(q^a_i\right)^2\ ,
\end{equation}
$V(q^a_i)$ being a gauge invariant potential, such as the quadratic one
indicated. In fact, upon dimensional reduction the pure Yang-Mills
Lagrangian density (\ref{eq:LYM}) leads to quartic terms
in the potential, which may be ignored without spoiling gauge invariance
in 0+1 dimensions. The advantage of the quadratic choice is that usual
harmonic oscillator techniques enable an explicit resolution, even at the
quantum level, of this gauge invariant system, in particular with the
identification of its gauge invariant physical spectrum.

\subsection{The relativistic scalar particle}
\label{Subsect5.2}

As a useful guide illustrating the discussion which is to follow
later on, let us present now in detail the analysis of constraints
for a interesting though simple enough system, namely the relativistic
scalar massive particle.\cite{JG1} When discussing that system, we shall 
take for the Minkowski spacetime metric the signature 
$\eta_{\mu\nu}={\rm diag}\,(-++...++)$
in a $D$-dimensional spacetime, $\mu,\nu=0,1,2\cdots,D-1$. Furthermore,
this system possesses two well known action principle formulations,
one leading to linear equations of motion, the other to nonlinear equations.
We shall consider here the linear formulation and indicate its relation
to the nonlinear one where appropriate.

\vspace{10pt}

\noindent\underline{The action principle}

\vspace{5pt}

Wishing to construct a manifestly spacetime Poincar\'e covariant
formulation of the particle's trajectories, one has to consider
its spacetime history in terms of a parametrized world-line $x^\mu(\tau)$
spanning some initial and final spacetime positions $x^\mu_i$ and $x^\mu_f$.
Nonetheless, the physics of the system should be independent not only
of the spacetime reference frame used, namely Poincar\'e invariant,
but it should also be independent of the world-line $\tau$ parametrization
used, namely invariant under arbitrary world-line reparametrizations or
diffeomorphisms. The latter include transformations
\begin{equation}
\tau\rightarrow\tilde{\tau}=\tilde{\tau}(\tau)\ \ ,\ \ 
x^\mu(\tau)\rightarrow\tilde{x}^\mu(\tilde{\tau})=x^\mu(\tau)
\label{eq:wltransf}
\end{equation}
which either preserve the world-line orientation, or reverse it.
Clearly, the former class of transformations defines an ensemble
of small gauge transformations, whereas the latter one an ensemble
of large gauge transformations. In particular, the quotient of
the group of world-line diffeomorphisms by its connected identity
homotopy component is isomorphic to the group $\Z_2$ of two elements.
This quotient group is also known as the modular group.

Hence, given these two general requirements, the action of the system
should be both a spacetime and a world-line scalar. Form the latter
point of view, the spacetime coordinates $x^\mu(\tau)$ are nothing
but scalar ``field" degrees of freedom on the world-line. One way to
construct a world-line scalar action is to couple in an invariant
manner these degrees of freedom to an intrinsic world-line metric
$g(\tau)=e^2(\tau)$, $e(\tau)$ being the intrinsic world-line einbein.
Consequently, the action reads as
\begin{equation}
S[x^\mu,e]=\int_{\tau_i}^{\tau_f}d\tau\ L(\dot{x}^\mu,e)\ ,
\end{equation}
with
\begin{equation}
L(\dot{x}^\mu,e)=\sqrt{g}\left[\frac{1}{2}g^{-1}\dot{x}^\mu\dot{x}^\nu
\eta_{\mu\nu}-\frac{1}{2}m^2\right]=
\frac{1}{2}|e|^{-1}\dot{x}^2-\frac{1}{2}|e|m^2\ .
\end{equation}
Here, $m>0$ stands for a parameter with the dimension of mass, which
will indeed turn out to correspond to the particle's mass. At this stage
however, it appears to play the role of a one-dimensional cosmological
constant for this metric theory, {\it i.e.\/}, a theory of gravity on the 
one-dimensional world-line. Note also that the requirement of Poincar\'e 
invariance, in particular under spacetime translations, forbids any term 
dependent on the coordinates $x^\mu$ rather than its $\tau$-derivatives. 
Finally, a dot above a quantity denotes a $\tau$-derivative, since the 
parameter $\tau$ is to be viewed as the time evolution parameter of the 
system's dynamics. Nevertheless, the actual physical time is the measurement 
of the time component $x^0$ of the particle's spacetime trajectory. It is 
the requirement of a manifestly Poincar\'e covariant formulation which 
necessitates the gauge symmetry in world-line reparametrizations, but this
latter symmetry also allows to obtain a description which, physically, is 
independent of the world-line parametrization. Only the gauge invariant 
relations between the components of the particle's trajectory 
$x^\mu(\tau)$ are physically relevant.

The above action provides the linear formulation of the system, since the
ensuing equations of motion are linear, as is easily established.
From the world-line point of view, Poincar\'e invariance defines an
internal global symmetry whose Noether charges are given by
\begin{equation}
P_\mu=\frac{\partial L}{\partial\dot{x}^\mu}\ \ \ ,\ \ \ 
M_{\mu\nu}=P_\mu x_\nu-P_\nu x_\mu\ ,
\end{equation}
for the particle's energy- and orbital angular-momentum, respectively.
In particular, the Euler-Lagrange equations of motion are nothing but the 
statement of the conservation of the Noether energy-momentum of the particle, 
$dP_\mu/d\tau=0$, for which we leave it as a straightforward exercise to 
construct the solutions given the above choice of boundary conditions.

On the other hand, under world-line diffeomorphisms, in addition to
the transformations (\ref{eq:wltransf}), the einbein variation is
\begin{equation}
\tilde{e}(\tilde{\tau})=\frac{d\tau}{d\tilde{\tau}}e(\tau)\ .
\end{equation}
In infinitesimal form, one then finds
\begin{equation}
\tilde{\tau}=\tau-\eta(\tau)\ \ ,\ \ 
\delta_\eta x^\mu=\tilde{x}^\mu(\tau)-x^\mu(\tau)=\eta\dot{x}^\mu(\tau)\ \ ,\ \ 
\delta_\eta e(\tau)=\frac{d}{d\tau}\left(\eta(\tau)e(\tau)\right)\ ,
\end{equation}
so that the algebra of small Lagrangian gauge symmetries is given by
\begin{equation}
\left[\delta_{\eta_1},\delta_{\eta_2}\right]=
\delta_{\eta_1\dot{\eta}_2-\eta_2\dot{\eta}_1}\ .
\end{equation}
Note that this algebra of Lagrangian diffeomorphisms is nonabelian.

\vspace{10pt}

\noindent\underline{The Hamiltonian formulation}

\vspace{5pt}

Turning to the Hamiltonian formulation, phase space is equipped with
the canonical brackets
\begin{equation}
\{x^\mu(\tau),P_\nu(\tau)\}=\delta^\mu_\nu\ \ \ ,\ \ \ 
\{e(\tau),\pi_e(\tau)\}=1\ ,
\end{equation}
with the conjugate momenta
\begin{equation}
P_\mu(\tau)=\frac{\partial L}{\partial\dot{x}^\mu(\tau)}\ \ \ ,\ \ \ 
\pi_e(\tau)=\frac{\partial L}{\partial\dot{e}(\tau)}=0\ ,
\end{equation}
hence leading to the primary constraint $\pi_e=0$. Note that this constraint
actually follows from the fact that $e(\tau)$ is a Lagrange multiplier
for a constraint, as the forthcoming analysis will confirm.
The canonical Hamiltonian is
\begin{equation}
H_0=\dot{x}^\mu P_\mu+\dot{e}\pi_e-L=\frac{1}{2}|e|\left[P^2+m^2\right]\ .
\end{equation}

The primary Hamiltonian is thus
\begin{equation}
H_*=\frac{1}{2}|e|\left[P^2+m^2\right]+u\pi_e\ ,
\end{equation}
$u$ being some {\sl a priori\/} unknown function. Consistent time evolution
of the primary constraint $\pi_e=0$ then requires
\begin{equation}
\dot{\pi}_e=\{\pi_e,H_*\}=-\left({\rm sign}\,e\right)\,
\frac{1}{2}\left[P^2+m^2\right]\ ,
\end{equation}
thus leading to the secondary constraint
\begin{equation}
\phi=\frac{1}{2}\left[P^2+m^2\right]=0\ .
\end{equation}
However, requiring the consistent time evolution of this second constraint
does to lead to any further condition, nor any restriction on the associated 
functions defining the contribution of the linear combination of the
constraints to the total Hamiltonian. Hence, the complete set
of constraints is given by $\pi_e=0$ and $\phi=0$, which are first-class since
\begin{equation}
\{\pi_e,\pi_e\}=0\ \ ,\ \ \{\pi_e,\phi\}=0\ \ ,\ \ \{\phi,\phi\}=0\ .
\end{equation}
The Hamiltonian dynamics is thus generated by the total Hamiltonian
\begin{equation}
H_T=\frac{1}{2}\left[|e|+\tilde{\lambda}\right]\left[P^2+m^2\right]+u\pi_e\ ,
\end{equation}
where $u(\tau)$ and $\tilde{\lambda}(\tau)$ are the associated Lagrange
multipliers. Correspon\-ding\-ly, the first-order action is
\begin{equation}
S[x^\mu,P_\mu;e,\pi_e;\tilde{\lambda},u]=\int_{\tau_i}^{\tau_f}d\tau
\left[\dot{x}^\mu P_\mu+\dot{e}\pi_e-\frac{1}{2}\left[|e|+\tilde{\lambda}\right]
\left[P^2+m^2\right]-u\pi_e\right]\ .
\end{equation}

However, this form makes it obvious that the $(e,\pi_e)$ sector may be 
decoupled altogether, since in fact the einbein $|e|$ indeed may be absorbed 
into the definition of the Lagrange multiplier for the first-class constraint
$\phi=0$, exactly as was anticipated above. Setting then $u=\dot{e}$ and
$\lambda=|e|+\tilde{\lambda}$, one finally reaches the basic Hamiltonian
formulation of the relativistic massive scalar particle in the form the
first-order action
\begin{equation}
S[x^\mu,p_\mu;\lambda]=\int_{\tau_i}^{\tau_f}d\tau
\left[\dot{x}^\mu P_\mu-\lambda\phi\right]=
\int_{\tau_i}^{\tau_f}d\tau\left[\dot{x}^\mu P_\mu-
\frac{1}{2}\lambda\left[P^2+m^2\right]\right]\ .
\label{eq:basicS}
\end{equation}
The $(e,\pi_e)$ sector has indeed been decoupled, leaving over only
one first-class constraint $\phi=(P^2+m^2)/2$ whose Lagrange multiplier
$\lambda$ should thus play the same role as that of the world-line einbein,
as shall be confirmed hereafter. Furthermore, phase space consists solely
of the sector of spacetime degrees of freedom $(x^\mu,P_\mu)$ with the
previous canonical Poisson brackets, while the first-class Hamiltonian $H$
vanishes identically, $H=0$, as befits any reparametrization invariant
theory since $H$ ought to be the generator of time reparametrizations.
Hence, since the total Hamiltonian is only given by the first-class
constraint, $H_T=\lambda\phi$, the latter should also correspond to the
generator of world-line reparametrizations in the Hamiltonian formulation,
an expectation confirmed below.

The Hamiltonian equations of motion are simply
\begin{equation}
\dot{x}^\mu=\lambda P^\mu\ \ ,\ \ \dot{P}_\mu=0\ \ ,\ \ P^2+m^2=0\ .
\end{equation}
Performing the Hamiltonian reduction of the momenta 
$P_\mu=\dot{x}_\mu/\lambda$, a substitution in (\ref{eq:basicS})
then finds
\begin{equation}
S[x^\mu;\lambda]=\int_{\tau_i}^{\tau_f}d\tau
\left[\frac{1}{2}\lambda^{-1}\dot{x}^2-\frac{1}{2}\lambda m^2\right]\ ,
\end{equation}
namely precisely the original Lagrangian action in the linear formulation,
with the Lagrange multiplier $\lambda$ playing now the role of the einbein
degree of freedom, except for the fact it is no longer the absolute value
of the einbein that appears in the action. Consequently, the Hamiltonian
formulation of the system is not invariant under the large gauge symmetries
of the original Lagrangian formulation. This gauge symmetry will have to be
enforced se\-pa\-ra\-te\-ly at the end of the analysis, whether at the 
classical level or after its canonical quantization.

Note that when also solving for the constraint $P^2+m^2=0$ after
the Hamiltonian reduction, namely with $\lambda=\sqrt{-\dot{x}^2}/m$, finally
the action reduces to
\begin{equation}
S[x^\mu]=-m\int_{\tau_i}^{\tau_f}d\tau\sqrt{-\dot{x}^2}\ .
\end{equation}
In fact, this action provides the nonlinear formulation of the same system,
in which the total world-line length between its boundary points is measured
this time in terms of the metric induced on the world-line by the
spacetime Minkowski metric in which the world-line is embedded through the
functions $x^\mu(\tau)$, rather than the intrinsic metric defined through
the choice of einbein $e(\tau)$ or $\lambda(\tau)$. The reader is invited
to consider the analysis of constraints star\-ting from this nonlinear
formulation. Among other results, it follows that the constraint
$\phi=[P^2+m^2]/2$ then appears immediately as a primary constraint,
with a vanishing canonical Hamiltonian, and that no further constraints arise.
Hence, the same basic Hamiltonian formulation as the one above is
recovered, providing an explicit illustration of the fact that the primary
and secondary character of constraints depends on the Lagrangian formulation
used, and that gauge invariant systems possess different though physically
equivalent Lagrangian and Hamiltonian formulations, but only a single basic
Hamiltonian one. Note however that the nonlinear formulation applies
only to massive particles, whereas the linear one remains valid even for
massless particles.

The sole first-class constraint of the system, $\phi=[P^2+m^2]/2=0$, is the
generator of a local Hamiltonian gauge symmetry of the system, which can but
only correspond to small world-line diffeomorphisms. To establish the
exact correspondence, let us consider the infinitesimal Hamiltonian gauge
transformations,
\begin{equation}
\delta_\epsilon x^\mu=\{x^\mu,\epsilon\phi\}=\epsilon P^\mu\ \ ,\ \ 
\delta_\epsilon P_\mu=0\ \ ,\ \ 
\delta_\epsilon\lambda=\dot{\epsilon}\ ,
\end{equation}
where $\epsilon(\tau)$ is an arbitrary infinitesimal function, which
must vanish at the end points, $\epsilon(\tau_{i,f})=0$, when enforcing
the above choice of boundary conditions for the particle's trajectory.
Given these expressions, the associated finite transformations are
readily found to be
\begin{equation}
{x'}^\mu(\tau)=x^\mu(\tau)+h(\tau)P^\mu(\tau)\ \ ,\ \ 
{P'}^\mu(\tau)=P^\mu(\tau)\ \ ,\ \ 
\lambda'(\tau)=\lambda(\tau)+\frac{dh(\tau)}{d\tau}\ ,
\label{eq:finitegauge}
\end{equation}
$h(\tau)$ being an arbitrary function such that $h(\tau_{i,f})=0$ when
enforcing the above boundary conditions, as may easily be confirmed by
checking the invariance of the action (\ref{eq:basicS}). Given these results,
we may now consider their relation to the reparametrization gauge symmetry
in the Lagrangian formulation.

First, notice that contrary to the Lagrangian diffeomorphism algebra
which is nonabelian, the Hamiltonian one is abelian, $\{\phi,\phi\}=0$,
on account of the antisymmetry property of Poisson brackets. Consequently,
these two algebraic structures are not identical, even though there is
a unique correspondence, but no identity, between the relevant transformations
of the degrees of freedom. Indeed, given the above finite
Hamiltonian reparametrization with parameter $h(\tau)$, the corresponding
finite Lagrangian reparametrization such that $\tau=f(\tilde{\tau})$
is constructed from the relation\cite{JG1}
\begin{equation}
h(\tau)=\int_\tau^{f(\tau)}d\tau'\,\lambda(\tau')\ ,
\end{equation}
with in particular $f(\tau_{i,f})=\tau_{i,f}$ and $h(\tau_{i,f})=0$ when
enforcing the boundary conditions $x^\mu(\tau_{i,f})=x^\mu_{i,f}$. In the
case of infinitesimal transformations such that $f(\tau)=\tau+\eta(\tau)$, 
this correspondence reduces to
\begin{equation}
\epsilon(\tau)=\lambda(\tau)\,\eta(\tau)\ ,
\end{equation}
possibly with the conditions $\epsilon(\tau_{i,f})=0=\eta(\tau_{i,f})$.
Hence, the advocated one-to-one correspondence but not necessarily
identity between Lagrangian and Hamiltonian small gauge symmetries is
established for this particular system. As a matter of fact, such a
correspondence remains valid for reparametrization invariant theories
in whatever dimension, thus including general relativity. In contradistinction
in the case of nonabelian Yang-Mills theories, this cor\-res\-pon\-den\-ce 
between the two classes of gauge symmetries becomes in fact an identity, since
the relevant gauge symmetries are then internal ones, independent of the
spacetime evolution of the system.

The total proper-time, or proper-length, of the particle's trajectory
for the specified boundary conditions,
\begin{equation}
\gamma=\int_{\tau_i}^{\tau_f}d\tau\,\lambda(\tau)\ ,
\end{equation}
is indeed also a gauge invariant quantity. In fact, it is the sole
gauge invariant degree of freedom in the Lagrange multiplier sector,
and characterizes the different metric structures that may be defined
on the world-line. Let us thus refer to it as the Teichm\"uller parameter
of the system.\cite{JG1} As such, it also appears explicitly in the solutions
to the equations of motion, given by
\begin{equation}
x^\mu(\tau)=x^\mu_i+\frac{\Delta x^\mu}{\gamma}\int_{\tau_i}^\tau\,d\tau'\,
\lambda(\tau')\ \ ,\ \ 
P^\mu(\tau)=\frac{\Delta x^\mu}{\gamma}\ \ ,\ \ 
\Delta x^\mu=x^\mu_f-x^\mu_i\ ,
\end{equation}
while the constraint $P^2+m^2=0$ requires that
\begin{equation}
\sqrt{-\left(\Delta x\right)^2}=m|\gamma|\ .
\end{equation}
Note how the choice of Lagrange multiplier indeed parametrizes the
freedom in the choice of world-line parametrization, hence the gauge
freedom of the system. Given this solution, the gauge invariant content
of the spacetime sector of the formulation may also be identified.
Thus the spacetime trajectory is given by
\begin{equation}
\vec{x}(x^0)=\vec{x}_i+\Delta\vec{x}\,\frac{x^0-x^0_i}{\Delta x^0}\ ,
\end{equation}
which is indeed a gauge invariant relation, independent of the choice
of world-line parametrization in $\tau$. Only the specific relation
between the physical time $x^0$ and the world-line evolution parameter
$\tau$ is gauge dependent and thus dependent on the choice of
Lagrange multiplier or world-line einbein, namely
\begin{equation}
x^0(\tau)=x^0_i+\frac{\Delta x^0}{\gamma}\,\int_{\tau_i}^\tau\,
d\tau'\,\lambda(\tau')\ .
\end{equation}

Note also that the above solution for the energy-momentum of the particle
shows that positive (respectively, negative) values for the quantity $\gamma$
correspond to the propagation forward (respectively, backward) in time of the
particle, namely in a quantum parlance, to a particle (respectively, 
antiparticle) as opposed to its antiparticle (respectively, particle). 
This brings us back to the
issue of large gauge symmetries, which are not realized in the Hamiltonian
formulation. Clearly, large world-line diffeomorphisms induce a change
of orientation in the world-line, hence a change of sign of the einbein
$e(\tau)$ or Lagrange multiplier $\lambda(\tau)$, and thus also of
the Teichm\"uller parameter $\gamma$. Consequently, the $\Z_2$ modular
group of orientation reversing diffeomorphisms modulo orientation preserving
ones, namely the class of large gauge transformations, acts on Teichm\"uller
space simply as $\gamma\rightarrow -\gamma$, thereby dis\-tin\-ghui\-shing a
particle description as opposed to its antiparticle.\cite{JG1} Invariance under
these modular transformations of the Teichm\"uller parameter thus needs to
be enforced when considering specific configurations of the unoriented
scalar particle. Positive energy solutions then propagate forward in time
with the modular invariant restriction $\gamma>0$.

\subsection{Gauge fixing, reduced phase space and Gribov problems}
\label{Subsect5.3}

\vspace{10pt}

\noindent\underline{Faddeev's reduced phase space}

\vspace{5pt}

The redundant degrees of freedom inherent to the gauge symmetries of
a constrained dynamics are certainly a challenge to the proper gauge invariant
quantization of such systems. {\sl A priori\/}, one possible approach would
be first to solve for the gauge constraints $\phi_\alpha$, and only then
quantize the reduced phase space degrees of freedom, which are certainly
then physical since no gauge symmetry freedom remains. Let us then introduce
a set of gauge fixing conditions $\Omega_\alpha=0$ whose number is equal to 
that of the first-class constraints $\phi_\alpha$ (note that this requires
the set of first-class constraints to be irreducible, namely locally
linearly independent), and which are such that
the matrix of brackets of this whole set of constraints is regular,
\begin{equation}
{\rm det}\,\{\Omega_\alpha,\phi_\beta\}\not{\!\!\approx}\ 0\ .
\label{eq:2ndclass}
\end{equation}
In other words, by introducing the additional conditions, the whole
set of constraints has been turned into second-class ones.

That such restrictions ``freeze" or fix the gauge symmetries of the system
may be seen from complementary points of view. First, consider arbitrary
infinitesimal gauge transformations of the gauge fixing conditions
\begin{equation}
\delta_\zeta\Omega_\alpha=\{\Omega_\alpha,\zeta^\beta\phi_\beta\}\ .
\end{equation}
Thus, if the condition (\ref{eq:2ndclass}) is met, there do not exist
infinitesimal gauge transformations leaving the gauge fixing conditions
invariant, namely the only solution to the equations 
$\delta_\zeta\Omega_\alpha=0$ is trivial, $\zeta^\alpha=0$. 
In other words, the conditions $\Omega_\alpha=0$ do 
indeed fix the gauge freedom, albeit for small infinitesimal gauge 
transformations only.
From an alternative point of view, consider now the time evolution of these
conditions,
\begin{equation}
\frac{d\Omega_\alpha}{dt}\approx\frac{\partial\Omega_\alpha}{\partial t}+
\{\Omega_\alpha,H\}+\{\Omega_\alpha,\phi_\beta\}\lambda^\beta\ .
\end{equation}
Thus once again, if the condition (\ref{eq:2ndclass}) is met, the requirement
that the gauge fixing conditions $\Omega_\alpha=0$ remain valid at all
times implies the unique determination of the Lagrange multipliers 
$\lambda^\alpha(t)$. Since these functions are known to parametrize the
gauge freedom of the system throughout its time history, it thus follows
that the conditions $\Omega_\alpha=0$ imply a specific choice for these
functions, namely a specific gauge fixing of the system.

Consequently, it appears that conditions $\Omega_\alpha=0$ such
that (\ref{eq:2ndclass}) is obeyed imply a gauge fixed formulation of
the system, in which all the redundant features inherent to such
symmetries are explicitly resolved. The latter is simply achieved by
working out the Dirac brackets associated to the condition (\ref{eq:2ndclass})
and to the second-class constraints $\Omega_\alpha=0=\phi_\alpha$.
The ensuing reduced phase space description based on these Dirac brackets
is known as Faddeev's reduced phase space formulation of gauge invariant 
theories.\cite{Fad} Only gauge invariant physical degrees of freedom remain
dynamical within such a formulation.

As an explicit illustration, let us consider again the relativistic scalar 
particle, with the gauge fixing condition
\begin{equation}
\Omega=x^0(\tau)-\left[x^0_i+\frac{\Delta x^0}{\Delta\tau}
\left(g(\tau)-\tau_i\right)+h_0(\tau)P^0(\tau)\right]\ ,
\label{eq:Omega}
\end{equation}
with of course $\Delta\tau=\tau_f-\tau_i$. Here, $g(\tau)$ 
(respectively, $h(\tau)$) is some arbitrary function such that 
$g(\tau_{i,f})=\tau_{i,f}$ (respectively, $h(\tau_{i,f})=0$). From previous 
results for this system, it
is clear that $g(\tau)$ parametrizes the gauge freedom in the choice of
world-line parametrization, whereas $h_0(\tau)$ parametrizes an arbitrary
finite small gauge transformation of this gauge fixing condition.
Consequently, the final reduced phase space description associated to
this choice should be independent of these two functions.

Since for a massive particle,
\begin{equation}
\{\Omega,\phi\}=P^0\neq 0\ \ ,\ \ 
\phi=\frac{1}{2}\left[P^2+m^2\right]=0\Rightarrow P^0=\eta\sqrt{\vec{P}^2+m^2}
\ \ ,\ \ \eta=\pm 1\ ,
\end{equation}
the two conditions $\Omega=0=\phi$ together indeed define a set of
second-class constraints. Solving then for the time component degrees
of freedom $x^0(\tau)$ and $P^0(\tau)$ from these two constraints,
the Dirac brackets for the space components are readily determined
to coincide with the original canonical brackets,
\begin{equation}
\{x^i(\tau),P_j(\tau)\}_D=\delta^i_j\ .
\end{equation}
The evolution of the system in terms of the physical
time $x^0$ rather than the world-line parameter $\tau$ is then generated
by the reduced Hamiltonian
\begin{equation}
H_{\rm reduced}=\eta\sqrt{\vec{P}^2+m^2}\ ,
\end{equation}
so that for any observable $F$ on this reduced phase space,
\begin{equation}
\frac{dF}{dx^0}=\frac{\partial F}{\partial x^0}+\{F,H_{\rm reduced}\}_D\ .
\end{equation}
All the gauge dependent features are indeed no longer involved in this
gauge fixed description of the system. Rather, they only appear in the
gauge dependent relations, namely the $\tau$ parametrization of the
physical time,
\begin{equation}
x^0(\tau)=x^0_i+\frac{\Delta x^0}{\Delta\tau}\left[g(\tau)-\tau_i\right]+
h_0(\tau)P^0(\tau)\ ,
\end{equation}
in which the energy degree of freedom is given by
\begin{equation}
P^0(\tau)=\eta\sqrt{\vec{P}^2(\tau)+m^2}\ \ ,\ \eta=\pm 1\ ,
\end{equation}
as well as the choice of world-line einbein and Teichm\"uller parameter,
\begin{equation}
\lambda(\tau)=\eta\frac{\Delta x^0}{\Delta\tau}
\frac{\dot{g}(\tau)}{\sqrt{\vec{P}^2(\tau)+m^2}}\ +\ \dot{h}_0(\tau)\ ,\ 
\gamma=\eta\frac{\Delta x^0}{\Delta\tau}\int_{\tau_i}^{\tau_f}d\tau\,
\frac{\dot{g}(\tau)}{\sqrt{\vec{P}^2(\tau)+m^2}}\ ,
\end{equation}
in ways that are totally in agreement with the role played by the functions
$g(\tau)$ and $h_0(\tau)$ and the gauge transformation properties of these
quantities (the quantity $P^0(\tau)$ is gauge invariant on its own already).
Note that for configurations solving the equations of motion, we always
have $\gamma=\Delta x^0/P^0$, which is indeed independent of the functions
$g(\tau)$ and $h_0(\tau)$, as it should.

\vspace{10pt}

\noindent\underline{Admissible gauge fixing and Gribov problems}

\vspace{10pt}

In order to properly assess\cite{JG1} what any gauge fixing procedure actually 
achieves, it is necessary to better understand the redundancy features inherent
to the gauge symmetry properties related to the first-class constraints.
For this purpose, let us consider the whole of phase space $\{z_A(t)\}$ 
together with the set of Lagrange multipliers $\lambda^\alpha(t)$, thus
defining a large space of functions. Small gauge transformations generated
by the first-class constraints directly act on that space, thereby organizing
it into a whole set of disjoint gauge orbits without any intersections.
The space of all such gauge orbits is nothing but the quotient of the
whole space $\{z_A,\lambda^\alpha\}$ by the action of the small local gauge
symmetry, and thus represents the ensemble of all physically distinct
gauge invariant configurations possibly accessible to the system throughout
its time evolution history. What one would in fact hope to be feasible
should be the dynamical description of the system, as well as its quantization,
on that quotient space of gauge orbits, rather than the space 
$\{z_A,\lambda^\alpha\}$ with all its inherent redundancy features related to 
the gauge transformations acting on it. In practice however, the topology 
and analytical properties of the space of gauge orbits are just too intricate
to contemplate such an approach towards gauge invariant dynamics.

Hence, the basic idea of gauge fixing is to identify within the original
space $\{z_A,\lambda^\alpha\}$ a subset chosen in such a way that each of
its elements is just one and only one representative for each of all the
possible gauge orbits of the system. Namely, any gauge fixing should in
effect implicitly define some gauge slicing of the space 
$\{z_A,\lambda^\alpha\}$ which would intersect each of its gauge orbits
once and only once. Provision can be made for those cases in which the
gauge slice intersects some gauge orbits more than once, but then in such
a manner that the gauge slice and its intersections are counted with
an orientation leading {\sl in fine\/} to an effective count of intersections
which still adds up to a single one. Clearly, when such a gauge fixing
is achieved, it is an admissible one, meaning that the dynamics of
the system reduced onto the gauge slice is totally equivalent to the
original dynamics formulated either on the space of gauge orbits or
equivalently within the space $\{z_A,\lambda^\alpha\}$ with proper
account for the gauge symmetries. Clearly, in order to assess whether
a given gauge fixing procedure is admissible, it is imperative first
to properly identify the space of gauge orbits of the system and its
characterization in terms of gauge invariant quantities constructed
from the variables $\{z_A,\lambda^\alpha\}$ and which may then serve as
coordinates parametrizing the space of gauge orbits.

In practice however, given some gauge fixing procedure, namely a restriction
on the variables $\{z_a,\lambda^\alpha\}$ which in effect ``freezes" the
gauge freedom of the system and leads thereby to an effective reduced
phase space formulation involving then physical degrees of freedom only
(Faddeev's reduced phase space being the archetype example), there is no
guarantee whatsoever that an admissible gauge fixing is achieved.
Indeed, any such gauge fixing procedure in effect singles out through some
gauge slicing a certain subset of the space $\{z_A,\lambda^\alpha\}$
for which no gauge freedom is left. This gauge slice intersects the
gauge orbits in a certain manner specific to the gauge fixing procedure,
which may be characterized in terms of a specific covering of the space
of gauge orbits.\cite{JG1} By covering is meant a certain domain or subset 
of that space, as well as some measure over that domain which represents the
possibly multiple degeneracy in the count of intersected orbits.
Due to the gauge invariance properties of the original formulation of
the system, this covering of the gauge orbits induced by some gauge
fixing procedure is all that is relevant for the characterization of
that gauge fixing. Two gauge fixing procedures leading to an identical
covering of the space of gauge orbits are thus said to be gauge equivalent.
The dynamical descriptions achieved through two gauge equivalent gauge
fixings are physically equivalent.

\vspace{5pt}

However, gauge fixing procedures leading to different coverings of the
space of gauge orbits imply dynamical descriptions of the system which
are not physically equivalent, even though they both are gauge invariant.
In particular, it is only for an admissible covering, namely a gauge
fixing which in effect singles out each of the gauge orbits once and only
once, that the associated gauge invariant dynamics is physically
equivalent to that of the original system.

Whenever a gauge fixing procedure is not admissible in this very specific
sense, it is said to suffer from a Gribov problem.\cite{Gribov} In fact, 
one ought to distinguish two types of Gribov problems.\cite{JG1} 
The Gribov problem of type I
is of a local character, and occurs whenever among the selected gauge orbits
some are selected more than once, thereby leading to an overcount of
the cor\-res\-pon\-ding physical configurations accessible to the dynamics.
Likewise, the Gribov problem of type II is of a global character, and
occurs whenever some of the gauge orbits are not selected by the gauge fixing
procedure, thereby forbidding the system to dynamically access some of its
physical configurations. In other words, a Gribov problem of type I occurs
whenever some gauge orbits are counted more than once (relative to the
others), while a Gribov problem of type II occurs when some orbits are
not counted at all. 

By definition, an admissible gauge fixing is one which does not suffer
a Gribov problem of either type. However, an arbitrarily chosen gauge 
fixing procedure may suffer a Gribov problem of type I or of type II,
or even of both types, in which case it is not admissible. Even though
any gauge fixing procedure leads to a gauge invariant formulation of the
system, when a Gribov problem arises the gauge invariant dynamics which is 
being described is no longer that of the original system, since it no longer
includes the same set of physically distinct configurations accessible
to the dynamics. In that sense, one may say that it is only an admissible
gauge fixing which is a physically correct gauge fixing, even though
any gauge fixing procedure with a Gribov problem leads nonetheless
to a gauge invariant description. Gauge invariance is not all there is
to gauge invariant systems!

Consequently, whenever considering a given gauge fixing procedure,
one must also determine whether it is admissible or not, namely
whether it suffers Gribov problems of type I and of type II. This issue
may be addressed only on a case by case basis.

As an illustration, let us consider Faddeev's reduced phase space gauge
fixing. It is often said in the literature that the condition 
(\ref{eq:2ndclass}) is a sufficient condition for an admissible gauge
fixing. However, this is not correct, and in fact (\ref{eq:2ndclass})
defines only a necessary condition for admissibility, but not 
necessarily a sufficient one. Indeed, as was discussed previously, 
(\ref{eq:2ndclass}) is necessary in order that a gauge fixing be achieved 
for which small infinitesimal gauge transformations are fixed. Nevertheless, 
this still allows the possibility of nontrivial small finite gauge 
transformations
leaving invariant the gauge fi\-xing conditions, as indeed established
by Gribov,\cite{Gribov} as well as large finite gauge transformations. 
Such a possibility amounts to a gauge slicing in which some
gauge orbits are intersected more than once, even though when accounting
for an oriented slicing the effective count of intersections may still be
acceptable.\cite{Gribov2} Furthermore, (\ref{eq:2ndclass}) does not 
guarantee either that all gauge orbits are included at least once. 
Hence, Faddeev's gauge fixing procedure is far from being protected from 
Gribov problems of either type, and in general does not lead to an 
admissible gauge fixing. The generic situation is indeed that Faddeev's 
reduced phase space approach is plagued by Gribov problems.\cite{JG1,JG4,JG5} 
Even for a system as simple as the relativistic
scalar particle, it is shown below that this gauge fixing procedure is
not admissible, casting doubt on all other instances where it is being
applied for reparametrization invariant theories, namely theories of the
gravitational interaction. In fact, this lack of admissibility also
applies to the quantization of nonabelian Yang-Mills theories.\cite{Gribov}

Other gauge fixing procedures have been developed over the years, the
main reason being that Faddeev's approach usually breaks manifest Poincar\'e
invariance in field theory. This has led to the so-called BRST-BFV
Hamiltonian formulation of gauge theories,\cite{JG1,BFV1} in which gauge 
fixing is achieved in a different manner. Nonetheless, the issue of Gribov 
problems arises within that framework as well,\cite{JG1,JG4,JG5} and needs to 
be assessed on a case by case basis, a problem which requires a comprehensive 
understanding of the structure of the space of gauge orbits.

Hence, Gribov problems are subtle, difficult but essential features which
must be addressed in order to establish the admissibility of a chosen gauge
fixing of any gauge invariant dynamics, whatever the gauge fixing procedure
being envisaged. Within the context of nonabelian Yang-Mills theories,
these issues do not affect any perturbative analysis of quantum properties,
since perturbation theory around the ground state amounts to a perturbation
within the neighbourhood of vanishing fields, so that the fixing of
the small in\-fi\-ni\-te\-si\-mal gauge symmetries should suffice. However, 
it is most likely that nonperturbative phenomena should be highly dependent
on Gribov problems, which must thus be avoided in order to gain a genuine
and physically correct understanding of such phenomena.

To conclude, let us reconsider the relativistic massive scalar particle.
Given the small finite gauge transformations (\ref{eq:finitegauge}) of 
the variables $x^\mu(\tau)$, $P^\mu(\tau)$ and $\lambda(\tau)$ as well as
the choice of boundary conditions on the spacetime coordinates, 
$x^\mu(\tau_{i,f})=x^\mu_{i,f}$, it is clear that the Teichm\"uller
parameter $\gamma$ defines the coordinate labelling the space of gauge
orbits within the space of Lagrange multipliers $\lambda(\tau)$. 
Furthermore, it is readily established\cite{JG1} that any gauge fixing 
procedure which induces an admissible gauge fixing in the latter space also
induces an admissible gauge fixing of the whole of the variables
$\{x^\mu(\tau),P^\mu(\tau),\lambda(\tau)\}$ of the system. This remark
thus provides a tool to assess the admissibility of gauge fixing
procedures: simply consider the inferred set of values for $\gamma$.
Given Faddeev's gauge fixing associated to the choice (\ref{eq:Omega}),
we found
\begin{equation}
\gamma=\eta\frac{\Delta x^0}{\delta\tau}\int_{\tau_i}^{\tau_f}\,d\tau\,
\frac{\dot{g}(\tau)}{\sqrt{\vec{P}^2(\tau)+m^2}}\ .
\end{equation}
Since this expression implies the upper bound
\begin{equation}
|\gamma|\le\frac{|\Delta x^0|}{m}\ ,
\end{equation}
it is clear that this gauge fixing procedure suffers a Gribov problem
of type~II. Furthermore, it also suffers a Gribov problem of type I,
on account of the degeneracy in $\gamma$ values as the system probes
all those physical configurations for which the function $\vec{P}^2(\tau)$ 
remains identical.\cite{JG1,JG5} Note that by construction of the gauge 
fixing procedure,
these Gribov problems do not affect the actual classical physical solution
to the equations of motion, since the gauge slice always selects the gauge
orbit to which that solution belongs, given the chosen boundary conditions.
Nevertheless, at the quantum level, these Gribov pro\-blems imply that the
physically correct quantum amplitudes are not obtained for Faddeev's
reduced phase space gauge fixing of this system,\cite{JG1,JG5} as may 
ea\-si\-ly be anticipated from the point of view of the path integral 
representation of quantum amplitudes.

\subsection{Dirac's quantization}
\label{Subsect5.4}

The canonical quantization of constrained dynamics, known as Dirac's 
quantization,\cite{Dirac} amounts to the canonical quantization
of the basic Hamiltonian formulation of such systems as it has been
developed in the above presentation.\cite{JG1} The space of quantum states 
$|\psi>$ is a representation space for the commutation relations of the basic
phase space degrees of freedom operators $\hat{z}_A$ in the Schr\"odinger
picture,
\begin{equation}
[\hat{z}_a,\hat{z}_B]=i\hbar\hat{C}_{AB}(\hat{z_A})\ ,
\end{equation}
with all the ensuing operator ordering issues whenever these brackets
are noncanonical. Time evolution of quantum states $|\psi,t>$ is generated 
through the Schr\"odinger equation,
\begin{equation}
i\hbar\frac{d}{dt}|\psi,t>=\hat{H}_T(t)\,|\psi,t>\ \ \ ,\ \ \ 
\hat{H}_T(t)=\hat{H}+\lambda^\alpha(t)\hat{\phi}_\alpha\ ,
\end{equation}
$\hat{H}$ and $\hat{\phi}_\alpha$ being the first-class Hamiltonian
and constraint operators, while $\lambda^\alpha(t)$ are the associated
Lagrange multipliers which still play their role as a parametrization of
the freedom in applying any gauge transformation as the system evolves 
in time through the space of quantum states. If the chosen operator ordering 
of composite quantities is anomaly free, namely if these operators retain 
their gauge symmetry properties at the quantum level,
\begin{equation}
[\hat{H},\hat{\phi}_\alpha]=i\hbar{{\hat{C}}_\alpha}{}^\beta\hat{\phi}_\beta
\ \ ,\ \ 
[\hat{\phi}_\alpha,\hat{\phi}_\beta]=i\hbar{{\hat{C}}_{\alpha\beta}}{}^\gamma
\hat{\phi}_\gamma\ ,
\end{equation}
with in particular the quantities ${{\hat{C}}_\alpha}{}^\beta$ and
${{\hat{C}}_{\alpha\beta}}{}^\gamma$, which could indeed be ope\-ra\-tors
themselves in the general case of an open algebra, standing to the left of 
the constraints in the r.h.s. of these commutation relations,
then the quantization of the system is consistent and compatible with
its classical gauge invariance properties. However, it could happen that
this is not possible, in which case the ensuing gauge anomaly terms
appearing as additional contributions of order at least $\hbar^2$
in the r.h.s. of these commutation relations render the physical 
interpretation of the quantum theory at least problematic, if not 
inconsistent altogether. In the following discussion, it is assumed that
a gauge covariant quantization has been achieved.

The same issue arises for gauge invariant physical observables,
namely for the operators to be associated to first-class quantities.
Here again, this gauge invariant status remains valid provided only
operator ordering makes it possible that the commutation relations
are still given by the correspondence principle,
\begin{equation}
[\hat{f},\hat{\phi}_\alpha]=i\hbar{{\hat{f}}_\alpha}{}^\beta\,\hat{\phi}_\beta
\ ,
\end{equation}
given the classical bracket $\{f,\phi_\alpha\}={f_\alpha}^\beta\phi_\beta$.
Examples of quantum physical observables are thus the first-class
Hamiltonian $\hat{H}$ and constraint $\hat{\phi}_\alpha$ operators.
Consequently, in the same way as at the classical level, physical 
obser\-va\-bles are defined as being the gauge equivalence classes of 
first-class operators under gauge transformations generated by the first-class
constraints. The constraints themselves belong to the trivial class.

So far, these issues are analogous to those that arise for the
canonical quantization of an ordinary system. For gauge invariant systems,
the additional feature is that of gauge transformations generated by the
first-class constraints, under which physical configurations should remain
invariant. Hence in the quantized system, gauge invariant physical states
are those quantum states which are left invariant by small finite gauge
transformations, namely which are annihilated by the first-class
contraint operators,
\begin{equation}
\hat{\phi}_\alpha\,|\psi,t>\ =\ 0\ .
\end{equation}
Note that in some cases, this requirement proves to be too restrictive
by not leaving over any state. A weaker condition, which in fact is
sufficient for a consistent physical interpretation, is that the
matrix elements of the constraint operators for physical states vanish
identically,
\begin{equation}
<\psi,t|\hat{\phi}_\alpha|\chi,t>=0\ .
\end{equation}
Provided the constraint algebra is anomaly free, it is clear that these
definitions of physical states are consistent with the dynamics, namely
the physical character of a state is preserved under time evolution induced
through the Schr\"odinger equation and the first-class total Hamiltonian
operator $\hat{H}_T$. In effect, $\hat{H}$ and $\hat{\phi}_\alpha$
are then commuting operators on the subspace of physical states.
In particular, the constraints themselves define gauge equivalence classes
of physical observables of vanishing value for physical states.

When the above programme is completed, one says that Dirac's quantization
of a constrained system has been achieved. However, this still leaves
open the issue of the quantum dynamics of such systems, namely the
description of the time dependency of the system which amounts to the
understanding of the physical properties of the evolution operator
associated to its Schr\"odinger equation. Clearly, the potential
difficulty is that when considering the time propagation of quantum
states, a proper representation of the actual physical content of
the system should include as the sole contributing intermediate states 
only one quantum physical state for each of the possible gauge orbits. 
However, the evolution operator based on the total Hamiltonian $\hat{H}_T$ 
does also propagate all gauge noninvariant states both as intermediate as 
well as external states. To put it within the framework of the path integral 
representation of the evolution operator in which a summation over
all possible configurations is effected, the actual physical content
should follow from an effective integration only over the space of
gauge orbits with an equal weight given to each orbit. Otherwise,
quantum states other than physical ones contribute to the propagator
as intermediate states. This is the specific issue which seems to require
some gauge fixing procedure, with its potential Gribov problems as
the generic difficulty to be addressed on a case by case basis.

It is often claimed that the gauge fixed path integral representation,
hence the gauge fixed quantization, is independent of the gauge fixing
procedure, whether for Faddeev's reduced phase space approach (the Faddeev
theorem\cite{JG1,Fad}) or the BRST-BFV approach (the Fradkin-Vilkovisky 
theorem\cite{JG1,BFV1}). However, such a statement is 
misleading,\cite{JG1,JG4,JG5} since what these theorems in
fact establish, and nothing more, is that the resulting path integral 
representations, and thus also the associated quantized formulations of 
the system, are gauge invariant. Indeed, what a gauge fixing procedure 
implies is a specific covering of the space of gauge orbits, 
so that gauge equi\-va\-lent gauge fixing procedures (whose induced coverings 
of gauge orbits are thus identical) do indeed lead to identical quantum 
formulations and path integral representations. However for gauge 
nonequivalent gauge fixings, thus inducing different coverings of gauge 
orbits, necessarily the corresponding quantized formulations of the system, 
even though each is gauge invariant, are themselves different and thus not 
physically nor gauge equivalent, thereby leading necessarily to different 
path integral representations, since throughout its time history the system
then explores a different set of physical configurations which is left
accessible to it through the gauge fixing procedure. It is only for the 
class of admissible gauge fixings that the correct quantized and path 
integral formulation of the system is achieved. Any other nonadmissible 
gauge fixing leads to a different quantized system, albeit always a gauge 
invariant one.

As an explicit example of Dirac's quantization, let us consider the
relativistic scalar particle. Its canonical quantization is defined
by the Heisenberg algebra commutation relations, in the Schr\"odinger
picture,
\begin{equation}
[\hat{x}^\mu,\hat{P}_\nu]=i\hbar\delta^\mu_\nu\ \ ,\ \ 
{\hat{x}_\mu}^\dagger=\hat{x}_\mu\ \ ,\ \ 
{\hat{P}_\mu}^\dagger=\hat{P}_\mu\ ,
\end{equation}
while the first-class Hamiltonian and constraint are
\begin{equation}
\hat{H}=0\ \ ,\ \ \hat{\phi}=\frac{1}{2}\left[\hat{P}^2+m^2\right]\ .
\end{equation}
In the configuration space representation of the Heisenberg algebra,
quantum states are thus described by their wave function $\psi(x^\mu;\tau)$,
which solves the Schr\"odinger equation
\begin{equation}i\hbar\frac{\partial\psi(x^\mu;\tau)}{\partial\tau}=
\frac{1}{2}\lambda(\tau)\left[-\hbar^2\partial^2_x+m^2\right]
\psi(x^\mu;\tau)\ ,
\end{equation}
$\lambda(\tau)$ being the einbein Lagrange multiplier. Hence, physical states,
defined to be annihilated by the reparametrization constraint
$\hat{\phi}|\psi,\tau>=0$, or
\begin{equation}
\left[-\hbar^2\partial^2_x+m^2\right]\,\psi(x^\mu;\tau)=0\ ,
\end{equation}
are independent of the world-line parameter $\tau$, as they should indeed.
Hence, once recovers the fact that the single quantum relativistic scalar
particle's dynamics is governed by the Klein-Gordon equation for its
wave function in configuration space.

Turning then to the quantum dynamics issue, it should be such that for
the physical gauge invariant states, their causal unitary quantum evolution
operator be given by Feynman's propagator for a scalar field. However,
unless an admissible gauge fixing of the above formulation is effected,
this is certainly not the result which one obtains from the above Dirac
quantization of the relativistic particle. For instance, following
Faddeev's reduced phase space approach based on the gauge fixing
condition (\ref{eq:Omega}), one may show\cite{JG1,JG4,JG5} that the correct 
Feynman propagator
indeed does not follow, but rather that it suffers precisely the Gribov 
problems of type I and II which were already described previously in
relation to that choice of gauge fixing, namely a bounded integration over 
the space of gauge orbits characterized by the finite
range of Teichm\"uller parameter values $\gamma$ as well as a nonuniform
integration measure over those gauge orbits that are accounted for because
of the degeneracy in the obtained values for $\gamma$.
In contradistinction, when an admissible gauge fixing is possible
(as is the case for this system within the BRST-BFV 
approach\cite{JG1,JG4,JG5}), then the
Feynman propagator is indeed readily recovered, provided the role
of large gauge transformations is also properly accounted for, as shall
be discuss hereafter. Hence, this simple example confirms the fact that the
issue of admissibility and Gribov problems is by no means a trivial and
irrelevant one, since otherwise the correct gauge invariant physical
content of the system is not recovered.

\subsection{Klauder's physical projector:\\
gauge invariant quantum dynamics without gauge fixing}
\label{Subsect5.5}

Given all the difficulties surrounding gauge fixing and Gribov problems,
it is legitimate to ask whether any gauge fixing is at all a necessity.
In fact, it is possible, entirely within Dirac's quantization scheme
and nothing more, to circumvent the problem simply by not addressing it,
which is most welcome given its most than intricate subtleties! Recall
that at the classical level, the analysis of constraints proceeded from
the idea\cite{Klaud1} that when starting from initial data lying within the
constraint hypersurface, namely starting from an initial physical state,
time evolution must be such that at each time increment one is projected
back onto the constraint hypersurface, even though {\sl a priori\/} it could
be that the complete dynamics generated by the total Hamiltonian could
pull the physical trajectory away from the physical subspace. Hence,
the identification of the set of constraints makes sure that physical
trajectories stay within the physical subspace. 

As commented above, this is precisely the issue which faces canonical 
quantization and quantum dynamics induced by the quantum evolution 
ope\-ra\-tor associated to the Schr\"odinger equation, which is thus given 
by the time-ordered exponential
\begin{equation}
U(t_2,t_1)=T\,e^{-\frac{i}{\hbar}\int_{t_1}^{t_2}dt\,\hat{H}_T(t)}\ \ ,\ \ 
\hat{H}_T(t)=\hat{H}+\lambda^\alpha(t)\hat{\phi}_\alpha\ .
\label{eq:gaugeevolution}
\end{equation}
{\sl A priori\/}, this operator could be such that given some initial physical
state, its time evolved product would no longer lie within the physical 
subspace. However, in order to make sure that physical states stay within
that subspace, it would suffice to project them back onto that subspace
using an appropriate physical projection operator after each time 
increment.\cite{Klaud1}

In fact, such a physical projector may readily be constructed.\cite{Klaud1} 
Since the constraints $\hat{\phi}_\alpha$ are the generators of small gauge
transformations, small finite global symmetry transformations on the space 
of quantum states are obtained from the operators
\begin{equation}
G(\theta^\alpha)=e^{-\frac{i}{\hbar}\theta^\alpha\hat{\phi}_\alpha}\ ,
\label{eq:symmetryG}
\end{equation}
$\theta^\alpha$ being the associated symmetry group parameters.
Such an operator does indeed appear for each time step in the above
total evolution operator (\ref{eq:gaugeevolution}), the parameters
then being the values $\lambda^\alpha(t)$ at that time step, which
effect an arbitrary symmetry transformation as the system evolves in time
through the first-class Hamiltonian $\hat{H}$.

Note that on account of the definition of physical states, the value
(namely either the eigenvalue or the expectation value) of the operators
$G(\theta^\alpha)$ acting on physical states is always unity, expressing
the gauge invariance of these states. In particular, the same property
shows that when considered for such states, in effect the complete
evolution operator (\ref{eq:gaugeevolution}), which otherwise also propagates
nonphysical states, reduces to the unitary operator
\begin{equation}
e^{-\frac{i}{\hbar}(t_2-t_1)\hat{H}}\ ,
\end{equation}
irrespective of the choice for the Lagrange multipliers $\lambda^\alpha(t)$.

In order to define now the physical projector, clearly it suffices to
consider all the small finite symmetry transformations $G(\theta^\alpha)$
summed over the space of all such transformations, namely
\begin{equation}
\proj=\int\left[dU(\theta^\alpha)\right]\ 
e^{-\frac{i}{\hbar}\theta^\alpha\,\hat{\phi}_\alpha}\ ,
\end{equation}
where $[dU(\theta^\alpha)]$ stands for the normalized group invariant measure
over the manifold of small finite symmetry transformations (in the case
of a compact Lie group, this is the Haar measure). Consequently, one has
\begin{equation}
\proj^2=\proj\ \ \ ,\ \ \ \proj^\dagger=\proj\ ,
\end{equation}
which are indeed the properties characteristic of a projection operator.
Given this construction, it should be clear that acting with $\proj$ on
any quantum state, all its gauge noninvariant components are averaged
out through the group integration, leaving over only its gauge invariant
physical component. Hence, $\proj$ is indeed the physical projector
of the system.\cite{Klaud1} Furthermore, whenever of application, large
gauge transformations may also be included in its construction, so that
a truly gauge invariant projector onto all physical states invariant
under small as well as large gauge transformations is achieved.

This operator may now be used to construct the physical propagator or
evolution operator of the quantized system, which thus only propagates
physical gauge invariant states, namely
\begin{equation}
\begin{array}{r c l}
U_{\rm phys}(t_2,t_1)&=&U(t_2,t_1)\proj=
\proj\,U(t_2,t_1)\proj=\proj\,e^{-\frac{i}{\hbar}(t_2-t_1)\hat{H}}\,\proj\\
 & & \\
&=&\proj\,e^{-\frac{i}{\hbar}(t_2-t_1)\proj\,\hat{H}\proj}\,\proj\ ,
\end{array}
\end{equation}
where the different expressions follow from the properties of $\proj$
as well as the fact that the operators $\hat{H}$ and $\hat{\phi}_\alpha$
commute on the physical subspace. Clearly, this physical evolution
operator obeys the usual and necessary unitary and involution properties
characteristic of such an operator generating time evolution.
Furthermore, given its very last representation, it obviously propagates
physical states only, both as external as well as intermediate states.
The physical projector together with this physical propagator thus
provide the complete answer to the issue of the genuine gauge invariant
physical quantum dynamics of a gauge invariant system. Nonetheless,
the construction is entirely set only within Dirac's quantization
framework, without any need for any gauge fixing procedure of any sort,
thereby avoiding from the outset the difficult issue of Gribov problems.
In fact, the quantum dynamical formulation is gauge invariant by
construction, and in effect amounts precisely to an dynamics over the
space of gauge orbits with an admissible covering. The physical projector
approach to the gauge invariant dynamics is necessarily void of any Gribov
problem,\cite{JG6} maintains manifest Poincar\'e covariance when present,
and avoids all the technical difficulties of functional determinants,
ghost contributions and the like following from any gauge fixing procedure.

In addition, the method also extends\cite{Klaud1} to second-class constraints, 
avoiding the difficulties often raised by the quantization of Dirac brackets,
as well as to reducible or nonregular constraints. Furthermore,
following the usual time slicing route, it is also possible to set up
path integral representations of matrix elements of the physical
evolution operator and physical observables, leading to convenient
calculational methods complementary to the quantum operator ones.

As with all the general discussion of constrained systems, the above
programme for the construction of the physical projector must be
considered and developed on a case by case basis. Its explicit definition
often requires specifications. For example,\cite{Klaud1} if the spectrum 
of a constraint,
say $\hat{\phi}$, is continuous in the neighbourhood of its zero eigenvalue,
the projector is not normalizable and corresponds to a projector density
that could be defined as follows
\begin{equation}
\proj_0=\lim_{\delta\rightarrow 0}\frac{1}{2\delta}\,
\proj[-\delta<\hat{\phi}<\delta]\ ,
\end{equation}
where the operator $\proj[-\delta<\hat{\phi}<\delta]$ stands for the
projector onto the subspace spanned by those states whose $\hat{\phi}$
eigenvalue lies within the shown interval. For instance, 
if $\hat{\phi}=\hat{q}$, $\hat{q}$ being the position operator of the
Heisenberg algebra, one has
\begin{equation}
\proj_0=\lim_{\delta\rightarrow 0}\frac{1}{2\delta}\int_{-\delta}^\delta\,
dq\,|q><q|=\lim_{\delta\rightarrow 0}\frac{1}{2\delta}
\int_{-\infty}^\infty\,d\xi\,e^{i\xi\hat{q}}\,\frac{\sin(\delta\xi)}{\pi\xi}
=\int_{-\infty}^\infty\,\frac{d\xi}{2\pi}\,e^{i\xi\hat{q}}\ ,
\end{equation}
where in the third expression the integral representation of the step
function is introduced, leading to the last expression which is indeed
nothing but the $\delta(q)$ function in operator form. Note that the
exponentiated operator $e^{i\xi\hat{q}}$ is the generator of
translations in the position, the associated group parameter $\xi$ being thus
integrated over with a normalization of the integration measure such that
$\proj_0$ is a nonnormalizable operator density.

As a particular example, the constraint $\hat{P}^2+m^2=0$ of the parametrized
relativistic scalar particle does possess a continuous spectrum including
the zero eigenvalue, so that the physical projector for that system
follows such a construction. Furthermore, since the first-class Hamiltonian
$\hat{H}=0$ for that reparametrization invariant system vanishes, the
physical evolution operator $U_{\rm physical}(\tau_f,\tau_i)$ in fact
coincides with the physical projector, hence
\begin{equation}
U_{\rm physical}(\tau_f,\tau_i)=\proj_0=\int_{-\infty}^\infty\,
\frac{d\gamma}{2\pi}e^{-\frac{1}{2}i\gamma\left(\hat{P}^2+m^2\right)}\ ,
\end{equation}
where the symmetry parameter $\gamma$ is nothing else but the
Teichm\"uller coordinate of the space of gauge orbits. Note that this
physical evolution operator is independent of the world-line $\tau$
coordinate, as it should on account of the gauge invariance
of the formulation. Furthermore, the integration over the space of gauge
orbits is indeed admissible, since each of the gauge orbits are accounted
for with an equal relative weight, hence once and only once. No gauge fixing
has been effected, but nevertheless the correct gauge invariant quantum
dynamics is readily obtained through the physical projector.

In particular, let us consider the configuration space matrix elements of 
this operator, which should thus be in direct correspondence with the
Feynman propagator for a scalar field theory. Thus,
\begin{equation}
<x^\mu_f|U_{\rm physical}(\tau_f,\tau_i)|x^\mu_i>=
\int_{(\infty)}\frac{d^Dp^\mu}{(2\pi)^D}\,e^{i\Delta x\cdot p}
\int_{-\infty}^\infty\frac{d\gamma}{2\pi}\,
e^{-\frac{1}{2}i\gamma(p^2+m^2)}\ .
\end{equation}
So far however, we have only accounted for the small gauge symmetries,
namely those that preserve the world-line orientation and are generated
by the first-class constraint $\hat{\phi}$. However, one should still
enforce gauge invariance under large gauge transformations reversing
the world-line orientation, and use the corresponding physical projector.
This projector is obtained by restricting the $\gamma$ range of integration
to the real positive axis, for the reasons having been discussed previously 
in that respect. Hence finally, the full gauge invariant spacetime propagator 
of the unoriented relativistic scalar particle is given by
\begin{equation}
\int_{(\infty)}\frac{d^Dp^\mu}{(2\pi)^D}\,e^{i\Delta x\cdot p}
\int_0^\infty\frac{d\gamma}{2\pi}\,e^{-\frac{1}{2}i\gamma(p^2+m^2)}=
\frac{1}{\pi}\int_{(\infty)}\frac{d^Dp^\mu}{(2\pi)^D}\,
e^{i\Delta x\cdot p}\,\frac{i}{p^2+m^2-i\epsilon}\ ,
\end{equation}
a result which indeed, up to the normalization factor $1/\pi$, coincides
exactly with the Feynman propagator for a scalar field with its
canonical normalization. The proper projection onto the physical subspace
of quantum states has been achieved in a most straightforward manner,
without any gauge fixing nor ghost system whatsoever.\cite{JG6}

So far, the physical projector approach has been applied to some well-known
integrable systems, as well as to some of the issues surrounding the
problems of quantum gravity.\cite{Klaud4} For example, it has been 
applied\cite{JG3,JG7} to the gauge invariant mechanical models in 0+1 
dimensions described previously, with Lagrangian
\begin{equation}
L=\frac{1}{2g^2}\left[\dot{q}^a_i+f^{abc}\lambda^Bq^c_i\right]^2-
\frac{1}{2}\omega^2\left(q^a_i\right)^2\ ,
\end{equation}
in the cases of the gauge groups SO(2) and SO(3). It could prove to be
of interest to extend the analysis to all compact Lie algebras,
as well as quartic coupling interactions as they arise from the dimensional
reduction of the matrix formulation of M-theory. Coherent state techniques
for nonabelian groups would certainly be of relevance, as well as
the general methods of dynamical integrable systems. The nonperturbative
solution of the Schwinger model (a U(1) gauge invariant field theory
coupled to a massless fermion in 1+1 dimensions) has also been recovered
through the physical projector without the necessaity of any gauge 
fixing.\cite{JG8} Likewise, the physical projector has been applied\cite{JG9} 
to the quantization of the U(1) invariant Chern-Simons theory, one of the 
simplest topological quantum field theories in which only a finite number 
of gauge invariant states, dependent only on the differential topology of 
the spacetime manifold but not its geometry, is to be projected out 
from an infinite set of quantum states.

These successes thus bode well for the relevance of this recent
approach towards the quantization of constrained dynamics. It would
certainly be a worthwhile project to extend its use to quantum field
theories in a perturbative quantization and determine what this implies for
a modification in the Feynman rules as usually derived within some gauge 
fixed framework. However and most undoubtedly, it is in the
nonperturbative realm of quantum phenomena hithertoo not fully
comprehended that this new method offers the widest prospects for original 
new results and insights into the dynamics of gauge invariant theories.

\section{Chern-Simons Quantum Field Theory}
\label{Sect6}

Even though general relativity has not been addressed to any extent so far,
let us briefly consider one of the issues surrounding a formulation of
quantum gravity using the Einstein-Hilbert action as a reference.
Presumably, some consistent definition of pure quantum gravity would
provide a specific meaning to the otherwise formal path integral
representation which effects a summation over all configurations of
the system, namely over all the possible geometrical structures associated
to a spacetime manifold of given topology and differential structure,
\begin{equation}
\int\left[{\cal D}g_{\mu\nu}\right]\,
e^{i\frac{1}{\kappa}\int d^nx\,\sqrt{g}R}\ ,
\end{equation}
where $g_{\mu\nu}$ stands for the metric tensor, $\kappa$ for a normalization
proportional to Newton's constant, and $R$ for the Riemann scalar curvature
of the con\-si\-de\-red metric. The definition of such a path integral requires
specification, if only to avoid double counting of configurations that
are diffeomorphic equivalent under spacetime reparametrizations, the local
gauge symmetry of general relativity.

In these terms, it thus appears that pure quantum gravity would be a theory 
whose physical properties are not only independent of the spacetime coordinate
system, but more importantly, independent of any spacetime geometrical
data.\cite{Wit1,Wit2} Thus, pure quantum gravity would be a system whose 
physical observables are only dependent on the topological and differentiable
spacetime structures, namely genuine diffeomorphic topological invariants.
It is by following ideas such as these that Witten has uncovered the
existence of so-called Topological Quantum Field Theories (TQFT), whose
quantum observables consist only of (diffeomorphic) topological 
invariants.\cite{Wit1,Wit2,TQFT}
Even though in their field theory formulation such systems possess an
infinite number of degrees of freedom, their actual gauge invariant
physical content is that of a finite number of quantum states in direct
relation to the diffeomorphic topological data of the manifold on which
the fields are considered. Two large classes of such models have been
identified,\cite{TQFT} namely TQFT's whose formulation requires metric data 
but whose gauge invariance is so large that their physics is independent
of the geometry nevertheless, and TQFT's whose formulation does not
require a metric structure on the base manifold. Such TQFT's are also
in direct relation to theories of the general relativistic or of the
Yang-Mills type through the character of the gauge symmetries that they 
possess. These classes of quantum field theories have grown into a topic
of great interest in mathematical physics, with applications
within fundamental physics of potential great relevance.

One such example is that of pure Chern-Simons theories on a manifold of 2+1 
dimensions.\cite{Wit2}
In fact, pure gravity in that spacetime dimension may be brought within
such a framework, the Yang-Mills symmetry being then based on a noncompact
Lie group.\cite{Wit3} The case of compact Lie groups is directly relevant 
to quite a number of fields in pure mathematic and theoretical 
physics.\cite{Wit2,Goldin} By lack
of time and space, here the full account of the application of the physical
projector quantization of the U(1) Chern-Simons theory as it was presented
at the Workshop is not reproduced, referring the interested reader
to the original publication for details.\cite{JG2bis,JG9} Only a few 
general comments will be provided.

In terms of the same notations as introduced previously for any Yang-Mills
theory, the 2+1 dimensional pure Chern-Simons action reads
\begin{equation}
S[A^a_\mu]=N_k\,\int_{\R\times\Sigma}\,dx^0dx^1dx^2\,
\epsilon^{\mu\nu\rho}\left[A^a_\mu F^a_{\nu\rho}+\frac{1}{3}f^{abc}
A^a_\mu A^b_\nu A^c_\rho\right]\ ,
\end{equation}
where the gauge coupling constant has been absorbed into the normalization
of the gauge fields $A^a_\mu$, while $N_k$ is some normalization factor
for the action, the index $k$ anticipating the fact that this quantity
needs to take a quantized value. Finally, $\epsilon^{\mu\nu\rho}$ stands
for the totally antisymmetric tensor such that $\epsilon^{012}=+1$.
Furthermore, the topology of the three-dimensional manifold on which the
fields are considered is that of the direct product of the real line $\R$,
the coordinate $x^0$ being considered as the time evolution parameter,
with a compact Riemann surface $\Sigma$ of given topology, the case of
a two-torus being the simplest and also of relevance. The reason for this
specific choice of topology is that the dynamics will be considered
from the Hamiltonian point of view. 

Note that the definition of the action is independent of any metric
structure. In particular, the index $\mu=0,1,2$ is not to be raised nor
lowered in different expressions. In addition, the action is locally gauge
invariant by construction. Under small gauge transformations, the action
only changes by a local surface term without consequence. For large gauge
transformations however, it changes by a constant shift which is
proportional to the winding number of the corresponding gauge transformation.
This is one reason why the normalization factor $N_k$ needs to be quantized
in the quantum theory, since otherwise the quantity $e^{iS}$ being
summed over in the path integral is not single-valued under large gauge
transformations,

As is generic for TQFT's, the above action is in fact dependent
only on the diffeomorphic topological invariant data of the underlying
three-dimensional manifold. This is best demonstrated from the 
Euler-Lagrange equations of motion which read
\begin{equation}
F^a_{\mu\nu}=0\ .
\end{equation}
The classical solutions are thus indeed nothing but the gauge equivalence
classes of flat gauge connections on the considered three-dimensional 
manifold. It is well known that this modular space is characterized in purely 
topological terms, namely the $G$-holonomies of all noncontractible cycles
within the manifold. Hence for the chosen topology $\R\times\Sigma$,
this set of solutions is a finite dimensional space of continuous variables. 
Pure quantum Chern-Simons theory is thus
nothing but quantum mechanics of the modular space of flat gauge connections,
indeed a genuine purely diffeomorphic topological system.\cite{Wit2} Among the
infinite set of degrees of freedom $A^a_\mu$, there remains only a finite
number of gauge invariant physical degrees of freedom. From this fact alone,
one may {\sl a priori\/} only conclude that the physical states of the 
quantized system span a discrete infinite vector space. However, as we
shall see, as a consequence of the compactness of phase space because of
large gauge transformations, the actual quantum space of gauge invariant
states itself is also finite dimensional. Quantum pure Chern-Simons theory
possesses only a finite number of quantum physical states.

Note that the above action is already in Hamiltonian form. Indeed, all
time derivatives of the fields appear linearily in the Lagrangian density.
Consequently, the configurations $A^a_\mu$ are in fact already the
phase space degrees of freedom, whose brackets are directly identified
from the action to be
\begin{equation}
\{A^a_1(x^0,\vec{x}\,),A^b_2(x^0,\vec{y}\,)\}=
\frac{1}{2N_k}\delta^{ab}\,\delta^{(2)}(\vec{x}-\vec{y}\,)\ ,
\end{equation}
while the first-class Hamiltonian vanishes identically, $H=0$, as
befits the Hamiltonian of any reparametrization invariant theory,
and the first-class constraints, generators of the small Yang-Mills
gauge transformations, are
\begin{equation}
\phi^a=-2N_k\left[\partial_1A^a_2-\partial_2A^a_1-f^{abc}A^b_1A^c_2\right]\ ,
\end{equation}
whose Lagrange multipliers are $A^a_0$. A direct calculation then confirms
the gauge algebra generated by these quantities
\begin{equation}
\{\phi^a(x^0,\vec{x}\,),\phi^b(x^0,\vec{y}\,)\}=f^{abc}\phi^c(x^0,\vec{x}\,)\,
\delta^{(2)}\left(\vec{x}-\vec{y}\,\right)\ .
\end{equation}

To make this brief discussion as simple as possible, let us now particularize
to the U(1) gauge symmetry and the two-torus topology, $\Sigma=T_2$. In such
case, given a local trivialization of the torus based on choice of basis
of the homology group of 1-cycles, a Fourier mode analysis of the periodic
fields $A_{1,2}$ is natural. One then finds that under small gauge 
transformations only the nonzero Fourier modes are varied, leaving invariant
the Fourier zero modes $A_{1,2}(x^0)$. In contradistinction, large U(1) gauge
transformations, falling themselves into U(1) homotopy classes because of
the torus topology, leave invariant the nonzero modes while the
zero modes are then shifted by integer multiplies of $2\pi$, depending
on the homotopy class of the gauge transformation. Consequently, it
follows that the actual gauge invariant phase space is nothing but the
zero mode sector taking its values on a two-torus itself, namely a compact
phase space. This is not a type of phase space encountered in usual
Hamiltonian dynamics, whose quantization requires specific methods
which have been designed for such a purpose within the framework of
so-called geometric quantization, discussed in Prof. S.T. Ali's lectures
in these Proceedings.
However, through the use of the physical projector, it remains possible
to use the straightforward methods of canonical quantization, and then
enforce invariance under large gauge transformations through the physical
projector, and thereby in effect quantize a compact phase space.

Physical phase space being compact, it is to be expected that the number of
gauge invariant physical states itself be finite, since each quantum state
occupies a specific quantum of volume of phase space. Because of the
nor\-ma\-li\-za\-tion of the above brackets in terms of the factor $N_k$, 
this fact translates in the quantization of that factor as
\begin{equation}
N_k=\frac{\hbar}{4\pi}\,k\ ,
\end{equation}
$k$ being an arbitrary integer. However, the same requirement follows
from the construction of the quantum operators generating large gauge
transformations of quantum states. Through the physical projector
which sums over both all small and all large gauge transformations,
a finite set of quantum physical states and their explicit wave
function representations is readily identified. One then recovers exactly
the states that have been constructed through an admissible
gauge fixing procedure of the same theories. Futher details may be
found in the original publication and references therein.\cite{JG2bis,JG9}

In conclusion, the physical projector is capable of identifying among an
infinite number of quantum states the finite subset of physical states
in the case of a TQFT, and this within a framework which is simply that
of Dirac's canonical quantization of constrained dynamics, without the
necessity of any gauge fixing procedure whatsoever, thereby avoiding from 
the outset the potential difficulties of Gribov problems.

This rather simple system provides a nontrivial example of the relevance
of purely topological features to the physics of gauge invariant
quantum field theories. Extended to theories of gravity along the
ideas mentioned in the introduction to this section, such a situation
gives some credence to the suggestion\cite{Wit1,Wit2,Wit3} that quantum 
topology and its generic finite number of quantum states is at the basis of 
actual pure quantum gravity,
whereas geometry, and thus in particular quantum gravity in which
geometrical concepts acquire their phy\-si\-cal meaning through some
mechanism akin to that of spontaneous symmetry breaking, enters the picture
through the introduction of matter interactions, thereby leading also
to an infinite number of quantum states. It could be that the path
towards a theory of quantum gravity coupled to all other fundamental
interactions goes first through the restricted framework of purely 
topological quantum gauge field theories followed by their coupling to the
dynamics of interacting ones.

\section{The Closed Bosonic String}
\label{Sect7}

As another illustration of the plausible great relevance of topology to
the quest for a quantum geometric framework for the unification of all
quantum interactions and particles, in the remainder of these notes we
shall briefly discuss the quantization of bosonic strings on a Minkowski
spacetime.\cite{Pol,GSW,JG10} Most of the considerations to be presented 
extend to fermionic
and superstrings as well. More specifically, among many such indications,
here we shall only address the features of so-called T-duality that
arise whenever a closed string theory is propagating within a spacetime
of which some of its spatial dimensions have been compactified into
a torus geometry. As shall be discussed, the to\-po\-lo\-gy of these spaces
is such that geometries whose radii are inversely related to one another
through some fundamental length scale become physically indistinguishable,
suggesting that in the realm of a theory of quantum gravity coupled to
matter, the pointwise concepts of differentiable continuous manifolds have 
to be extended at short-distance by some new concepts at the basis of
quantum geometry. Presumably, quantum geometry is where quantum interactions,
quantum particles and quantum topology meet in fundamental physics.

As in our discussion of the relativistic scalar particle, the choice
for the signature of the Minkowski metric is 
$\eta_{\mu\nu}={\rm diag}\,(-+...+)$, $\mu,\nu=0,1,2\cdots,D-1$,
$D$ being the spacetime dimension. The spacetime coordinates of the
string world-sheet embedded into spacetime are denoted 
$x^\mu(\tau,\sigma)=x^\mu(\xi)$, $\xi^\alpha=(\tau,\sigma)$, $\alpha=0,1$,
being the dimensionless world-sheet coordinates with $\tau$ considered
as the time evolution parameter and $\sigma$ restricted to the interval
$0\le\sigma\le\pi$. Thus in the world-sheet, $\tau$ (respectively, $\sigma$) 
is time-like (respectively, space-like). In these notes, we shall only
consider the closed bosonic string, all quantities then being periodic
in $\sigma$ with periodicity $\pi$.

For the same reasons as in the case of the relativistic scalar particle,
the action of the system must be a spacetime scalar for the Poincar\'e
group, as well as a world-sheet scalar for world-sheet reparametrizations.
In particular, the redundancy in the physical time $x^0$ and the world-sheet
time evolution parameter $\tau$ is to be resolved through the latter gauge 
invariance properties
of the system. Consequently, the most natural choice for the action
principle is to measure the total area of the string world-sheet swept out
between two initial and final configurations. This area may be measured either
in terms of the metric induced on the world-sheet by the ambiant spacetime
Minkowski metric, or else by some additional intrinsic world-sheet metric.
The first choice leads to the Nambu-Goto action with its nonlinear equations
of motion, and the second choice to the Polyakov action and its linear
equations of motion. Only the Nambu-Goto action will be discussed
presently, leaving it as an exer\-ci\-se to develop the same analysis
for the Polyakov action. In either case, one is in fact dealing, from the 
world-sheet point of view, with a two-dimensional theory of quantum 
gravity coupled to a collection of scalar fields.

\subsection{The nonlinear Nambu-Goto action}
\label{Subsect7.1}

Given the Minkowski spacetime line element $ds^2=\eta_{\mu\nu}dx^\mu dx^\nu$,
the induced world-sheet line element reads
\begin{equation}
ds^2=\gamma_{\alpha\beta}\,d\xi^\alpha\,d\xi^\beta\ \ ,\ \ 
\gamma_{\alpha\beta}=\partial_\alpha x^\mu\,\partial_\beta x^\nu\,
\eta_{\mu\nu}\ .
\end{equation}
The signature of this induced metric $\gamma_{\alpha\beta}$ being $(-+)$,
the local reparametrization invariant area element is 
$d^2\xi\,\sqrt{-{\rm det}\,\gamma_{\alpha\beta}}$. Hence, causal
string propagation is defined through the Nambu-Goto action
\begin{equation}
S[x^\mu]=\frac{-1}{2\pi\alpha'}\int_{\tau_1}^{\tau_2}d\tau\int_0^\pi
d\sigma\sqrt{-{\rm det}\,\gamma_{\alpha\beta}}=
\int_{\tau_1}^{\tau_2}d\tau\int_0^\pi\,d\sigma{\cal L}
\left(\dot{x},x'\right)\ ,
\end{equation}
with
\begin{equation}
{\cal L}(\dot{x},x')=\frac{-1}{2\pi\alpha'}
\sqrt{\left(\dot{x}\cdot x'\right)^2-\dot{x}^2{x'}^2}\ .
\end{equation}
In these expressions, a dot above a quantity stands for a derivative
with respect to $\tau$, as usual, while a prime stands for a derivative
with respect to $\sigma$. Furthermore, the coefficient $\alpha'$ has
the dimensions of a length squared, measures the string tension and
is related to the so-called Regge slope. This is the parameter which sets
the physical scale of the system, for instance the mass scale of string
oscillating modes. Since, as we shall see, the closed bosonic string
always includes a massless spin 2 mode, it would be natural to associate
the string tension $\alpha'$ to the Planck scale.

\vspace{10pt}

\noindent\underline{Remark}

\vspace{5pt}

Introducing an intrinsic world-sheet metric $g_{\alpha\beta}$ of
signature $(-+)$, the Polyakov action reads
\begin{equation}
S[x^\mu,g_{\alpha\beta}]=\frac{-1}{4\pi\alpha'}
\int_{\tau_1}^{\tau_2}d\tau\int_0^\pi d\sigma\sqrt{-{\rm det}\,
g_{\alpha\beta}}\,g^{\alpha\beta}\,\partial_\alpha x^\mu\partial_\beta x^\nu
\eta_{\mu\nu}\ .
\end{equation}
Note that the usual Einstein-Hilbert action, proportional to
\begin{equation}
\frac{1}{4\pi}\int d^2\xi\,\sqrt{-g}\,R^{(2)}\ ,
\end{equation}
is not included, since this latter contribution measures the Euler 
characteristic of the world-sheet, hence is a surface term that does not
affect the equations of motion. Furthermore, a world-sheet cosmological
term is not included either,
\begin{equation}
\frac{-1}{4\pi\alpha'}\int d^2\xi\,\sqrt{-g}\,\mu^2\ ,
\end{equation}
since at the classical level there exists a solution to the associated
equations of motion provided only $\mu^2=0$. Hence, this two-dimensional 
theory of gravity is in fact Weyl invariant in the absence of such a term, 
namely invariant under local changes in the scale factor of the metric,
$g_{\alpha\beta}(\xi)\rightarrow e^{\chi(\xi)}\,g_{\alpha\beta}(\xi)$.
This is the symmetry which at the quantum level restricts
the spacetime dimension to the critical value $D=26$, when the Polyakov action
is quantized in a manifestly reparametrization invariant 
manner.\cite{Pol,GSW,JG10,JG11}

Finally, note how describing the propagation of the bosonic string
in a curved spacetime of background metric $G_{\mu\nu}(x^\mu)$ is
readily achieved, by direct substitution of $\eta_{\mu\nu}$ in the above
expressions. The Polyakov action is then the natural starting point
for a study of the low energy effective field theory description of
strings coupled to background fields.\cite{Pol,GSW}

\vspace{10pt}

Let us restrict to the Nambu-Goto action. Since from the world-sheet
point of view, one is dealing with a local field theory possessing as
internal symmetry the Poincar\'e group of Minkowski spacetime,
it follows from Noether's theorem that there exist currents and charges
associated to spacetime translations and rotations which are conserved
for the classical con\-fi\-gu\-ra\-tions solving the equations of motion.
The energy-momentum Noether current and charge are
\begin{equation}
P^\alpha_\mu=\frac{\partial{\cal L}}{\partial(\partial_\alpha x^\mu)}\ \ ,\ \ 
P_\mu=\int_0^\pi d\sigma\,P^{\alpha=0}_\mu\ ,
\end{equation}
with
\begin{equation}
\begin{array}{r c l}
P^0_\mu&=&\frac{-1}{2\pi\alpha'}\,
\frac{1}{\sqrt{(\dot{x}\cdot x')^2-\dot{x}^2 {x'}^2}}\,
\left[(\dot{x}\cdot x')x'_\mu-{x'}^2\dot{x}_\mu\right]\ ,\\
 & & \\
P^1_\mu&=&\frac{-1}{2\pi\alpha'}\,
\frac{1}{\sqrt{(\dot{x}\cdot x')^2-\dot{x}^2 {x'}^2}}\,
\left[(\dot{x}\cdot x')\dot{x}_\mu-\dot{x}^2x'_\mu\right]\ .
\end{array}
\end{equation}
The angular-momentum current and charge are
\begin{equation}
M^\alpha_{\mu\nu}=P^\alpha_\mu x_\nu - P^\alpha_\nu x_\mu\ \ ,\ \ 
M_{\mu\nu}=\int_0^\pi d\sigma M^{\alpha=0}_{\mu\nu}\ .
\end{equation}
Any solution to the equations of motion is thus such that
\begin{equation}
\partial_\alpha P^\alpha_\mu=0\ \ ,\ \ 
\partial_\alpha M^\alpha_{\mu\nu}=0\ \ ,\ \ 
\frac{dP_\mu}{d\tau}=0\ \ ,\ \ 
\frac{dM_{\mu\nu}}{d\tau}=0\ .
\end{equation}

Given the fact that the Lagrangian density is dependent only on the
$\xi^\alpha$ derivatives of the coordinates $x^\mu(\xi)$ (because of
the required invariance under spacetime translations), the Euler-Lagrange
equations of motion are nothing but the conservation equations for the
energy-momentum Noether currents,
\begin{equation}
\partial_\alpha P^\alpha_\mu=0\ .
\end{equation}
These equations are to be accompanied by boundary conditions. One set of
boundary conditions specifies the initial and final string configurations,
while the boundary conditions in $\sigma$ (which are required since the
two-dimensional world-sheet is compact, in contradistinction with
ordinary field theories which are required to vanish at infinity) are
nothing but the periodicity requirement in $\sigma\rightarrow \sigma+\pi$.
Hence,
\begin{equation}
x^\mu(\tau_{1,2},\sigma)=x^\mu_{1,2}(\sigma)\ \ ,\ \ 
x^\mu(\tau,\sigma+\pi)=x^\mu(\tau,\sigma)\ .
\end{equation}
It is at this point that other choices of boundary conditions in $\sigma$
are possible, such as for open strings, or more generally for strings
ending of fixed branes leading to so-called $Dp$-branes.\cite{Pol} 
None of these shall be addressed here.

It thus appears that this set of equations of motion is highly nonlinear,
and thus difficult to solve explicitly. In addition, all the above
quantities are not independent, but in fact obey the following constraints,
\begin{equation}
\left[P^0_\mu\pm\frac{\partial_\sigma x_\mu}{2\pi\alpha'}\right]^2=0\ \ ,\ \ 
\left[P^1_\mu\pm\frac{\partial_\tau x_\mu}{2\pi\alpha'}\right]^2=0\ ,
\end{equation}
as may be checked by direct calculation. In fact, since $P^0_\mu$ is
nothing else than the momentum conjugate to $x^\mu$, the first pair
of constraints are primary constraints for the Hamiltonian formulation,
and are thus expected to be the generators for small world-sheet
reparametrizations as is indeed confirmed by the explicit Hamiltonian
analysis. On the other hand, the second pair of constraints is related 
to the first through large world-sheet reparametrizations which exchange
the $\tau$ and $\sigma$ coordinates while preserving the world-sheet
orientation, such as
\begin{equation}
\tilde{\tau}=\tau_1+\frac{\tau_2-\tau_1}{\pi}\sigma\ \ ,\ \ 
\tilde{\sigma}=\frac{\pi}{\tau_2-\tau_1}(\tau_2-\tau)\ .
\end{equation}
In the same way as for the relativistic scalar particle, it is thus possible
to consider oriented strings theories which are required to be invariant
under both small and large orientation preserving 
world-sheet reparametrizations, and unoriented strings invariant under
both orientation preserving and reversing world-sheet diffeomorphisms.

Since by construction the system is invariant under small world-sheet
reparametrizations, namely a small local gauge symmetry, it follows that
the general solution to the equations of motion involves arbitrary
functions of $\xi^\alpha$, in fact two such functions related to the
two independent coordinates $\xi^\alpha=(\tau,\sigma)$. In order
to construct explicit solutions, it is thus necessary to first fix
this large gauge freedom, and then afterwards eventually restore it
if necessary.

\subsection{Conformal gauge fixing}
\label{Subsect7.2}

In order to specify some gauge fixing of the system, let us consider
the quantities $\partial_\tau x^\mu$ and $\partial_\sigma x^\mu$.
Clearly, they define the vectors tangent to the world-sheet associated
to the $(\tau,\sigma)$ coordinate system set-up on that manifold.
Hence, it is always possible, by an appropriate local change of coordinates,
to bring these two tangent vectors to be locally perpendicular
with respect to the spacetime Minkowski metric, namely $\gamma_{01}=0$,
and then by a local rescaling of each of the coordinates $\tau$ and $\sigma$
to set the Lorentz invariant length of each of these vectors to be identical 
up to their sign since one is time-like and the other space-like, namely
$\gamma_{00}+\gamma_{11}=0$. In terms of the coordinates, we thus have
\begin{equation}
\dot{x}\cdot x'=0\ \ ,\ \ \dot{x}^2+{x'}^2=0\ \ 
\Longleftrightarrow \left(\dot{x}\pm{x}'\right)^2=0\ .
\label{eq:conformalgaugefixing}
\end{equation}
However, given this geometrical description of this choice of coordinates,
there still remains quite some large gauge freedom, namely choices of
coordinates $\xi^\alpha$ which locally amount to a local change of scale
for the induced metric $\gamma_{\alpha\beta}$,
while the orthogonality conditions are preserved. In the case of an
euclidean signature metric, such transformations in two dimensions
correspond to conformal or analytic transformations. Hence in the
present content, the above choice of gauge fixing is also called conformal 
gauge fixing, even though the term pseudo-conformal would be more fitting
given the Minkowski si\-gna\-tu\-re $(-+)$ on the world-sheet. Note however
that the conformal gauge is yet not a complete gauge fixing, since conformal
transformations inducing a local change of scale in the induced metric
$\gamma_{\alpha\beta}$ are still possible. This gauge redundancy will
be resolved in the light-cone gauge to be discussed in the next section.

Once the conformal gauge fixing (\ref{eq:conformalgaugefixing}) effected,
the equations of motion become linear, since one finds
\begin{equation}
P^0_\mu=\frac{\dot{x}_\mu}{2\pi\alpha'}\ \ \ ,\ \ \ 
P^1_\mu=-\frac{x'_\mu}{2\pi\alpha'}\ ,
\end{equation}
thus leading to the Klein-Gordon equations of $D$ free massless scalar 
fields in two dimensions,
\begin{equation}
\left[\partial^2_\tau-\partial^2_\sigma\right]\,x^\mu(\tau,\sigma)=0\ .
\end{equation}
However, these equations still have to be supplemented with the
gauge fixing conditions $(\dot{x}\pm x')^2=0$. Consequently, the general
solution is always of the form
\begin{equation}
x^\mu(\tau,\sigma)=x^\mu_L(\tau+\sigma)+x^\mu_R(\tau-\sigma)\ ,
\end{equation}
namely a linear superposition of left- and right-moving modes
propagating along the two disjoint branches of the
two-dimensional light-cone and noninteracting
with one another. This decoupling of massless chiral modes in two dimensions
is generic to all closed string theories, and put to great advantage in
the construction of string theories in different spacetime 
dimensions.\cite{Pol,GSW}

A simple Fourier mode analysis of the system then readily leads to the
following general solution in the conformal gauge,
\begin{equation}
\begin{array}{r l}
x^\mu(\tau,\sigma)=\sqrt{2\alpha'}\left[\right.&
q^\mu+\alpha^\mu_0(\tau-\sigma)+\bar{\alpha}^\mu_0(\tau+\sigma)\\
 & \\
 &\left. + \frac{1}{2}i{\sum_n}'\frac{1}{n}\left(\alpha^\mu_n 
e^{-2in(\tau-\sigma)}+\bar{\alpha}^\mu_ne^{-2in(\tau+\sigma)}\right)\right]\ ,
\end{array}
\end{equation}
where the summation runs over all positive and negative integers $n$
except for $n=0$ as indicated by the prime on the summation symbol, while
the different constant factors are integration constants such that
\begin{equation}
{\alpha^\mu_n}^*=\alpha^\mu_{-n}\ \ ,\ \ 
{\bar{\alpha}^{\mu*}_n}=\bar{\alpha}^\mu_{-n}\ \ ,
\alpha^\mu_0=\frac{1}{2}\sqrt{2\alpha'}P^\mu=\bar{\alpha}^\mu_0\ .
\end{equation}
Furthermore, the conformal gauge fixing conditions translate into the
constraints
\begin{equation}
L_n=0\ \ ,\ \ \bar{L}_n=0\ ,
\end{equation}
where
\begin{equation}
L_n=\frac{1}{2}\sum_m\alpha^\mu_{n-m}\alpha_{m\mu}\ \ ,\ \ 
\bar{L}_n=\frac{1}{2}\sum_m\bar{\alpha}^\mu_{n-m}\bar{\alpha}_{m\mu}\ .
\end{equation}
In particular, the zero mode contraints $L_0=0=\bar{L}_0$ are equivalent to
\begin{equation}
\frac{1}{2}\alpha' M^2=N+\bar{N}\ \ ,\ \ N=\bar{N}\ ,
\end{equation}
with the excitation level number quantities
\begin{equation}
N=\sum_{n=1}^\infty \alpha^\mu_{-n}\alpha_{n\mu}\ \ ,\ \ 
\bar{N}=\sum_{n=1}^\infty\bar{\alpha}^\mu_{-n}\bar{\alpha}_{n\mu}\ .
\end{equation}
As we shall see later on, the quantities $L_n$ and $\bar{L}_n$ are nothing 
but the generators of the remaining reparametrization invariance, namely 
conformal symmetry, in the conformal gauge, known as the Virasoro generators. 
In particular, the sum of the Virasoro zero modes $L_0+\bar{L}_0=0$ leads 
to the mass spectrum of the solutions, while their difference, 
$L_0-\bar{L}_0=0$, expresses the invariance of the system under constant 
shifts in the $\sigma$ coordinate, $\sigma\rightarrow\sigma+\sigma_0$, since 
the string is closed in that direction of the world-sheet.

Note that from the expressions for $N$ and $\bar{N}$, it is not obvious
that all oscillating solutions to the equations of motion have a positive 
definite mass, given the semi-definite signature of the Minkowski metric.
This issue may be resolved only by solving all these Virasoro constraints,
or equivalently, by completely fixing the conformal gauge freedom remaining
in the conformal gauge. This is the purpose of the next section.

The fact that the Virasoro constraints are related to conformal 
transformations is readily established. As indicated previously,
the constraints $[P^0_\mu\pm\dot{x}_\mu/2\pi\alpha']^2=0$ are the generators
of world-sheet reparametrizations. In the conformal gauge, their expression
is proportional to $(\dot{x}\pm x')^2=0$, namely the gauge fixing conditions
themselves, which in terms of their Fourier modes in $\sigma$ coincide
with the Virasoro quantities $L_n=0=\bar{L}_n$. Hence, the reason
why we still have the Virasoro constraints to enforce in the conformal
gauge is that this gauge fixing is not yet complete. Indeed,
(pseudo)conformal reparametrizations $\tilde{\xi}=\tilde{\xi}(\xi)$ such that,
\begin{equation}
\frac{\partial\tilde{\tau}}{\partial\tau}=
\frac{\partial\tilde{\sigma}}{\partial\sigma}\ \ ,\ \ 
\frac{\partial\tilde{\tau}}{\partial\sigma}=
\frac{\partial\tilde{\sigma}}{\partial\tau}\ ,
\end{equation}
leave the conformal gauge fixing conditions invariant, by inducing only
a local rescaling of the induced metric, as may easily be checked. Note that
these relations also imply that
\begin{equation}
\left[\partial^2_\tau-\partial^2_\sigma\right]\tilde{\tau}=0\ \ ,\ \ 
\left[\partial^2_\tau-\partial^2_\sigma\right]\tilde{\sigma}=0\ ,
\end{equation}
namely once again the massless Klein-Gordon equations.

\subsection{Light-cone gauge fixing}
\label{Subsect7.3}

A complete gauge fixing of the system requires some further condition
in addition to the conformal ones, $(\dot{x}\pm x')^2=0$. Since conformal
transformations also obey the free massless Klein-Gordon equation on the
world-sheet, as has just been established, a complete gauge fixing would
be achieved by setting some linear combination of the spacetime coordinates
$x^\mu$ equal to some combination of the world-sheet coordinates 
$(\tau,\sigma)$. However, since $\sigma$ is free to be shifted by an arbitrary
amount while $x^\mu$ is then invariant, only a combination involving $\tau$
may be envisaged.

In order to give a geometrical interpretation to such a gauge fixing,
let us consider a specific constant spacetime vector $n^\mu$. In terms
of this vector, the condition that the combination $n_\mu x^\mu$ takes
a constant value determines a specific hyperplane of dimension $(D-1)$
in spacetime, perpendicular to the direction of $n^\mu$. This hyperplane
should intersect the string world-sheet, an occurrence that may be associated
to a specific value of $\tau$ as a function of the value for the constant
$n_\mu x^\mu$. Hence, let us consider the additional gauge fixing condition
\begin{equation}
n_\mu x^\mu(\tau,\sigma)=2\alpha' n_\mu P^\mu\,\tau\ ,
\label{eq:gaugefixinglc}
\end{equation}
where the coefficient in the r.h.s. multiplying $\tau$ is identified
from the expression for the total energy-momentum $P^\mu$.
A further constant term $\sqrt{2\alpha'}n_\mu q^\mu$ could also be added
to the r.h.s. of this condition, but may always be reabsorbed into
a redefinition of the value $\tau=0$ by a constant shift, which is
a local world-sheet symmetry. That this additional condition indeed leads
to a complete gauge fixing in combination with the conformal gauge fixing
conditions is readily established as follows.

For a given value of $\tau$, the condition (\ref{eq:gaugefixinglc})
identifies a specific curve lying within the world-sheet as being the
line of intersection of the world-sheet with the hyperplane defined
by (\ref{eq:gaugefixinglc}). Since this curve is associated to a constant
value for $\tau$, the curve is parametrized in $\sigma$ in a certain manner. 
However, as the value for $\tau$ changes, this line of intersection also 
changes accordingly, specifying the parametrization in $\tau$ of the 
world-sheet. Given now the conformal gauge fixing conditions 
$(\dot{x}+x')^2=0$, the parametrization in $\sigma$ for each of the lines
of intersection is then also uniquely specified, thereby singling out from
among the whole set of conformal reparametrizations 
$(\tilde{\tau},\tilde{\sigma})$ the unique world-sheet parametrization
for which all three gauge fixing conditions are met.

Clearly, such a gauge fixing is no longer manifestly spacetime Poincar\'e
invariant, given the role of the constant vector $n^\mu$. Hence, only the
little group, namely the subgroup of the Lorentz group leaving this
constant vector in\-va\-riant, is still a manifest symmetry of the formulation
of the system. Among all possible choices for the vector $n^\mu$, it proves
convenient to work with a light-like one which, by an appropriate space
rotation may always be taken to be
\begin{equation}
n^\mu=\frac{1}{\sqrt{2}}\,(1,0,\cdots,-1)\ \ \ ,\ \ \ n^2=0\ .
\end{equation}
In such a case, the little group is isomorphic to the euclidean group
E($D-2$)$_n$, with as subgroup the set SO($D-2$)$_n$ of all rotations in the
space directions perpendicular to the light-like vector $n^\mu$. Properties
of physical observables under this latter symmetry group are readily 
identified from the space indices
\begin{equation}
i=1,2,\cdots,D-2\ ,
\end{equation}
carried by diverse quantities. This manifest symmetry will suffice for
our purposes.

Hence within this gauge fixing known as the light-cone gauge fixing
of string theory, it proves useful to introduce the following notations
for any two spacetime vectors $u^\mu$ and $v^\mu$,
\begin{equation}
u^\pm=\frac{1}{\sqrt{2}}\left(u^0\pm u^{D-1}\right)\ ,\ 
u^i\ ,\ i=1,2,\cdots,D-2\ ,\ 
u\cdot v=-u^+v^--u^-v^++u^iv^i\ ,
\end{equation}
so that the light-cone gauge fixing conditions now read
\begin{equation}
(\dot{x}\pm x')^2=0\ \ \ ,\ \ \ x^+=2\alpha' P^+\,\tau\ .
\end{equation}
When written out, these conditions imply that the actual physical degrees
of freedom are the transverse string coordinates $x^i(\tau,\sigma)$ as well
as the zero modes $q^-$ and $P^+$, while all other degrees of freedom,
namely $x^\pm(\tau,\sigma)$ (except for the previous two zero modes)
are gauge degrees of freedom expressed in terms of the physical ones.

When considered in terms of the explicit solutions to the equations of
motion, the set of physical modes is thus
\begin{equation}
q^i\ ,\ P^i\ ;\ \alpha^i_{n\ne 0}\ ,\ \bar{\alpha}^i_{n\ne 0}\ ;\
q^-\ ,\ P^+\ ,
\end{equation}
with the mode expansion
\begin{equation}
\begin{array}{r l}
x^i(\tau,\sigma)=\sqrt{2\alpha'}\left[\right.&q^i+
\left(\alpha^i_0+\bar{\alpha}^i_0\right)\tau\\
 & \\
&\left. +\frac{1}{2}i{\sum_n}'\frac{1}{n}
\left(\alpha^i_ne^{-2in(\tau-\sigma)}+
\bar{\alpha}^i_ne^{-2in(\tau+\sigma)}\right)\right]\ ,
\end{array}
\end{equation}
and the relations
\begin{equation}
\alpha^i_0=\frac{1}{2}\sqrt{2\alpha'}P^i=\bar{\alpha}^i_0\ .
\end{equation}
As to the other two longitudinal components $x^\pm(\tau,\sigma)$, when
expressed in similar mode expansions, one has the relations
\begin{equation}
q^+=0\ ,\
\alpha^+_n=\frac{1}{2}\sqrt{2\alpha'}P^+\delta_{n,0}=\bar{\alpha}^+_n\ ;
\alpha^-_n=\frac{2}{\sqrt{2\alpha'}P^+}\,L^\perp_n\ ,\
\bar{\alpha}^-_n=\frac{2}{\sqrt{2\alpha'}P^+}\,\bar{L}^\perp_n\ ,
\end{equation}
with the transverse Virasoro generators
\begin{equation}
L^\perp_n=\frac{1}{2}\sum_m\alpha^i_{n-m}\alpha^i_m\ \ ,\ \ 
\bar{L}^\perp_n=\frac{1}{2}\sum_m\bar{\alpha}^i_{n-m}\bar{\alpha}^i_m\ ,
\end{equation}
$q^-$ and $P^+$ being the only two other independent and physical degrees
of freedom. 

Note that the above expressions do indeed solve the Virasoro constraints
$L_n=0=\bar{L}_n$ of the conformal gauge, thus demonstrating that gauge
fixing has been completed. Furthermore in this form, it appears now
obvious that the mass spectrum is indeed positive definite, since one
readily determines, from the relation $M^2=-P^2=2P^+P^--P^iP^i$,
\begin{equation}
\frac{1}{2}\alpha' M^2=N^\perp+\bar{N}^\perp\ \ \ ,\ \ \ 
N^\perp=\bar{N}^\perp\ ,
\end{equation}
with of course
\begin{equation}
N^\perp=\sum_{n=1}^\infty\,\alpha^i_{-n}\alpha^i_n\ \ ,\ \ 
\bar{N}^\perp=\sum_{n=1}^\infty\,\bar{\alpha}^i_{-n}\bar{\alpha}^i_n\ ,
\end{equation}
and in which the level matching condition is again the expression of the
invariance of the closed string dynamics and spectrum under constant
shifts in the $\sigma$ coordinate, $\sigma\rightarrow\sigma+\sigma_0$.

\vspace{10pt}

It would now be possible to work out the Hamiltonian formulation of the
system, first in the general setting in which all constraints are
identified and classified in terms of their first- or second-class character,
and then following either the conformal or the light-cone gauge fixings.
The details of such an analysis are left as a useful exercise for the
interested reader.\cite{JG10,JG11} Let it suffice to say here that conjugate 
momenta are nothing but the $\alpha=0$ components of the Noether 
energy-momentum current,
\begin{equation}
x^\mu(\tau,\sigma)\ \ \ \ ,\ \ \ \ 
\pi_\mu(\tau,\sigma)=\frac{\partial{\cal L}}{\partial\dot{x}^\mu(\tau,\sigma)}
=P^0_\mu(\tau,\sigma)\ ,
\end{equation}
while their canonical brackets
\begin{equation}
\{x^\mu(\tau,\sigma),\pi_\nu(\tau,\sigma')\}=\delta^\mu_\nu\,
\delta(\sigma-\sigma')\ ,
\end{equation}
translate into the following brackets for the modes defining the solutions
to the equations of motion in the conformal gauge
\begin{equation}
\{\sqrt{2\alpha'}q^\mu,P^\nu\}=\eta^{\mu\nu}\ \ ,\ \ 
\{\alpha^\mu_n,\alpha^\nu_m\}=-in\eta^{\mu\nu}\delta_{n+m,0}=
\{\bar{\alpha}^\mu_n,\bar{\alpha}^\nu_m\}\ .
\end{equation}
Furthermore, the system of constraints reduces to the two first-class 
primary constraints
\begin{equation}
\phi_\pm=\frac{1}{2}\pi\alpha'
\left[\pi^\mu\pm\frac{\partial_\sigma x^\mu}{2\pi\alpha'}\right]^2\ ,
\end{equation}
which are the generators of small world-sheet reparemetrizations,
while the first-class Hamiltonian density vanishes identically, ${\cal H}=0$,
thus implying that the total Hamiltonian of the system is given by
\begin{equation}
\int_0^\pi\,d\sigma\left[\lambda^+\phi_++\lambda^-\phi_-\right]\ ,
\end{equation}
$\lambda^\pm(\tau,\sigma)$ being the associated Lagrange
multipliers. The conformal gauge corresponds to the choice 
$\lambda^+=1=\lambda^-$, in which the Hamiltonian equations of motion
become equivalent to those of free massless scalar fields $x^\mu(\tau,\sigma)$
on the world-sheet, hence leading back to the solutions given previously,
as well as the mode brackets listed above. In terms of these quantities, in 
then also follows that the Virasoro generators obey the algebra,
\begin{equation}
\{L_n,L_m\}=-i(n-m)L_{n+m}\ \ \ ,\ \ \ 
\{\bar{L}_n,\bar{L}_m\}=-i(n-m)\bar{L}_{n-m}\ ,
\end{equation}
which is indeed that of the conformal algebra in two dimensions.
In particular, time translations are generated by the total Hamiltonian
in the conformal gauge,
\begin{equation}
H=2[L_0+\bar{L}_0]\ ,
\end{equation}
while the generator for constant translations in $\sigma$ is $L_0-\bar{L}_0$,
each of these quantities thus being in direct correspondence with the
expressions for the mass spectrum $\alpha'M^2/2$ and the level matching
conditions $N=\bar{N}$.

Finally, the light-gauge is obtained through Faddeev's reduced phase
space approach by introducing two further gauge fixing conditions,
namely $x^+=2\alpha' P^+\tau$ and $\pi^+=P^+/\pi$, and determining then
the corresponding Dirac brackets. The above representation of the system,
with in particular its physical mode degrees of freedom, is then readily
recovered, as the interested reader may easily verify as a useful exercise
of his understanding of constrained dynamics.

\subsection{Dirac's conformal quantization}
\label{Subsect7.4}

\vspace{5pt}

\noindent\underline{Fundamental operator algebra}

\vspace{10pt}

Given the Hamiltonian formulation of the system within the conformal
gauge, its Dirac quantization is defined by the set of basic commutation
relations for its mode degrees of freedom,
\begin{equation}
[\sqrt{2\alpha'}q^\mu,P^\nu]=i\eta^{\mu\nu}\ ,\ {q^\mu}^\dagger=q^\mu\ ,\
{P^\mu}^\dagger=P^\mu\ ,
\end{equation}
\begin{equation}
[\alpha^\mu_n,\alpha^\nu_m]=n\eta^{\mu\nu}\delta_{n+m,0}=
[\bar{\alpha}^\mu_n,\bar{\alpha}^\nu_m]\ ,\ 
{\alpha^\mu_n}^\dagger=\alpha^\mu_{-n}\ ,\
{\bar{\alpha}^{\mu\dagger}_n}=\bar{\alpha}^\mu_{-n}\ .
\end{equation}
The zero mode algebra is nothing but the Heisenberg algebra, for which,
given a particle interpretation to be associated to the string spectrum,
one chooses a momentum eigenstate basis representation, hence
\begin{equation}
P^\mu\,|p^\mu>=\,p^\mu\,|p^\mu>\ \ ,\ \ 
<p|p'>=\delta^{(D)}(p-p')\ .
\end{equation}
The nonzero mode algebra is the tensor product over all positive integers
$n\ge 1$ of Fock space algebras, for which we shall use the Fock space
representation with vacuum $|\Omega>$ annihilated by all operators 
$\alpha^\mu_n$ and $\bar{\alpha}^\mu_n$,
\begin{equation}
\alpha^\mu_n\,|\Omega>=0\ \ ,\ \ 
\bar{\alpha}^\mu_n\,|\Omega>=0\ ,
\end{equation}
the negative moded operators $\alpha^\mu_{-n}$ and $\bar{\alpha}^\mu_{-n}$,
$n\ge 1$, being the creation ope\-ra\-tors. Hence, the basis of all quantum
states is spanned by the Fock vacua $|\Omega;p>$ as well as all their
Fock excitations. In particular, note that because of the appearence of
the Minkowski metric in the nonzero mode Fock space algebras, the space
of quantum states includes negative norm states, such as
$\alpha^\mu_{-n}|\Omega;p>$, $n\ge 1$, whose norm is proportional to
$\eta^{\mu\nu}$ up to the function $\delta^{(D)}(p-p')$. A consistent
causal and quantum unitary theory requires however that no negative norm
state contributes to physical amplitudes. One may only hope that this issue
of negative norm states and physical consistency is to be resolved through
the existence of the gauge symmetries in world-sheet reparametrizations.

\vspace{10pt}

\noindent\underline{Physical states}

\vspace{10pt}

At the classical level, physical configurations in the conformal gauge
are identified through the Virasoro constraints. At the quantum level,
these quantities being composite, their actual definition requires a choice
of operator ordering for which one chooses of course the usual normal
ordering for Fock space creation and annihilation ope\-ra\-tors, inclusive
now of the zero mode ope\-ra\-tors with the $q^\mu$ operators always brought
to the left of all $P^\mu$ operators. Given this choice of normal ordering,
the Virasoro operators read
\begin{equation}
L_n=\frac{1}{2}\sum_m\,:\alpha^\mu_{n-m}\alpha_{m\mu}:\ \ ,\ \ 
\bar{L}_n=\frac{1}{2}\sum_m\,:\bar{\alpha}^\mu_{n-m}\bar{\alpha}_{m\mu}:\ .
\end{equation}
Clearly, normal ordering only affects the zero mode Virasoro operators
$L_0$ and $\bar{L}_0$, hence the physical spectrum of the theory.

In order to define physical states, it turns out that requiring all the
Virasoro operators to vanish is too strong a restriction, whereas it
suffices to only require that all positive moded operators actually vanish.
Indeed, such a condition is tantamount to requiring the Virasoro constraints
in a weak sense, namely that the matrix elements of all Virasoro operators
vanish for physical external states, which is also sufficient from the quantum
Virasoro algebra point of view to be discussed hereafter. Hence, gauge
invariant physical states of the closed bosonic string are defined by the
following set of operator constraints
\begin{equation}
L_n\,|\psi_{\rm phys}>=0\ \ ,\ \ 
\bar{L}_n\,|\psi_{\rm phys}>=0\ \ ,\ \ n\ge 1\ ,
\end{equation}
as well as
\begin{equation}
\left[L_0+\bar{L}_0-2a\right]\,|\psi_{\rm phys}>=0\ \ ,\ \ 
\left[L_0-\bar{L}_0\right]\,|\psi_{\rm phys}>=0\ ,
\end{equation}
where the constant $a$ stands for the unknown normal ordering constant
that arises for the Virasoro zero modes, and which must be identical for
the left- and right-moving sectors of the theory.. Note that these two
zero mode conditions are equivalently expressed as
\begin{equation}
\frac{1}{2}\alpha' M^2=N+\bar{N}-2a\ \ ,\ \ N=\bar{N}\ \ ,
\end{equation}
with the excitation level number operators
\begin{equation}
N=\sum_{n=1}^\infty\,\alpha^\mu_{-n}\alpha_{n\mu}\ \ ,\ \ 
\bar{N}=\sum_{n=1}^\infty\,\bar{\alpha}^\mu_{-n}\bar{\alpha}_{n\mu}\ .
\end{equation}
Thus in particular, if the normal ordering subtraction constant $a$
happens to be strictly positive, the physical spectrum will include tachyonic
states, beginning with the physical Fock vacuum. 

\vspace{10pt}

\noindent\underline{Poincar\'e and conformal algebras}

\vspace{10pt}

In order to allow a consistent interpretation of quantum string excitations
in terms of relativistic quantum particle states, it is necessary that 
the Poincar\'e algebra be realized on the space of states, and in particular
on the subspace of physical states. Thus, one needs to explicitly check
whether the Poincar\'e algebra is recovered for the quantum Noether charges
$P^\mu$ and $M^{\mu\nu}$, a fact which is readily established. Furthermore,
since it is straightforward to verify that the Poincar\'e generators commute
with all Virasoro operators, the Poincar\'e algebra is also obtained
for the subspace of gauge invariant physical states. Hence, at each
excitation level $N=\bar{N}$, all physical quantum states span specific
irreducible Poincar\'e representations of definite mass $M^2$ and ``spin"
values, the latter being characterized in terms of the spatial rotation
subgroup of the corresponding little group which is SO($D-2$) for a massless
particle and SO($D-1$) for a massive particle.

As far as the conformal algebra is concerned, it follows from normal
ordering in the Virasoro zero modes that the conformal algebra acquires
a conformal anomaly or central extension, hence leading to the Virasoro
algebra with central charge $c=D$,
\begin{equation}
\begin{array}{r c l}
\left[L_n,L_m\right]&=&(n-m)L_{n+m}+\frac{1}{12}D(n^3-n)\delta_{n+m,0}\ ,\\
 & & \\
\left[\bar{L}_n,\bar{L}_m\right]&=&(n-m)\bar{L}_{n+m}
+\frac{1}{12}D(n^3-n)\delta_{n+m,0}\ .
\end{array}
\end{equation}
Note that because of these algebraic relations, when solving for the
physical constraints $L_{n\ge 1}=0=\bar{L}_{n\ge 1}$, it suffices to
solve only for the modes $n=1,2$, since all other modes may then be recovered
through these commutations relations.

\vspace{10pt}

\noindent\underline{The no-ghost theorem}

\vspace{10pt}

Finally, we have to address the issue of the possibility of negative norm
physical states, namely the fact that among all quantum states which obey
the physical Virasoro constraints, there may remain some states of 
strictly negative 
norm, spelling disaster for the physical consistency of these theories.
The no-ghost theorem\cite{Pol,GSW,JG10} establishes that, at tree level, 
the absence of any physical state of negative norm requires that
\begin{itemize}
\item[.] $a\le 1$;
\item[.] if $a=1$: $D\le 26$;
\item[.] if $a<1$: $D< 26$.
\end{itemize}
Even though establishing this general result is not straightforward, one
may explicitly check that such conditions are indeed necessary by solving
the phy\-si\-cal state conditions for the first few excitations levels, which
in itself is also a worthwhile exercise.

Furthermore, when one then considers one-loop corrections to quantum
string amplitudes, one quickly comes to realize that quantum unitarity also
requires\cite{Pol,GSW} the exact value $D=26$, hence also $a=1$ given 
the above statement
of the no-ghost theorem. More specifically, when these two conditions
$D=26$ and $a=1$ are met, any physical state is given by the sum of a strictly
positive norm physical states, as well as a zero norm physical state,
which itself then decouples from any physical amplitude either as an
external or as an intermediate state. Consequently, normal ordering of
operators and quantum unitarity of the manifestly Poincar\'e convariant
conformal gauge quantization of bosonic strings requires the critical
spacetime dimension $D=26$ for a physically consistent interpretation
of string excitations as being relativistic quantum particle states of
definite mass and spin.

Note that the charaterization of physical states in terms of components
of strictly positive and vanishing norm is also that which arises within
the Gupta-Bleuler quantization of quantum electrodynamics. In that case,
Gauss' law (the first-class constraint generating the local internal U(1)
gauge symmetry) is imposed for positive moded components of the gauge field,
with the consequence that physical quantum photon states are given by
the superposition of a strictly positive norm component corresponding to a
transverse photon polarization state, and a zero norm component corresponding
to a longitudinal photon polarization. The above characterization in the
string case is thus an extension of this result to higher spin massless 
as well as massive states.

When both $D=26$ and $a=1$, one finds that the physical 
ground state is at zero excitation level, $N=0=\bar{N}$, and corresponds
to the Fock vacuum $|\Omega;p>$ such that $\alpha' m^2/2=-2$, hence a
tachyonic scalar particle. At the first excitation level $N=1=\bar{N}$,
one has a collection of strictly positive norm massless physical states,
one such state corresponding to a massless graviton with 299 independent
physical polarization components, another to a massless antisymmetric
tensor state of 276 components, and finally a scalar known as the dilaton
with a single polarization state, thus leading to a total of 576 positive norm
physical states. Likewise, it is possible to identify all such positive
norm physical states at higher excitation levels. Note that all physical
states lie along so-called Regge trajectories, namely linear trajectories
relating the $\alpha' m^2$ and spin values of these states, the string tension
$\alpha'$ indeed playing the role of the Regge slope, and the subtraction
constant $a=1$ that of the intercept for the lowest lying Regge trajectory.

It is quite remarkable that the spacetime physical spectrum
of the quantized closed bosonic string includes a massless spin 2 state,
the quanta of a metric field usually associated to gravitational interactions
in a field theory setting. Indeed, when considering the low energy effective
interactions (in comparison to the energy scale set by the string tension
$\alpha'$) of these states, their effective action is precisely that of
the low energy graviton modes of ge\-ne\-ral relativity expanded around 
Minkowski spacetime.\cite{Pol,GSW} It is thus perfectly consistent to 
identify these
closed string states with the gravitons of the gra\-vi\-ta\-tio\-nal 
interaction. Had we quantized the open bosonic string, in a likewise manner 
we would have uncovered massless spin 1 states whose low energy effective 
interactions are those of massless Yang-Mills gauge bosons! In other words, 
and this is indeed a generic feature of all string theories, it appears that 
the world-sheet symmetries, in the present instance those under world-sheet
reparametrizations, translate at the level of the spacetime spectrum into
the usual local gauge symmetries and their bosonic carriers of interactions
which have proved to provide the basic physics principle for a quantum
field theory description of all fundamental interactions and particles.
Undoubtedly, there is some profound lesson to be gathered from such a
nontrivial result. Many more such fascinating convergences of basic facts
have been uncovered within string theories, thus suggesting that this 
framework may well have brought us to the brink of the long sought-for 
formalism for a fundamental unification of all quantum interactions and 
matter. Only time will tell, through the work of the quantum geometers of 
the XXI$^{\rm st}$ century.

\subsection{Light-cone quantization}
\label{Subsect7.5}

In the light-cone gauge, the quantized system is defined by the commutation
relations
\begin{equation}
\begin{array}{r l}
\left[\sqrt{2\alpha'}q^-,P^+\right]=-i\ ,&\
\left[\sqrt{2\alpha'}q^i,P^j\right]=i\delta^{ij}\ ,\\
 & \\
\left[\alpha^i_n,\alpha^j_m\right]=n\delta^{ij}&\delta_{n+m,0}=
\left[\bar{\alpha}^i_n,\bar{\alpha}^j_m\right]\ ,
\end{array}
\end{equation}
including the by now usual hermiticity properties of these operators. 
Consequently,
the Fock space representation of this algebra is based on Fock vacua
$|\Omega;p^i,p^+>$ which are normalized eigenstates of the momentum operators
$P^+$ and $P^i$ and which are annihilated by the positive moded operators
$\alpha^i_{n\ge 1}$ and $\bar{\alpha}^i_{n\ge 1}$, the action of the creation
operators $\alpha^i_{-n}$ and $\bar{\alpha}^i_{-n}$, $n\ge 1$, spanning the
remainder of the Fock space basis. 

Given this algebra, it follows that all these quantum states are physical,
are of strictly positive norm, and that they correspond solely to
transverse string excitation modes. Finally, the mass spectrum of these
states is given by
\begin{equation}
\frac{1}{2}\alpha' M^2=N^\perp+\bar{N}^\perp-2a\ \ \ ,\ \ \ 
N^\perp=\bar{N}^\perp\ ,
\end{equation}
with the excitation level number operators $N^\perp$ and $\bar{N}^\perp$
defined as previously in terms of the transverse creation and annihilation
operators $\alpha^i_n$ and $\bar{\alpha}^i_n$ only. Here, $a$ stands
again for the required normal ordering subtraction constant that arises
for the transverse Virasoro zero modes $L^\perp_0$ and $\bar{L}^\perp_0$.
In the same manner as before, normal ordering is defined by bringing all
position and creation operators, $q^-$, $q^i$, $\alpha^i_{-n}$
and $\bar{\alpha}^i_{-n}$, to the left of all momentum and annihilation 
ones, $P^+$, $P^i$ and $\alpha^i_n$, $\bar{\alpha}^i_n$, $n\ge 1$.

Clearly, this characterization of physical states coincides with that
reached within the above conformal gauge quantization for those physical
states of strictly positive norm. However, in contradistinction with the
latter quantization, the present one no longer possesses a manifest Poincar\'e
covariant formulation, and one is forced to check whether the Poincar\'e
algebra is rea\-li\-zed nonetheless on the space of quantum states, albeit in
a nonlinear fashion. Since only the little group E($D-2$)$_n$ of the 
light-like vector $n^\mu$ used to define the light-cone gauge is still a 
manifest spacetime symmetry of this quantization, one needs to check whether 
the commutation relations which involve the operators $M^{-i}$ and $M^{0(D-1)}$
take the values required by the Poincar\'e algebra. In turns out that it is
only for the commutators $[M^{-i},M^{-j}]=0$ that this requirement is
not necessarily met, leading to the critical values\cite{Pol,GSW,JG10}
\begin{equation}
D-2=24\ \ \ ,\ \ \ a=1\ ,
\end{equation}
again in agreement with the critical values required in the conformal
gauge for a consistent manifestly Poincar\'e covariant quantization of
the bosonic string.

Even though this result is by no means trivial to establish, once
again it is possible to show that it is necessary by working out the
first few excitation levels of the system. Consider thus the states
at $N^\perp=1=\bar{N}^\perp$, namely
\begin{equation}
\alpha^i_{-1}\,\bar{\alpha}^j_{-1}\,|\Omega;p^+,p^i>\ ,
\end{equation}
whose mass is such that $\alpha' m^2/2=2(1-a)$. There are thus $(D-2)^2$
such states. However, spacetime covariant properties of this collection
of states are obtained only if they are massless, since otherwise they
should belong to some spin representation of SO($D-1$) rather than SO($D-2$)
which clearly is impossible. Consequently, one must have $a=1$, implying 
that the physical ground state is tachyonic. Furthermore, an heuristic 
$\zeta$-function evaluation of the infinite series defining the normal 
ordering constant $a$ then also leads to the value $D-2=24$. 

Given these critical values, it then follows that the above states at level 
$N^\perp=1=\bar{N}^\perp$ correspond to a massless spin 2 graviton with 
299 polarization states, an antisymmetric tensor with 276 states and a 
scalar with a single polarization state, hence a total of $576=24^2$ 
physical states, in complete agreement with the count in the conformal gauge.

As a matter of fact, it is possible to introduce the partition function
of the system which counts the number of positive norm physical states
at each excitation level, and study further fascinating properties
of this simple string theory. However, we shall refrain from presenting
these considerations here.

As a conclusion concerning the quantization of the closed bosonic string,
let us point out that these conformal and light-cone gauge fixing
procedures are affected by Gribov problems,\cite{JG10,JG11} which, however, 
may be circumvented in the actual construction of physical amplitudes. 
Furthermore, when including then the proper Faddeev-Popov or BFV Hamiltonian 
ghost systems following from these gauge fixings, one may check that the
total conformal algebra is recovered at the quantum level
provided once again the critical conditions $D=26$ and $a=$ are 
imposed.\cite{Pol,GSW,JG10} Indeed,
it is reparametrization and conformal invariance which guarantees the
quantum consistency of the system, so that the conformal algebra should
better not be affected by an anomalous central extension contribution
when properly including all relevant degrees of freedom. In that respect,
it would certainly be quite interesting to apply the physical projector
approach free of any gauge fixing procedure to, first, the bosonic string,
and then to all its supersymmetric cousins in all dimensions ranging
from the critical $D=10$ one down to $D=4$ or even $D=2$.

\section{Toroidal Compactification of the Closed Bosonic String}
\label{Sect8}

As the above discussion has established, the closed bosonic spectrum
includes a massless spin 2 graviton. In fact, if one ignores the problem
raised by the presence of a tachyonic state which may be projected out for
theories of physical relevance, it has been established\cite{Pol,GSW} that 
the perturbative expansion of string theory, when properly renormalized, is 
indeed a finite one. In other words, string theory defines a perturbative 
finite quantum theory for quantum gravity coupled to other interactions and 
matter degrees of freedom, including dynamical geometric degrees of freedom. 
This rather remarkable conclusion has been one of the strongest motivations 
to pursue this framework as a possible formulation for the problem of the 
ultimate unification.

Given this fact and the suggestion made previously that topology is also
called to play a fundamental role in a physically relevant and consistent
formulation of quantum geometry, to conclude these notes let us discuss
the simplest example which shows that our usual geometrical concepts have
to be extended when considered from the string theory point of view.
Some em\-bo\-die\-ment of the sought-for principles of quantum geometry must
already be lying hidden behind the properties of these theories.
For this purpose, this section briefly considers how the previous
discussion of the closed bosonic string is modified even when only one
of the spatial dimensions is compactified into a circle geometry.

\subsection{Toroidal compactification in field theory}
\label{Subsect8.1}

To begin with, let us consider a simple free massless scalar field 
$\phi(x^\mu,y)$ evolving over a spacetime manifold of which one of the 
spatial dimensions is compactified into a circle of radius $R$,
\begin{equation}
y\ \equiv y+2\pi R\ ,
\end{equation}
while the remaining spacetime dimensions define a Minkowski spacetime of
some given dimension. Since the massless Klein-Gordon equation reads
\begin{equation}
\left[\partial^2_x+\partial^2_y\right]\phi(x^\mu,y)=0\ ,
\end{equation}
a Fourier mode expansion of the field over the compactified direction,
\begin{equation}
\phi(x^\mu,y)=\sum_n\phi_n(x^\mu)\,e^{i\frac{n}{R}y}\ ,
\end{equation}
implies that from the lower dimensional point of view the modes 
$\phi_n(x^\mu)$obey the massive Klein-Gordon equation
\begin{equation}
\left[\partial^2_x-m^2_n\right]\phi_n(x^\mu)=0\ \ ,\ \ 
m_n=\frac{|n|}{R}\ .
\end{equation}
Consequently, upon compactification of a field theory onto a compactified
space, in the lower dimensional description there appear infinite
so-called Kaluza-Klein (KK) towers of massive states whose mass values are
determined by the momentum values of the field along the compactified
directions. In the decompactification limit $R\rightarrow\infty$, these
massive KK towers coalesce back into the massless modes of the initial
massless field in the complete spacetime which recovers its Minkowski
geometry. {\sl A contrario\/}, in the compactification limit $R\rightarrow 0$,
the KK towers become infinitely massive, and thus decouple altogether
from the lower dimension dynamics, since the extra dimension then collapses
to a point in that limit.

These features of dimensional compactification are generic to any field
theory. For instance given some vector field $A_M=(A_\mu,\phi)$ in the higher 
dimensional theory, the compactified theory will include vector, $A_\mu$, 
and scalar, $\phi$, components each of which possesses a similar KK mode
expansion in terms of massless and massive states associated to the
momentum values of the original fields along the compactified directions.
The same clearly applies to a two-index tensor $A_{MN}$, leading
to KK towers of tensor, vector and scalar states in the lower dimensional
theory, $A_{\mu\nu}$, $A^{(1)}_\mu$, $A^{(2)}_\mu$ and $\phi$.

Furthermore, within the context of a pure general relativity theory in the
higher dimensional spacetime, which is thus invariant under arbitrary 
coordinate transformations in that space, whenever the compactified space 
possesses some continuous symmetry, as for example the U(1) symmetry of a 
circle or the SO($n$) symmetry of a sphere, the KK reduction implies the 
appearance of massless gauge bosons associated to the corresponding local 
internal gauge symmetries. Indeed, one is then free to perform at each
lower dimensional spacetime point a different coordinate redefinition
within the compactified space, without changing the physics of the system.
In other words, there does appear an internal local Yang-Mills gauge symmetry,
the internal space being nothing but the compactified directions of the
initial theory. This is the basis of the original Kaluza-Klein programme
for the fundamental unification of gravity with all other Yang-Mills
interactions within a purely geometric and field theory framework. Given that
consistent string theories need to be formulated in higher dimensional
spacetimes, this programme recovers suddenly new relevance and
urgency within this new formalism which supersedes that of ordinary
field theory, inclusive of quantum dynamics.

\subsection{Toroidal compactification in string theory}
\label{Subsect8.2}

Given the large critical spacetime dimensions for consistent quantum
string theories, spatial compactification is a natural approach towards
the four-dimensional physical world. Let us thus reconsider our previous
discussion of the closed bosonic string, but this time with one the
space components, say $\mu=25$, compactified into a circle of radius 
$R$.\cite{Pol,GSW} Correspondingly, in the conformal or light-cone gauge, 
one has the mode expansion
\begin{equation}
\begin{array}{r l}
x^{25}(\tau,\sigma)=\sqrt{2\alpha'}\left[\right.&q^{25}+
\alpha^{25}_0(\tau-\sigma)+\bar{\alpha}^{25}_0(\tau+\sigma)\\
 & \\
&\left. +\frac{1}{2}i{\sum_n}'\frac{1}{n}\left(\alpha^{25}_n
e^{-2in(\tau-\sigma)}+\bar{\alpha}^{25}_ne^{-2in(\tau+\sigma)}\right)\right]\ .
\end{array}
\end{equation}
However, given the possibility of winding configurations 
\begin{equation}
x^{25}(\tau,\sigma+\pi)=x^{25}(\tau,\sigma)+2\pi R m\ ,
\end{equation}
of integer winding number $m$ around the compactified circle, the zero modes
$\alpha^{25}_0$ and $\bar{\alpha}^{25}_0$ need no longer be equal, but they
actually differ by a multiple of the winding number. Given that the
quantum momentum operator $P^{25}$ is then also quantized as $n/R$, $n$
being the momentum quantum, the zero modes take the values
\begin{equation}
\alpha^{25}_0=\frac{1}{2}\sqrt{2\alpha'}
\left[\frac{n}{R}-\frac{R}{\alpha'}m\right]\ \ ,\ \ 
\bar{\alpha}^{25}_0=\frac{1}{2}\sqrt{2\alpha'}
\left[\frac{n}{R}+\frac{R}{\alpha'}m\right]\ .
\end{equation}
The mass spectrum and level matching conditions are modified accordingly,
\begin{equation}
\frac{1}{2}\alpha' M^2=N+\bar{N}+\frac{1}{2}
\left[\left(\frac{\sqrt{\alpha'}}{R}n\right)^2+
\left(\frac{R}{\sqrt{\alpha'}}m\right)^2\right]-2\ \ ,\ \ 
N-\bar{N}=nm\ ,
\end{equation}
while the other Virasoro conditions defining quantum physical states remain
unaffected since only the zero modes are modified by the winding state
contributions.

In the $(n=0,m=0)$ sector, the lowest lying physical state is
the ta\-chyo\-nic ground state $|\Omega;p>$ at level $N=0=\bar{N}$.
At level $N=1=\bar{N}$, one finds again the massless symmetric graviton,
antisymmetric tensor and scalar for the strictly positive norm
physical states obtained from
\begin{equation}
\alpha^\mu_{-1}\bar{\alpha}^\nu_{-1}\,|\Omega;p>\ ,
\end{equation}
but new massless vector and scalar states also appear, namely those
related to
\begin{equation}
\alpha^\mu_{-1}\bar{\alpha}^{25}_{-1}\,|\Omega;p>\ \ ,\ \ 
\alpha^{25}_{-1}\bar{\alpha}^\mu_{-1}\,|\Omega;p>\ \ ,\ \ 
\alpha^{25}_{-1}\bar{\alpha}^{25}_{-1}\,|\Omega;p>\ ,
\end{equation}
which are consequence of the circle compactification
(as is most straightforwardly established in the light-cone gauge).
These are the massless KK states that follow the compactification
of the original symmetric, antisymmetric and dilaton degrees of freedom, in 
exactly the same manner as discussed above in the case of the compactification
of such fields from higher dimensions. The fact that these states have to 
be massless is consequence of the U(1) symmetry of the circle
compactification, which translates into the U(1)$\times$U(1) local Yang-Mills
gauge invariance for the compactified theory, since the higher dimensional
string theory includes a gravitational sector. This conclusion is in full
accord with the Kaluza-Klein programme briefly outlined above.

Besides these states, there also exist the towers of KK states with $n\ne 0$,
as well as the sector of winding states $m\ne 0$, the latter being a feature
totally specific to the compactification of closed string theories, and
totally absent from the field theory discussion in which point-like rather
than string-like objects are being described in a quantized formulation.
In particular, the contribution of winding states to the mass spectrum
is a measure of the energy required to stretch and wind a closed string
around the compactified dimension.

In fact, this is not the only difference with the field theory case.
In the decompactification limit $R\rightarrow\infty$, the towers
of KK states do indeed coalesce back into the continuous spectrum of
momentum eigenstates for the com\-pac\-ti\-fied direction. However at the same
time the winding states $m\ne 0$ become infinitely massive and decouple
from the theory. {\sl A contrario\/}, in the com\-pac\-ti\-fi\-ca\-tion limit
$R\rightarrow 0$, the towers of KK states become infinitely massive as
they do in the field theory case and thus decouple, but this time the
winding states $m\ne 0$ coalesce into a continuum of states with a vanishing
contribution to the mass spectrum. In other words, even when the compactified
dimension has degenerated into a single point, there is still a trace of
that extra dimension in the spectrum of the compactified theory. The notion
of the dimension of spacetime appears to be no longer an absolute geometrical 
nor topological concept within the context of string theory.

This interchange of the decompactification and compactification limits
is in fact valid even for whatever finite value of the circle radius $R$.
Indeed, given the above expressions for the mass spectrum, it is clear that
this spectrum, and in fact the whole of the string dynamics, remains totally
invariant under the transformation
\begin{equation}
n\leftrightarrow m\ \ \ ,\ \ \ 
\frac{\sqrt{\alpha'}}{R}\leftrightarrow\frac{R}{\sqrt{\alpha'}}\ ,
\end{equation}
in which momenta and winding states are exchange as well as the values
of the radius with the value $\alpha'/R$. This symmetry is known as
T-duality and again is totally specific to string theories.\cite{Tdual} 
In fact, this is quite a fascinating symmetry, since it exchanges small 
distance with large distance physics, and calls into question our usual 
concepts of local geometry and topology. Note that this a symmetry of the 
quantized theory, and should thus be one of the expected manifestations of a
quantum geometry of spacetime, with its accompanying quantum gravitational
interactions as fluctuations in the quantum geometry. If not the topology,
the compactified geometry is modified under T-duality without any consequence
whatsoever for the quantum dynamics of string theory.

The existence of T-duality also suggests that within quantum geometry
there ought to exist some smallest distance scale beyond which physical
process may no longer be probed, being equivalent then to physical processes
at the inverse distance scale. This smallest distance scale is thus set by 
the self-dual point under T-duality, associated to the compactification radius
\begin{equation}
R=\sqrt{\alpha'}\ .
\end{equation}
In fact at the self-dual point, the U(1)$\times$U(1) gauge symmetry of the
com\-pac\-ti\-fied theory for a generic value of $R$ is enhanced to a 
SU(2)$\times$SU(2) Yang-Mills symmetry, whose rank is still that of the
generic U(1)$\times$U(1) Yang-Mills symmetry. For instance at the massless 
level, the self-dual value implies the following extra states
\begin{equation}
\begin{array}{r|r|r|r}
n & m & N & \bar{N} \\
\hline
1 & 1 & 1 & 0 \\
1 & -1 & 0 & 1 \\
-1 & 1 & 0 & 1 \\
-1 & -1 & 1 & 0 \\
2 & 0 & 0 & 0 \\
-2 & 0 & 0 & 0 \\
0 & 2 & 0 & 0 \\
0 & -2 & 0 & 0 
\end{array}
\end{equation}
The first four collections of states include four massless vector ones,
which together with the two massless vectors of the generic case, combine into
the six massless gauge bosons of the SU(2)$\times$SU(2) Yang-Mills symmetry
in 25-dimensional Minkowski spacetime. Likewise, the first four collections
of states include also four massless scalars as do the last four collections,
leading to a total of nine massless scalars when the generic massless scalar
is also accounted for. These scalar states fit into the $(3,3)$ Higgs
representations under SU(2)$\times$SU(2). In other words, at 
$R=\sqrt{\alpha'}$, the system possesses the SU(2)$\times$SU(2) Yang-Mills 
symmetry which gets hidden by spontaneous Higgs symmetry breaking whenever 
the compactification radius $R$ takes a different value.\cite{Pol,GSW}

Hence, this simplest example of spatial compactification of string
theory already points to quite fascinating quantum geometric properties
realized within the realm of quantized string theory, which presumably
are nothing but some facets of what a formulation and understanding of 
quantum geometry has to offer with regards to the hidden secrets for
the physics of quantum gravity unified with all other quantum interactions 
and particles.

\section{Conclusions}
\label{Sect9}

The principle aim of these notes has been to provide a brief outline,
restricted to bosonic degrees of freedom only, of the
relativistic and quantum concepts that are at the basis of our present
understanding of all fundamental quantum interactions and elementary
particles. The general considerations that have led during the XX$^{\rm th}$
century to the identification of relativistic quantum Yang-Mills gauge field
theories as the appropriate framework for a consistent causal and
quantum unitary description of relativistic quantum point-particles
and their interactions have been recalled. The same convergence of
ideas centered onto the fundamental concept of the local gauge
symmetry principle applies to the gravitational interaction, which,
when des\-cri\-bed within general relativity and its extensions all based
on the dynamics of the geometry of spacetime, has been successful so far
only at the classical level, while a full-fledged theory for quantum
gravity is still eluding us. It appears that the physicist of the 
XXI$^{\rm st}$ century has arrived at the cross-roads of the three fundamental
paths that have guided him during the previous one, and which may be
characterized in terms of the three fundamental constants $c$, $\hbar$ and
$G_N$. It seems that in spite of the amazing successes of the marriage
of $c$ with $\hbar$, it is close to impossible to force these sets of ideas 
to happily live within a {\sl m\'enage \`a trois\/}. Some new paradigm
of geometrical and topological concepts is most probably called for within
the realm of the quantum gravitational interaction coupled to all other
quantum interactions and particles.

As another but complementary aim of these notes, the general issues surrounding
the quantization of constrained systems, which include all possible gauge
invariant theories based on a field theory formulation, have been described,
providing the basic tools necessary for such a study in general. In particular,
having shown that the potential difficulties which follow from gauge fixing
procedures for such theories are often unavoidable, an alternative and recent
approach based on a physical projector\cite{Klaud1} onto the gauge invariant 
quantum configurations of such systems and free of the necessity of gauge 
fixing, has been advocated as a powerful new tool with which to address these 
difficult issues, especially with regards to nonperturbative aspects of 
strongly interacting Yang-Mills theories.

Yang-Mills, and more generally local gauge invariant theories have also
shown that topological features, either of spacetime or of the field
configuration space, do play a fundamental role in the proper understanding
of such interactions. With the discovery of topological quantum field 
theories,\cite{Wit1,Wit2,Wit3} void of any genuine dynamics but not of any 
quantum physics nonetheless, it is conceivable that pure quantum gravity 
could be the physics of quantum to\-po\-lo\-gy rather than of spacetime 
geometry, and that it is by coupling quantum topology to matter and 
interactions that the quantum geometric properties of spacetime should arise,
local relativistic quantum field theories with gauge invariances being
their appropriate low energy effective description.

As one illustration among possibly many others that have not been
discussed at the Workshop, some of these issues have briefly been
touched on within the context of bosonic string theory. Specific 
fascinating new features having to do with the gravitational sector
of such systems and its interplay with the geometry and topology of spacetime
have been described in the simplest terms available. Many more such
issues have arising within that context, such as for example the possible
noncommutative character of spacetime itself within string theory.\cite{SW}

It is equipped with this understanding of the world of the fundamental
quantum interactions and particles, and the role played by topology
within the relativistic gauge invariant quantum field theoretic framework 
describing this world today, that the physicist of the XXI$^{\rm st}$ century
in quest of the ultimate unification is to set out into the unchartered
territory towards a truly genuine formulation and understanding of what
quantum geometry will turn out to be, the final unification of the 
relativistic quantum and the relativistic continuum, the completed 
symphony of the three constants $c$, $\hbar$ and $G_N$ which have guided us 
already through the three fundamental conceptual revolutions of 
XX$^{\rm th}$ century physics.

\section*{Acknowledgements}

Prof. John R. Klauder is gratefully acknowledged for his many relevant
comments concerning the contents of these lectures, and for his supportive
interest over the years in applications of the physical projector. Thanks 
also go out to Prof. M.N. Hounkonnou for his insistence to give these
lectures, and to all participants to the Workshop for their active 
participation, questions and discussions throughout the presentation of 
this material. 

It is hoped that this write-up will entice them to raise ever 
more in\-qui\-si\-ti\-ve issues, embark onto their own adventures into the 
quantum geometer's world of XXI$^{\rm st}$ century physics, and 
contribute with the mathematics and physics world community to the
completion of this unfinished symphony by uncovering with a definite african
beat some of its music scores, the full colours of its harmonies 
being Nature's own.


\clearpage

\begin{thebibliography}{99}

\bibitem{Wein1} S. Weinberg, {\sl What is Quantum Field Theory, and What
Did We Think It Is?\/}, Talk given at the Conference on the Historical 
Examination and Philosophical Reflections on the Foundations of Quantum 
Field Theory, Boston (Massachusetts, USA)  March 1-3, 1996, 
{\tt hep-th/9702027}.

\bibitem{Pol} For a review and references to the original literature, see,\\
J. Polchinski, {\sl String Theory\/} (Cambridge University Press,
Cambridge, 2000), 2 Volumes.

\bibitem{GSW} For a review and references to the original literature, see,\\
M.B. Green, J.H. Schwarz and E. Witten, {\sl Superstring Theory\/}
(Cambridge University Press, Cambridge, 1987), 2 Volumes.

\bibitem{Loop} For a review and references to the original literature, see for
example,\\
C. Rovelli, {\em Living Rev. Rel.\/}, 1 (1998), {\tt gr-qc/9710008}.

\bibitem{Wit1} E. Witten, {\em Comm. Math. Phys.\/} {\bf 117} , 353 (1988);\\
E. Witten, {\em Phys. Lett.\/} {\bf B206}, 601 (1988);\\
E. Witten, {\em Comm. Math. Phys.\/} {\bf 118}, 411 (1988);\\
J.M.F. Labastida, M. Pernici and E. Witten, {\em Nucl. Phys.\/} {\bf B310}, 
611 (1988).

\bibitem{Wit2} E. Witten, {\em Comm. Math. Phys.\/} {\bf 121}, 351 (1989).

\bibitem{TQFT} For a review and references to the original literature, see
for example,\\
D. Birmingham, M. Blau, M. Rakowski and G. Thompson, {\em Physics Reports\/}
{\bf 209}, 129 (1991).

\bibitem{Dirac} P.A.M. Dirac, {\sl Lectures on Quantum Mechanics\/} (Belfer
Graduate School of Science, Yeshiva University, New York, 1964).

\bibitem{Klaud1}
J.R. Klauder, {\em Ann. Phys.\/} {\bf 254}, 419 (1997);\\
J.R. Klauder, {\em Nucl. Phys.\/} {\bf B547}, 397 (1999);\\
J.R. Klauder, {\sl Quantization of Constrained Systems\/},
{\sl Lect. Notes Phys.\/} {\bf 572}, 143 (2001), {\tt hep-th/0003297}.

\bibitem{JG1} For a detailed discussion and references to the
original literature, see,\\
J. Govaerts, {\sl Hamiltonian Quantisation and Constrained Dynamics\/}
(Leuven University Press, Leuven, 1991).

\bibitem{Klaud2} J.R. Klauder, {\sl Beyond Conventional Quantization\/}
(Cambridge University Press, Cambridge, 2000).

\bibitem{JG2} J. Govaerts and V.M. Villanueva, {\em Int. J. Mod. Phys.\/} 
{\bf A15}, 4903 (2000).

\bibitem{JG2bis} J. Govaerts, Proc. First International
Workshop on Contemporary Pro\-blems in Mathematical Physics, 
31 October-5 November
1999, eds. J.~Govaerts, M.N.~Hounkonnou and W.A.~Lester, Jr. (World Scientific,
Singapore, 2000), pp. 244-259.

\bibitem{Klaud3}
J.R. Klauder and B.-S. Skagerstam, {\sl Coherent States: Applications in
Physics and Mathe\-ma\-ti\-cal Physics\/} (World Scientific, Singapore, 1985).

\bibitem{Feynman} R.P. Feynman and A.R. Hibbs, {\sl Quantum Mechanics and 
Path Integrals\/} (McGraw-Hill Book Company, New York, 1965).

\bibitem{CCR} For a discussion and references, see for example,\\
M. Florig and S.J. Summers, {\em Proc. London Math. Soc.\/} {\bf 80}, 451 (2000),
{\tt math-ph/0006011};\\
G. Sardanashvily, {\sl Nonequivalent Representations of Nuclear Algebra of
Canonical Commutation Relations. Quantum Fields\/}, {\tt hep-th/0202038};\\
A. Corichi, J. Cortez and H. Quevedo, {\sl On the Relation between Fock and
Schr\"odinger Representations for a Scalar Field\/}, {\tt hep-th/0202070};\\
A. Iorio, G. Lambiase and G. Vitiello, {\sl Hopf Algebra, Thermodynamics
and Entanglement in Quantum Field Theory\/}, {\tt quant-ph/0207040}.

\bibitem{Wein2} S. Weinberg, {\sl The Quantum Theory of Fields\/}
(Cambridge University Press, Cambridge, 1995), 3 Volumes.

\bibitem{PS} M.E. Peskin and D.V. Schroeder, {\sl An Introduction to
Quantum Field Theory\/} (Perseus Books Publishing, Cambridge, Massachusetts,
1995).

\bibitem{IZ} C. Itzykson and J.-B. Zuber, {\sl Quantum Field Theory\/}
(McGraw-Hill Book Company, New York, 1980).

\bibitem{Ramond} P. Ramond, {\sl Field Theory: a Modern Primer\/}
(Benjamin-Cummings Pu\-bli\-shing, Reading, Massachusetts, 1981).

\bibitem{HV}
M.J.G. Veltman, {\em Nucl. Phys.\/} {\bf B7}, 637 (1968);\\
G. 't Hooft, {\em Nucl. Phys.\/} {\bf B35}, 167 (1971 );\\
G. 't Hooft and M.J.G. Veltman, {\em Nucl. Phys.\/} {\bf B44}, 189 (1972 );\\
G. 't Hooft and M.J.G. Veltman, {\em Nucl. Phys.\/} {\bf B50}, 318 (1972 ).

\bibitem{FJ} L. Faddeev and R. Jackiw, {\em Phys. Rev. Lett.\/} {\bf 60}, 
1692 (1988);\\
J. Govaerts, {\em Int. J. Mod. Phys.\/} {\bf A5}, 3625 (1990).

\bibitem{JG3}
J. Govaerts and J.R. Klauder, {\em Ann. Phys.\/} {\bf 274}, 251 (1999).

\bibitem{Fad}
 L.D. Faddeev, {\em Theor. Math. Phys.\/} {\bf 1}, 1 (1970).

\bibitem{Gribov}
V.N. Gribov, {\em Nucl. Phys.\/} {\bf B139}, 1 (1978);\\
I.M. Singer, {\em Comm. Math. Phys.\/} {\bf 60}, 7 (1978).

\bibitem{Gribov2}
K. Fujikawa, {\em Prog. Theor. Phys.\/} {\bf 61}, 627 (1979);\\
P. Hirschfeld, {\em Nucl. Phys.\/} {\bf B157}, 37 (1979);\\
M.B. Halpern and J. Koplok, {\sl Nucl. Phys.\/} {\bf B132}, 239 (1978).

\bibitem{JG4}
J. Govaerts, {\em Int. J. Mod. Phys.} {\bf A4}, 173 (1989);\\
J. Govaerts, {\em Int. J. Mod. Phys.} {\bf A4}, 4487 (1989);\\
J. Govaerts and W. Troost, {\em Class. Quantum Grav.} {\bf 8}, 1723 (1991).

\bibitem{JG5} J. Govaerts, {\sl The Cosmological Constant of One-Dimensional 
Matter Coupled Quantum Gravity is Quantized\/}, 
preprint STIAS-02-002, {\tt hep-th/0202134}.

\bibitem{BFV1}
E.S. Fradkin and G.A. Vilkovisky, {\em Phys. Lett.\/} {\bf B55}, 224 (1975);\\
I.A. Batalin and G.A. Vilkovisky, {\em Phys. Lett.\/} {\bf B69}, 309 (1977);\\
E.S. Fradkin and T.E. Fradkina, {\sl Phys. Lett.\/} {\bf B72}, 343 (1978) 343;\\
I.A. Batalin and E.S. Fradkin, {\sl Rivista Nuovo Cimento\/} {\bf 9}, 1 (1986).

\bibitem{JG6} J. Govaerts, {\em J. Phys.\/} {\bf A30}, 603 (1997).

\bibitem{Klaud4}
J.R. Klauder, {\em J. Math. Phys.\/} {\bf 40}, 5860 (1999);\\
G. Watson and J.R. Klauder, {\em J. Math. Phys.\/} {\bf 41}, 8072 (2000);\\
J.R. Klauder, {\em J. Math. Phys.\/} {\bf 42}, 4440 (2001);\\
J.R. Klauder, {\em Class. Quant. Grav.\/} {\bf 19}, 817 (2002);\\
G. Watson and J.R. Klauder, {\em Class. Quant. Grav.\/} {\bf 19}, 3617 (2002).

\bibitem{JG7}
V.M. Villanueva, J. Govaerts and J.-L. Lucio-Martinez, {\em J. Phys.\/}
{\bf A33}, 4183 (2000).

\bibitem{JG8} G.Y.H. Avossevou and J. Govaerts, {\sl The Schwinger Model
and the Physical Projector: a Nonperturbative Quantization without Gauge
Fixing\/}, contribution to these Proceedings.

\bibitem{JG9}
J. Govaerts and B. Deschepper, {\em J. Phys.\/} {\bf A33}, 1031 (2000).

\bibitem{Wit3}
E. Witten, {\em Nucl. Phys.\/} {\bf B311}, 46 (1988);\\
J.H. Horne and E. Witten, {\em Phys. Rev. Lett.\/} {\bf 62}, 501 (1989);\\
E. Witten, {\em Nucl. Phys.\/} {\bf B323}, 113 (1989).

\bibitem{Goldin} See for example, and references therein,\\
G.A. Goldin and D.H. Sharp, {\em Phys. Rev. Lett.\/} {\bf 76}, 1183 (1996).

\bibitem{JG10}
J. Govaerts, Proc. $2^{\rm nd}$ Mexican School of Particles and 
Fields, Cuernavaca - Morelos (Mexico), 4-12 December 1986,
eds. J.-L. Lucio and A.~Zepeda (World Scientific, Singapore, 1987), pp. 247-442;\\
J. Govaerts, {\sl Nuclear Physics B (Proc. Suppl.)} {\bf 11}, 186 (1989);\\
J. Govaerts, Proc. Carg\`ese Summer Institute 1989 
``Particle Physics", Carg\`ese (Corsica, France), 18 July-4 August 1989,
eds. M. Levy, D.~Speiser, J.-L. Basdevant, R. Gastmans,
M. Jacob and J. Weyers (Plenum, New York, 1990), pp. 141-216.

\bibitem{JG11}
J. Govaerts, {\em Int. J. Mod. Phys.} {\bf A4}, 173 (1989).

\bibitem{Tdual} For a review and references to the original literature, see
for example,\\
A. Giveon, M. Porrati and E. Rabinovici, {\em Physics Reports\/} {\bf 244}, 
77 (1994).

\bibitem{SW}
N. Seiberg and E. Witten, {\em JHEP\/} {\bf 9909}, 032 (1999).

\end{thebibliography}
\end{document}